\documentclass{emulateapj}
\usepackage{apjfonts}
\usepackage{epsfig}
\usepackage{amsmath, amsthm, amssymb}

\newcommand{\beq}{\begin{equation}}
\newcommand{\eeq}{\end{equation}}
\newcommand{\bea}{\begin{eqnarray}}
\newcommand{\eea}{\end{eqnarray}}
\newcommand{\Msun}{$M_\sun$}
\providecommand{\abs}[1]{\lvert#1\rvert}

\slugcomment{Draft version - \today}
\begin{document}

\title{Aspherical Supernova Shock Breakout and the Observations of Supernova 2008D}
\author{Sean M. Couch\altaffilmark{1,2,3,4},  David Pooley\altaffilmark{5}, J. Craig Wheeler\altaffilmark{1},  and Milo\v s Milosavljevi\'c\altaffilmark{1,6} }

\altaffiltext{1}{Department of Astronomy, The University of Texas, Austin, TX, 78712.}
\altaffiltext{2}{Center for Astrophysical Thermonuclear Flashes, University of Chicago, Chicago, IL, 60637}
\altaffiltext{3}{Department of Astronomy \& Astrophysics, University of Chicago, Chicago, IL, 60637}
\altaffiltext{4}{NASA Earth and Space Science Fellow}
\altaffiltext{5}{Eureka Scientific, Inc., Austin, TX 78756}
\altaffiltext{6}{Texas Cosmology Center, The University of Texas, Austin, TX, 78712}

\righthead{ASPHERICAL SHOCK BREAKOUT}
\lefthead{COUCH ET AL.}

\begin{abstract}
Shock breakout is the earliest, readily-observable emission from a core-collapse supernova explosion.  Observing supernova shock breakout may yield information about the nature of the supernova shock prior to exiting the progenitor and, in turn, about the core-collapse supernova mechanism itself.  X-ray Outburst 080109, later associated with SN 2008D, is a very well-observed example of shock breakout from a core-collapse supernova.  Despite excellent observational coverage and detailed modeling, fundamental information about the shock breakout, such as the radius of breakout and driver of the light curve time scale, is still uncertain.  The models constructed for explaining the shock breakout emission from SN 2008D all assume spherical symmetry.  We present a study of the observational characteristics of {\it aspherical}  shock breakout from stripped-envelope core-collapse supernovae surrounded by a wind.  We conduct two-dimensional, jet-driven supernova simulations from stripped-envelope progenitors and calculate the resulting shock breakout X-ray spectra and light curves. The X-ray spectra evolve significantly in time as the shocks expand outward and are not well-fit by single-temperature and radius black bodies.  The time scale of the X-ray burst light curve of the shock breakout is related to the shock crossing time of the progenitor, not the much shorter light crossing time that sets the light curve time scale in spherical breakouts.  This could explain the long shock breakout light curve time scale observed for XRO 080109/SN 2008D.  We also comment on the distribution of intermediate mass elements in asymmetric explosions.


\end{abstract}

\section{Introduction}
\label{sec:intro}
\setcounter{footnote}{0}

The serendipitous discovery of XRO 080109, associated with SN 2008D, on 9 January 2008 \citep{Berger:08, Kong:08,Soderberg:08} has allowed us to view a stripped-envelope core-collapse supernova (SN) from its earliest stages, at or near the moment of shock breakout from the progenitor star.  The radiation burst associated with shock breakout is the first electromagnetic indicator of a supernova explosion \citep{Colgate:68,Colgate:74}.  Shock breakout occurs when radiation trapped in the vicinity of the supernova shock is able to escape ahead of the shock \citep{Klein:78, Ensman:92, Matzner:99}.  When this happens, the shock transitions from a radiation-mediated shock to a hydrodynamic shock \citep{Katz:09}.  SN 2008D was a normal Type Ib supernova \citep{Modjaz:09}, indicating a compact progenitor lacking a significant hydrogen envelope.  Shock breakout emission from such compact progenitors may retain more information about the nature and shape of the supernova driving mechanism than breakouts from larger progenitors with intact envelopes \citep[see, e.g.,][]{Couch:09}. 

The discovery of X-ray outburst (XRO) 080109 is described by \citet{Soderberg:08}.  The burst lasted about 500 seconds and reached a peak {\it Swift} XRT count rate of about 7 counts s$^{-1}$.  Based on a Comptonized, non-thermal emission model, Soderberg et al. conclude that the origin of the XRO is shock breakout at a radius of about $7\times10^{11}$ cm.  This radius is larger than that of the typical Wolf-Rayet star, and Soderberg et al. argue that this indicates the need for an optically-thick wind around the supernova progenitor.  Radio observations of SN 2008D, however, imply a wind mass-loss rate too low for the wind to be optically thick at a radius of $7\times10^{11}$ cm \citep{Soderberg:08, Chevalier:08}.  These conclusions were drawn assuming spherical symmetry and raise questions about the actual radius of shock breakout.

The X-ray spectrum of XRO 080109 can be fit reasonably well with a power law, a black body, or a combination of the two \citep{Modjaz:09}.  It is thus reasonable to consider both thermal and non-thermal sources of emission in attempting to explain the outburst.  \citet{Chevalier:08} posit that a thermal source with a black-body spectrum is plausible within the uncertainties of the observations.  \citet{WangX:08} argue, however, that bulk-Comptonization will scatter thermal photons to higher energies creating a power-law spectrum.  Although both thermal and non-thermal emission sources may be able to explain the shape of the spectrum, neither can account for the timescale of the X-ray light curve in a spherically-symmetric geometry.  The characteristic time for both emission types, measured as the full-width at half-maximum (FWHM) of the observed light curve is the light-crossing time of the progenitor star, $R / c$, assuming a spherically-symmetric shock breakout.  The FWHM of the XRO, about 100 s \citep{Soderberg:08, Modjaz:09}, indicates a spherical breakout radius of $\sim10^{12}$ cm, greater than any plausible progenitor radius and well above where the progenitor wind could be optically-thick.  This is an additional contradiction that is difficult to explain with a spherical shock breakout model.

In this work, we describe the characteristics of aspherical supernova shock breakout and compare these characteristics to the observations of XRO 080109/SN 2008D.  Compounding observational evidence indicates that core-collapse supernovae, especially those involving envelope-stripped progenitors (Type Ib/c), are not spherical \citep[see, e.g.,][]{Wang:08}.  In the particular case of SN 2008D, polarization measurements show that the explosion is not spherical, with dramatic asymmetries in the structure of some line-forming regions \citep{Maund:09, Gorosabel:08}. Theoretically, current models for the explosion mechanism of core-collapse SNe produce inherently aspherical shock waves \citep{Wheeler:00, Wheeler:02, Blondin:03, Burrows:06, Obergaulinger:06a, Buras:06, Burrows:07}.  Here we focus on core-collapse SNe driven by bipolar jets \citep[see][]{Khokhlov:99, Couch:09}, as may arise from a magneto-rotational mechanism \citep{Wheeler:00, Wheeler:02, Burrows:07}.  These models have features that may explain many of the observed features of core-collapse supernovae that indicate asymmetry \citep{Khokhlov:99, Wheeler:00, Wang:01, Hoflich:01, Wang:08, Couch:09}  The general features of aspherical shock breakout that we discuss, however, apply to arbitrarily aspherical shocks, not just those produced by bipolar jets.

The absence of spherical symmetry dramatically modifies the observational characteristics of shock breakout and subsequent stages of emission.  We assume black-body emission in our models and we apply a detector response function appropriate for the {\it Swift} XRT and account for X-ray absorption due to neutral matter along the line of sight so that we can make a direct comparison to the observations of XRO 080109.  We show that the timescale of the light curve is not set by the light crossing time of the progenitor star but by the {\it shock}-crossing time.  This can account for the length of the XRO associated with SN 2008D with a Wolf-Rayet star progenitor of reasonable parameters.  Further, we demonstrate that the spectral shape of aspherical shock breakouts is considerably different from that of a single-temperature, spherically-symmetric black-body even if thermal emission is assumed.

Recently \citet{Suzuki:10} have reported on their study of aspherical supernova shock breakout from blue supergiant progenitors.  \citet{Suzuki:10} present a semi-analytic method for calculating shock breakout light curves based on results of 2D hydrodynamic simulations of aspherical core-collapse supernovae.  They assume, as we do in this work, that the breakout emission is thermal and calculate bolometric breakout light curves.  \citet{Suzuki:10} find that the asphericity of the explosion and the angle from which the explosion is viewed determine the shapes of the resulting light curves.  We find a similar result in this work.  Our study, however, is targeted to explaining the observations of XRO 080109/SN 2008D and, as such, we calculate X-ray spectra and light curves that allow a direct comparison to the observations.  Also, our simulations are carried out in a more compact Wolf-Rayet progenitor star.

This paper is organized as follows.  In Section \ref{sec:simulations} we describe the hydrodynamic simulations of jet-driven SNe.  In Section \ref{sec:breakout} we describe our method of modeling the shock breakout emission from our simulations.  In Section \ref{sec:radModels}, we present the resulting spectra and light curves and compare our spectral and light curve models with the observations of SN 2008D.  We discuss our results and give our conclusions in Section \ref{sec:discussion}.

\section{Hydrodynamic Simulations}
\label{sec:simulations}

\begin{figure}
\centering
\plotone{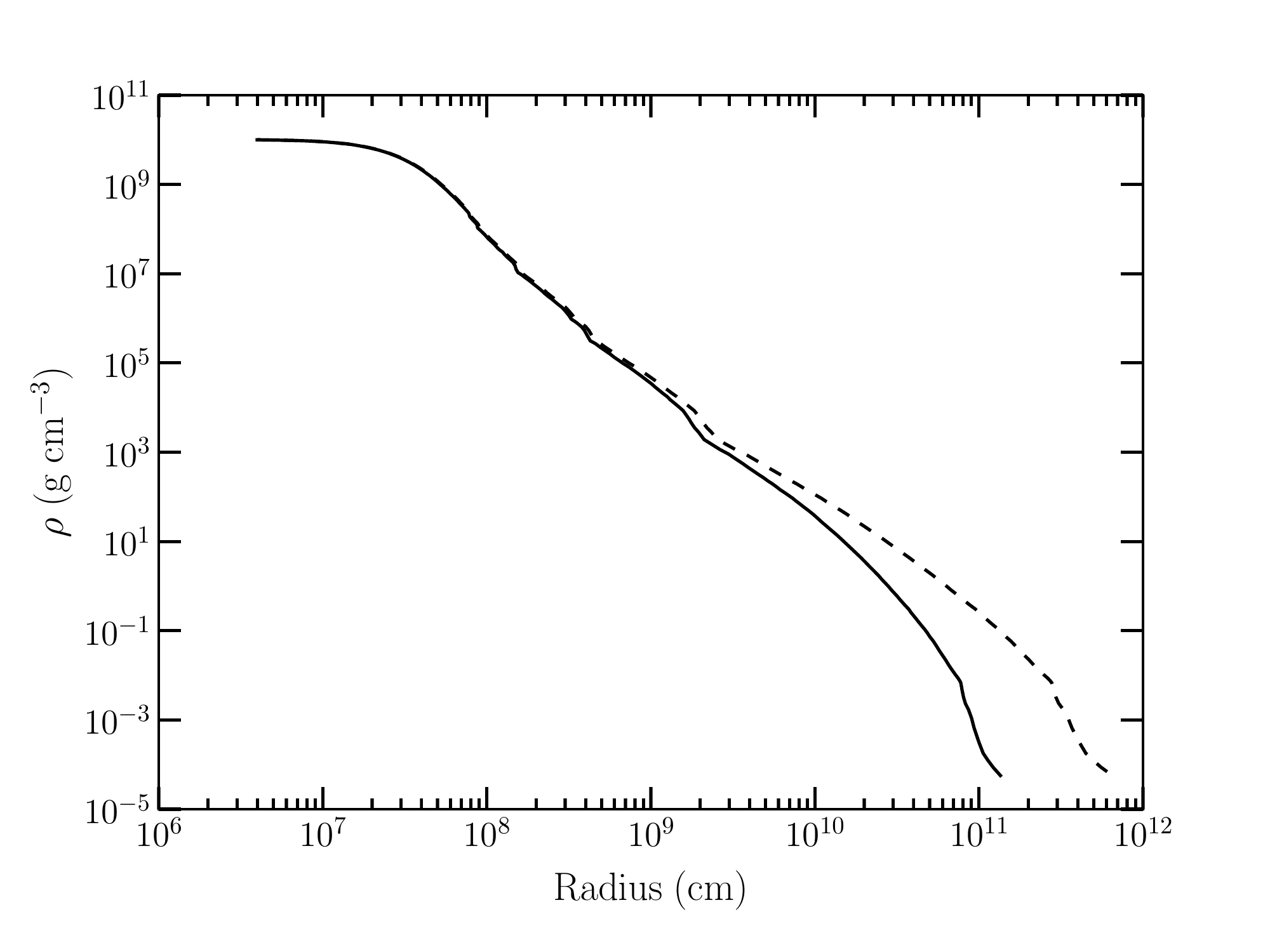}
\caption{Progenitor model density profiles.  Model m2r1 is shown as the solid line and model m7r6 is represented by the dashed line.  Model m7r6 is simply a stretched version of m2r1;  m7r6 is created by increasing the radius of model m2r1 according to $r_{\rm new} = r_{\rm orig} + 6.17\times10^{-8} r_{\rm orig}^{1.7}$.  This establishes a model with a more extended envelope but leaves the core mass and radius practically unchanged.  Model m2r1 has a mass of 2.5$M_\odot$ and a radius of $1.35\times10^{11}$ cm.  Model m7r6 has a mass of 6.8 \Msun and a radius of $6.5\times10^{11}$ cm.  The constant mass-loss rate wind is not shown.}
\label{fig:progenitors}
\end{figure}

We have carried out high-resolution hydrodynamic simulations of jet-driven core-collapse supernovae using the FLASH code, version 2.5 \citep{Fryxell:00}.  We use two progenitor model stars in our calculations.  The first is model s1c5a from \citet{Woosley:95}.  This model is a non-rotating, non-magnetic evolved helium star with a pre-supernova radius of $1.35\times10^{11}$ cm (1.93 $R_\odot$) and a pre-supernova mass of about 2.5 \Msun.  The second model used is a stretched version of model s1c5a.  For this model, we stretch the radial coordinates at each model grid point of s1c5a according to $r_{\rm new} = r_{\rm orig} + 6.17\times10^{-8} r_{\rm orig}^{1.7}$.  The physical variables, such as density and temperature, at each grid point are left unchanged. This function then leaves the core mass and radius approximately unchanged while extending the envelope of the progenitor.  The mass and radius of this model are 6.8 \Msun\ and  $6.5\times10^{11}$ cm (9.3 $R_\odot$), respectively.  Both progenitors are surrounded by a wind with a mass loss rate of $1.5\times10^{-5}\ M_\sun\ {\rm yr}^{-1}$ and wind velocity of 1000 km s$^{-1}$.  The wind is assumed to be spherically symmetric, which may not be the case for a rotating progenitor.  The transition from the progenitor model profile to the wind profile is made linearly over about eight computational zones.  Figure \ref{fig:progenitors} shows the density profiles of the two progenitor models.

We use an equation of state (EoS) that accounts for contributions to the internal energy and pressure from radiation and gas.  In the wind, the gas and radiation are not in thermal equilibrium and so radiation will not contribute to the pressure or internal energy, however our single-temperature code cannot correctly account for this using an EoS that calculates contributions from both radiation and gas.  Therefore, we initially set the temperature in the wind to a small value.  This is justified because for a strong shock the upstream temperature is unimportant to the downstream thermodynamics.  We track seven atomic species in our simulations:  $^4$He, $^{12}$C, $^{16}$O, $^{20}$Ne, $^{24}$Mg, $^{28}$Si, and $^{56}$Ni.    We employ the consistent multi-fluid advection scheme implemented in FLASH \citep{Plewa:99}.  We do not include nuclear burning.  Gravity is calculated using the multipole Poisson solver with $m=0$ and $\ell=1$.

All simulations are carried out in 2D cylindrical geometry.  In order to cover the enormous dynamic range from the inner regions of the progenitor star ($\sim 10^8$ cm) to the shock radius several minutes after shock breakout ($\sim 10^{15}$ cm), we have implemented a logarithmically-spaced cylindrical mesh.  This is achieved through a radially-dependent maximum level of refinement limiter.  This limiter requires that the grid spacing at radius $r$, $\Delta x$, not fall below some fraction of the radius $r$.  The grid spacing then takes the form $\Delta x > \eta N_{x}^{-1} r$, where $N_x$ is the number of zones per block in the $x$-direction and $\eta$ is a small number that sets the resolution scale.  In effect, $\eta N_x^{-1}$ is analogous to the minimum angular resolution in spherical geometry.  Additionally, we have set the maximum level of refinement anywhere on the grid to be time-dependent;  successively higher levels of refinement are dropped from the grid as the simulation proceeds.  This has the effect of dramatically increasing the Courant-limited time step at late times, allowing the calculations to be completed in a relatively small amount of computer time.  Each simulation described in this paper required approximately 3000 CPU hours to cover $10^5$ seconds of simulation time.  This also negated the need to re-map the simulation onto a new grid to continue the simulations to late times \citep[e.g.,][]{Couch:09, Kifonidis:03, Kifonidis:06}.

The jets that drive the explosions are introduced as time-dependent boundary conditions at the inner boundary of the grid where we inject two identical, oppositely-directed energetic flows.  In order to facilitate this, an essentially spherical inner hole is excised from the 2D cylindrical grid.  Within this hole, the hydrodynamic solution is not calculated.  A diode boundary condition was enforced at the edge of the hole \citep[see, e.g.,][]{Zingale:02}.  This boundary condition is equivalent to an outflow boundary condition when the flux into the hole is positive, but the flux out of the hole is always zero.  We include the gravitational effect of the mass initially residing within the hole as a Newtonian point-mass at the center of the grid, and compute the self-gravity of the gas on the grid.  The mass that flows into the hole is tracked and included in the calculation of the central point-mass gravitational potential.  The radius of the hole expands during the simulation, cutting out the smallest zones where the Courant condition is most limiting and ensuring that the hole radius is always resolved by a large number of zones as the maximum allowed refinement level is reduced.  The jet injection velocity, $v_{\rm jet}$, varies in time according to 
\beq
v_{\rm jet}(t) = 
\begin{cases}
v_{\rm max}, & \text{$t \leq 0.25\ t_{\rm jet} $}, \\
v_{\rm max} \frac{4}{3} (1 - t/t_{\rm jet}), & \text{$ 0.25\ t_{\rm jet} < t \leq t_{\rm jet}$}, \\
0, & \text{$t > t_{\rm jet}$},
\end{cases}
\eeq
where $v_{\rm max}$ is the maximum jet injection velocity and $t_{\rm jet}$ is the total jet injection time.

We ran a total of six simulations.  Two of these simulations are spherical, non-jet-driven, explosions for comparison to the jet-driven cases.  The spherical explosions are initiated in an identical manner to the jet-driven cases: injection of energetic material, except that the ``jet" opening half-angle is $\pi/2$.  For the four jet simulations, the opening half-angle of the jets is about $\pi/12$.  The parameters of the jets are listed in Table \ref{table:jets}.  The model name labeling scheme is m{\it M}r{\it R}[cold, hot], where {\it M} is the progenitor mass to the nearest solar mass, {\it R} is the progenitor radius in units of $10^{11}$ cm, and the cold or hot designates the jet parameters used, given in Table \ref{table:jets}.  For all simulations, the maximum extent of the grid is $10^{15}$ cm and the initial radius of the inner hole is $2\times10^8$ cm, roughly the radius of the iron core of the progenitor models used.  The ambient density and temperature at this inner radius for both progenitors is about $5.2\times10^6$ g cm$^{-3}$ and $3.3\times10^9$ K.  The jets are assumed to consist entirely of $^{56}$Ni to facilitate the tracking of the injected jet material.  We note, however, the jet parameters in some of our models, e.g., m7r6cold and m7r6hot, would predominantly freeze out into lighter nuclei (e.g., $^4$He) and not into iron group elements (see, e.g., Pruet et al. 2004). Our slower, denser jets would freeze out into the iron group.  The true resulting $^{56}$Ni fraction will then be a strong function of the proton fraction $Y_e$ in the jet that we do not attempt to model. The parameters of the simulations were chosen so that in every case, the injected jet mass is about $0.1 M_\odot$, a value similar to the $^{56}$Ni mass estimated from observations of SN 2008D \citep{Soderberg:08, Mazzali:08, Modjaz:09}.  Also, for each model, except m2r1hot and m2r1sph, the ratio of explosion energy to ejecta mass is about 0.8 $(10^{51}\ {\rm erg} / M_\sun)$, similar to the ratio estimated from measurements of the photospheric velocity of SN 2008D at maximum light \citep{Soderberg:08, Mazzali:08}.  Model m2r1hot and m2r1sph have slightly higher explosion energy to ejecta mass ratios of about 1.4 $(10^{51}\ {\rm erg} / M_\sun)$.  The jet parameters for models m2r1cold and m2r1hot approximately correspond to the jet parameters used in \citet{Couch:09} for their models v3m12 and v1m12, respectively.  There are 25 levels of refinement at the start of each simulation and the effective angular resolution, $\eta N_x^{-1}$, is $\pi/1024$.  The simulations are run until 10$^5$ seconds, long after shock breakout in all cases.

\begin{deluxetable}{lcccccc}
\tablewidth{0pt}
\tablecaption{Simulation parameters}
\tablehead{
\colhead{Model} & \colhead{$v_{\rm max}$}\tablenotemark{a} & \colhead{$\rho_{\rm jet}$}\tablenotemark{b} & \colhead{$T_{\rm jet}$}\tablenotemark{c} & \colhead{$t_{\rm jet}$}\tablenotemark{d} & \colhead{$M_{\rm tot}$}\tablenotemark{e} & \colhead{$E_{\rm tot}$}\tablenotemark{f}
}
\startdata
m2r1cold & $3.3$ & $25.0$ & $3.0$ & 2.00 & 0.10 & 0.8 \\
m2r1hot & $1.0$ & $70.0$ & $8.0$ & 2.00 & 0.12 & 1.4 \\
m2r1sph & $1.0$ & $70.0$ & $8.0$ & 0.08 & 0.09 & 1.4 \\
m7r6cold & $7.0$ & $3.0$ & $3.0$ & 8.00 & 0.10 & 3.8 \\
m7r6hot & $2.1$ & $7.0$ & $6.2$ & 8.00 & 0.07 & 3.7 \\
m7r6sph & $2.1$ & $7.0$ & $6.2$ & 0.32 & 0.07 & 3.7 \\
\enddata
\tablenotetext{a}{Maximum injection velocity of jets in units of $10^9$ cm s$^{-1}$.}
\tablenotetext{b}{Density of the injected material in units of $10^5$ g cm$^{-3}$.}
\tablenotetext{c}{Temperature of the injected material in units of $10^9$ K.}
\tablenotetext{d}{Total injection time in seconds.}
\tablenotetext{e}{Total mass injected in solar masses.}
\tablenotetext{f}{Total injected energy in units of $10^{51}$ erg.}
\label{table:jets}
\end{deluxetable}

\begin{figure*}
\centering
\begin{tabular}{ccc}
\includegraphics[width=2in]{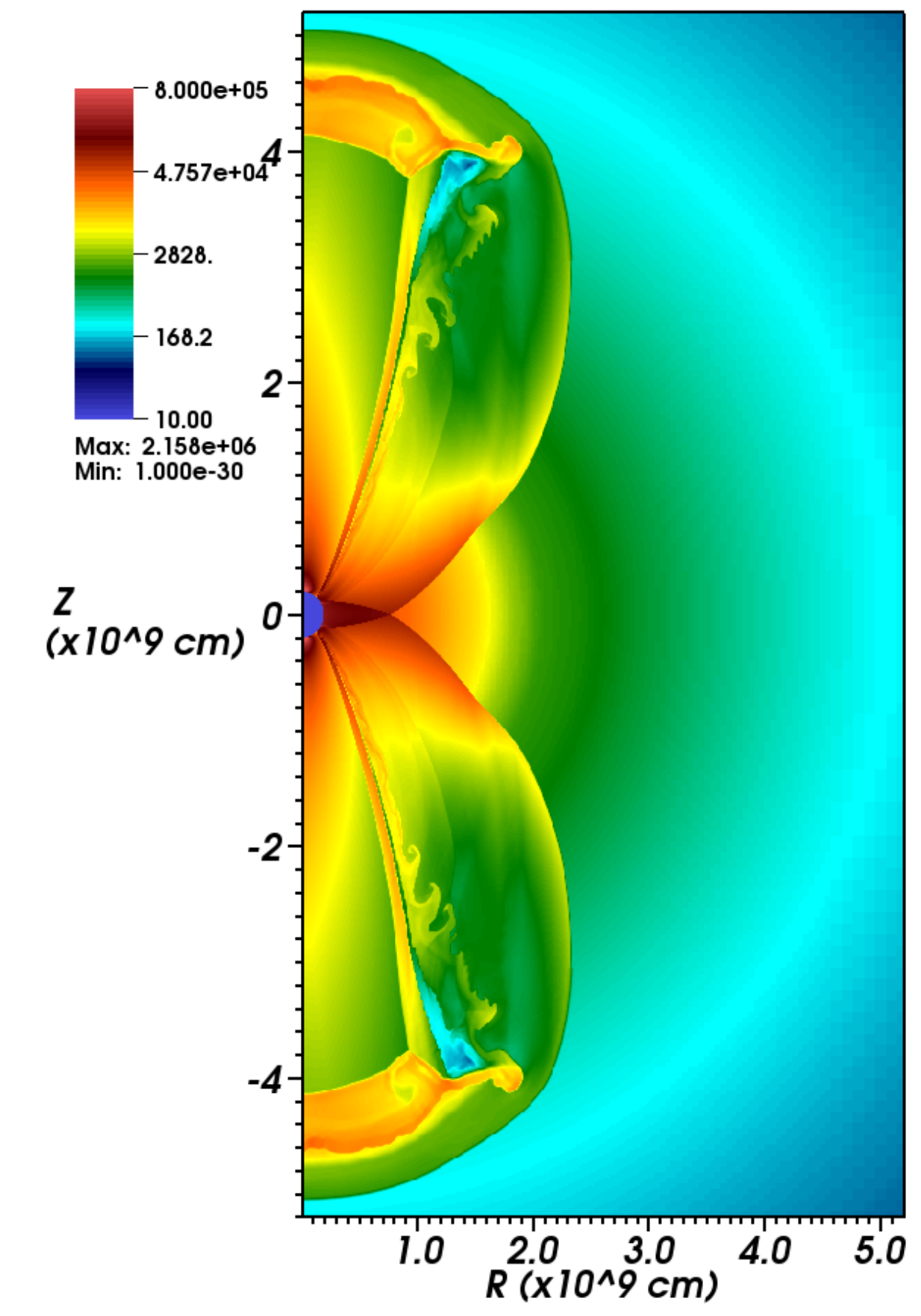} &
\includegraphics[width=2in]{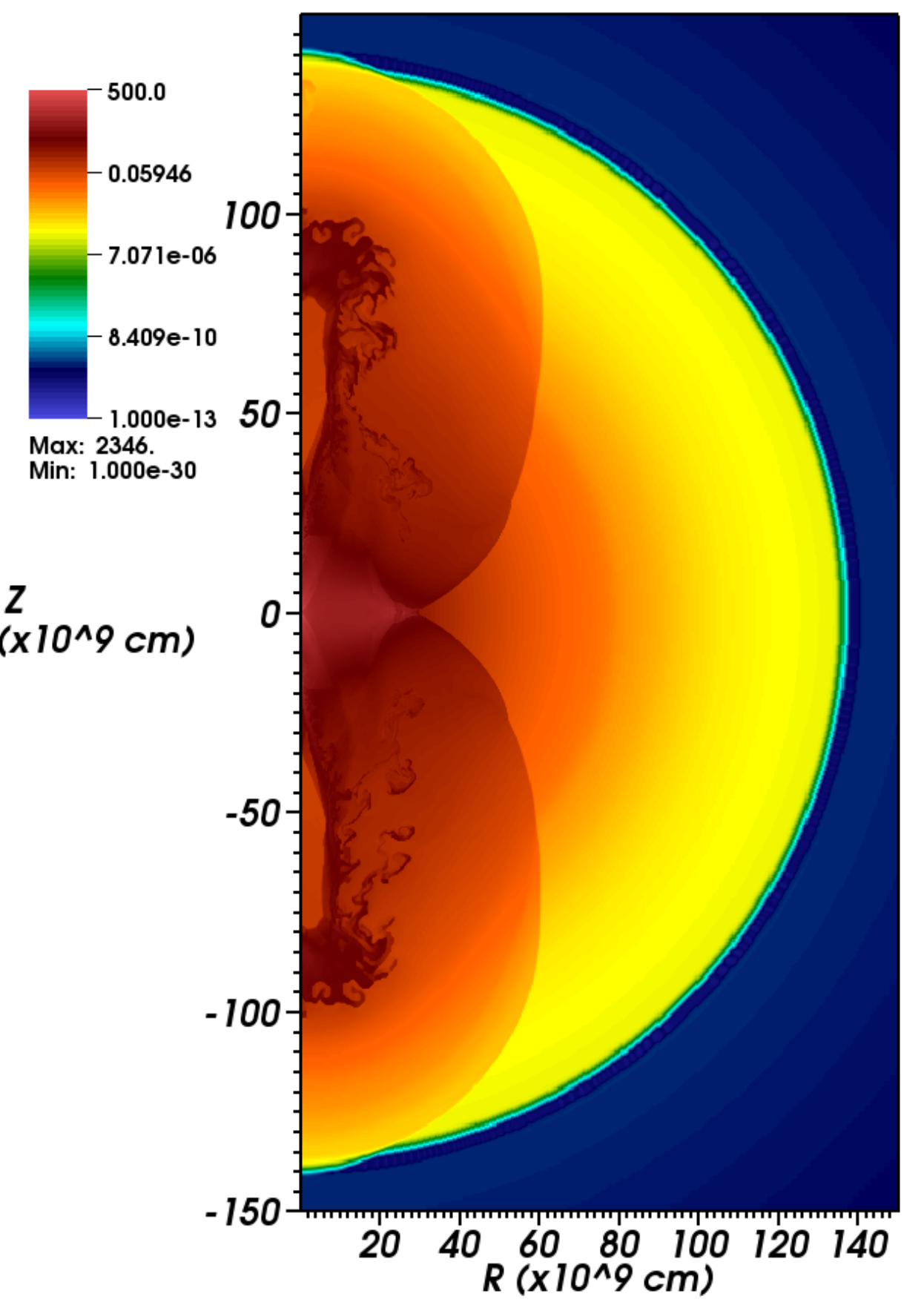} &
\includegraphics[width=2in]{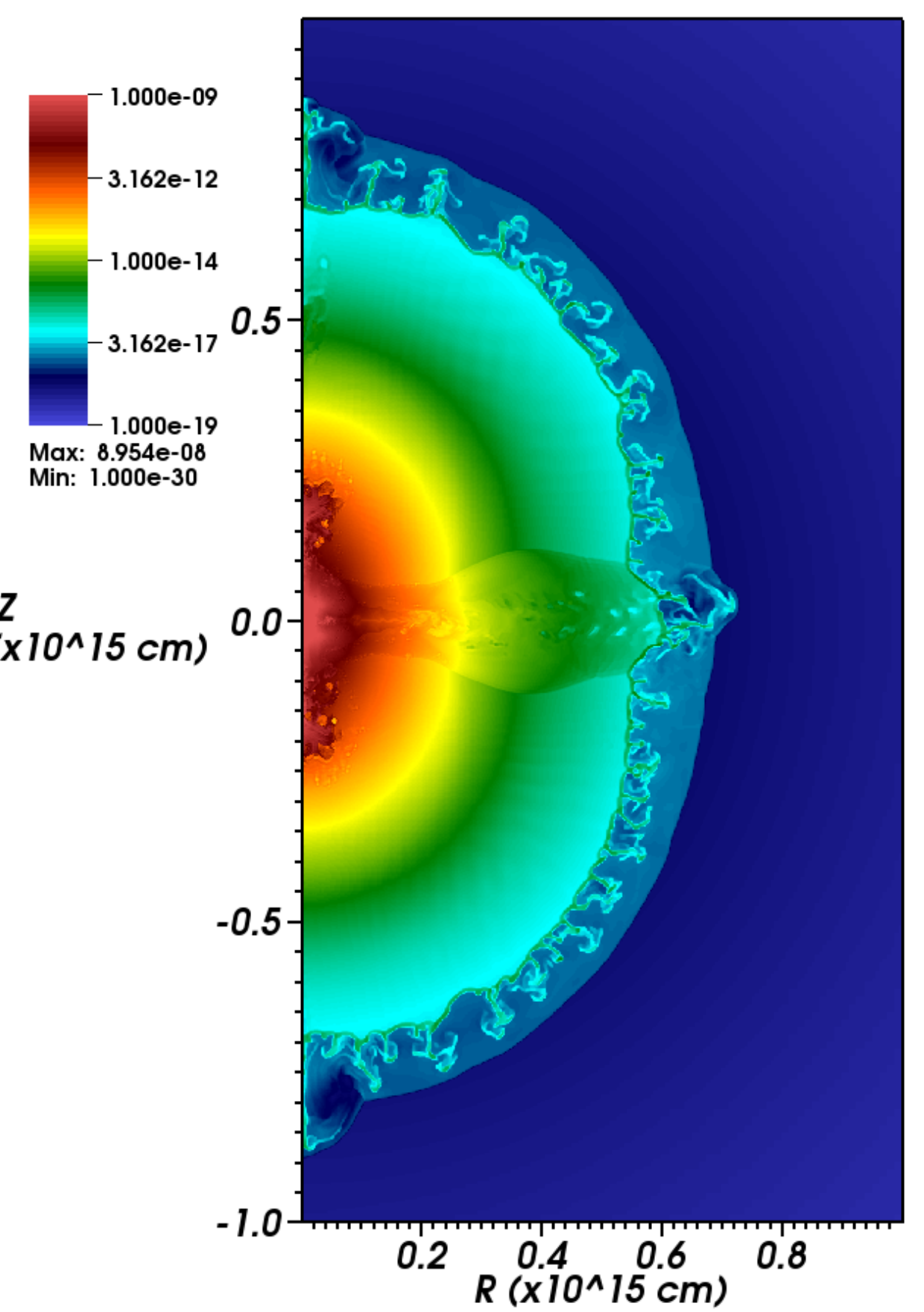}
\end{tabular}
\caption{Density plots for model m2r1cold at 2, 46, and $10^5$ seconds, from left to right.  The left panel shows m2r1cold at the time jet injection ceases.  The bipolar shocks are beginning to cross in the equator, establishing a hot, dense, outward-moving pancake of ejecta.  The edges of the jets present Kelvin-Helmholtz ripples while the contact discontinuity at the jet head is beginning to show growth of  Rayleigh-Taylor fingers.  The middle panel shows the simulation at the moment the polar shocks are erupting from the progenitor's surface.  As the shocks erupt from the surface, they accelerate to speeds approaching $0.5c$ and quickly sweep across the circumference of the star crossing again in the equatorial plane.  The right panel shows m2r1cold at the end of the simulation.  A thin, unstable shell has formed at the contact between the wind and ejecta and has broken up into several Rayleigh-Taylor fingers.  The shock structure is slightly prolate with an axis ratio of about 1.2.}
\label{fig:m2r1cold}
\end{figure*}

\begin{figure*}
\centering
\begin{tabular}{ccc}
\includegraphics[width=2in]{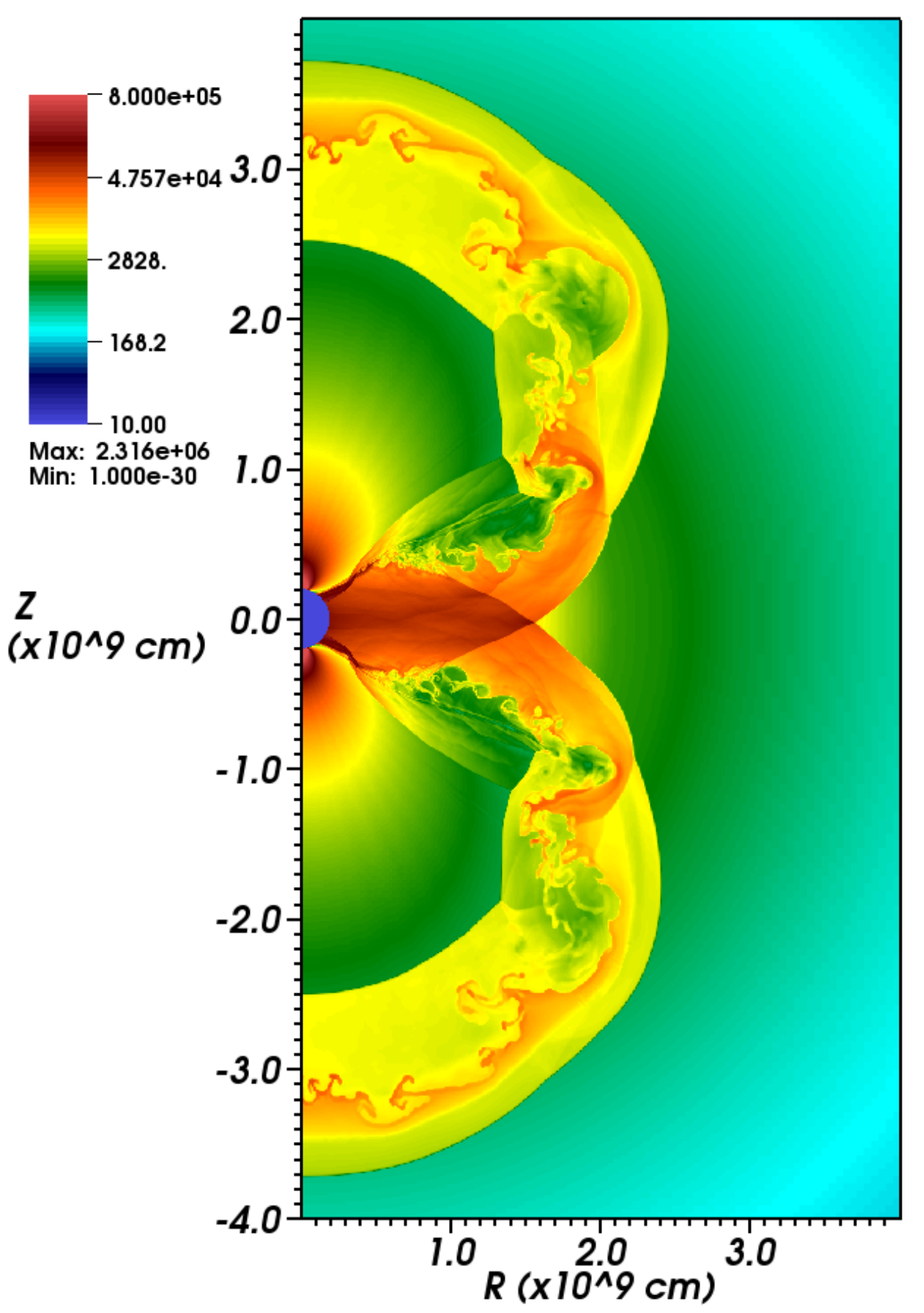} &
\includegraphics[width=2in]{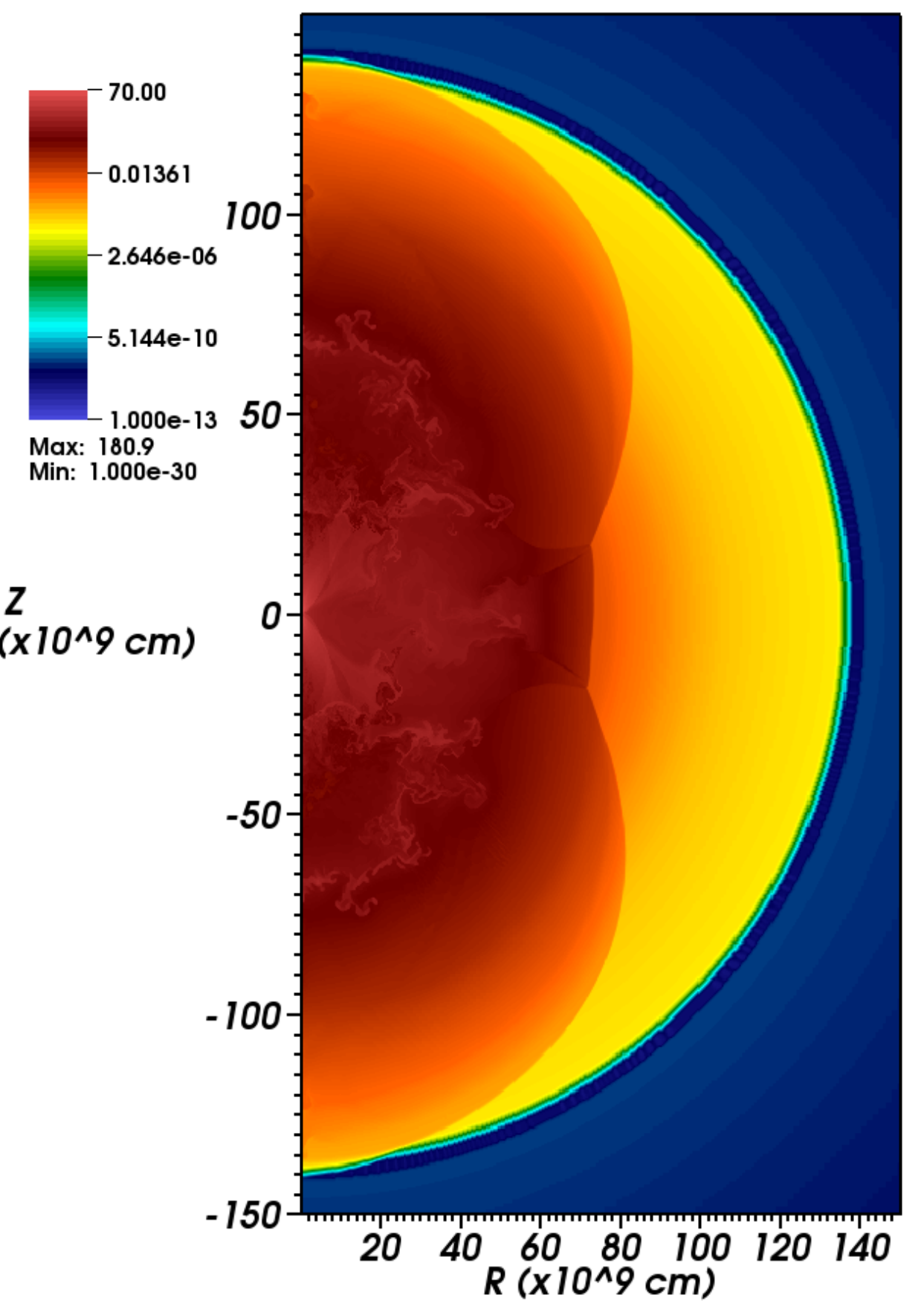} &
\includegraphics[width=2in]{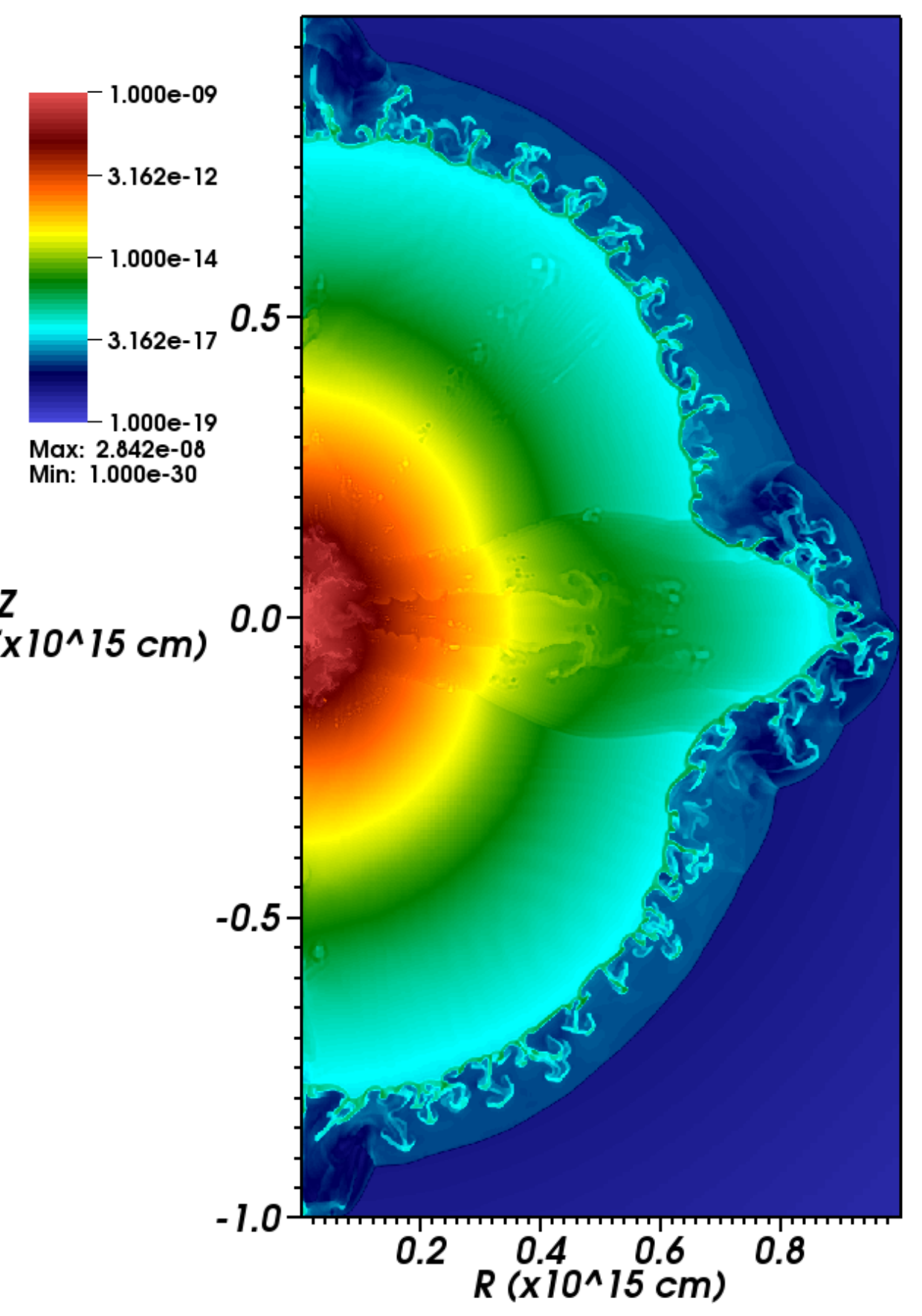}
\end{tabular}
\caption{Density plots for model m2r1hot at 2, 50, and 10$^5$ seconds, from left to right.  The left and middle panels clearly show the greater amount of lateral expansion of the jets as compared with m2r1cold (Fig. \ref{fig:m2r1cold}).  The left panel also shows the dense pancake of ejecta formed in the equator by the crossing bipolar shocks, as well as the very unstable nature of the contacts between the jet flow and the star.  The middle panel shows m2r1hot at the first instance the shocks erupt from the progenitor surface.  The shocks sweep across the progenitor circumference at speeds nearing $0.5c$ and cross again in equator.  The right panel shows m2r1hot at the end of the simulation.  The contact surface between the wind and the ejecta shows significant growth of RT fingers.  The shock structure at the end of the simulation is close to spherical.}
\label{fig:m2r1hot}
\end{figure*}

\begin{figure*}
\centering
\begin{tabular}{ccc}
\includegraphics[width=2in]{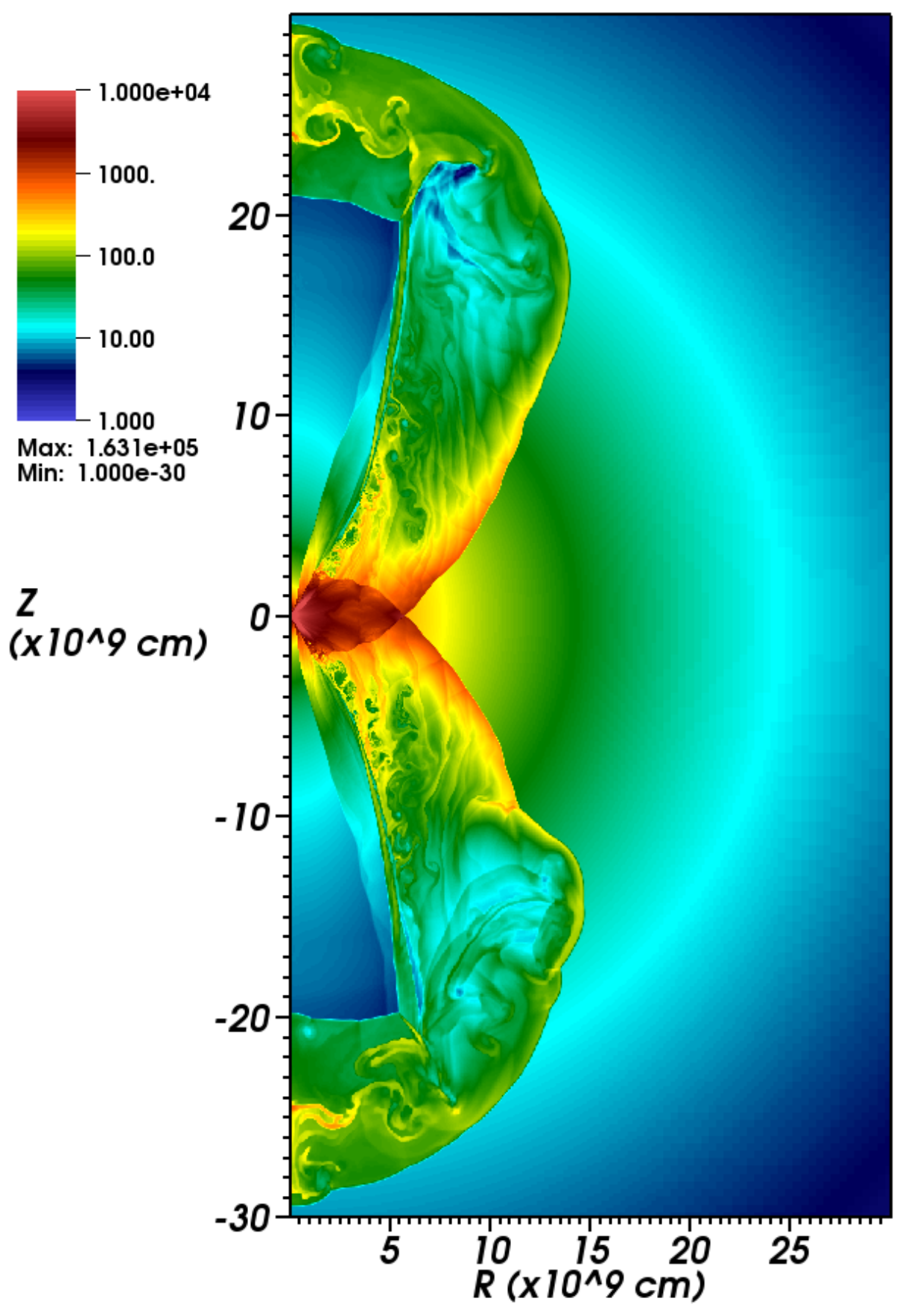} &
\includegraphics[width=2in]{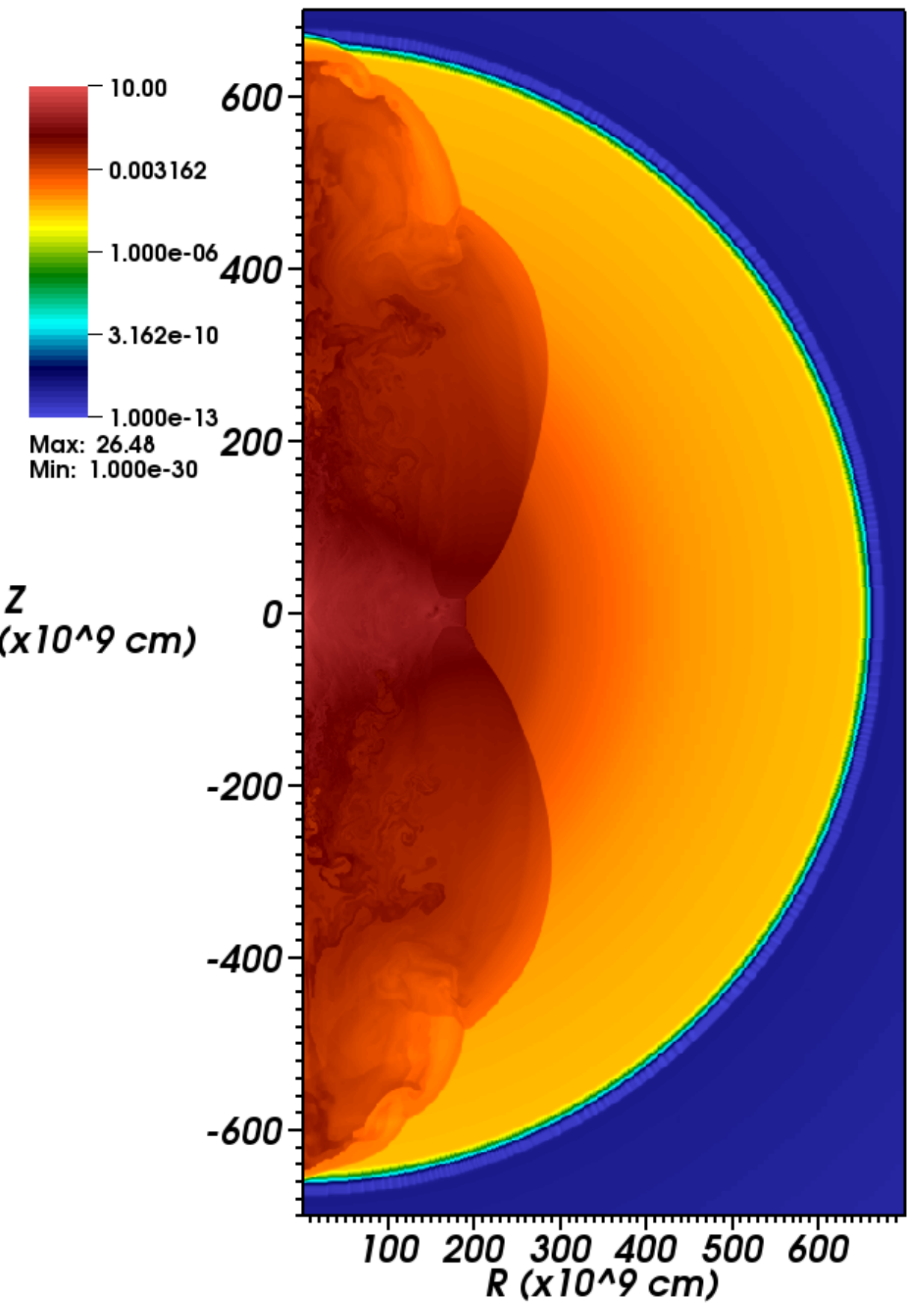} &
\includegraphics[width=2in]{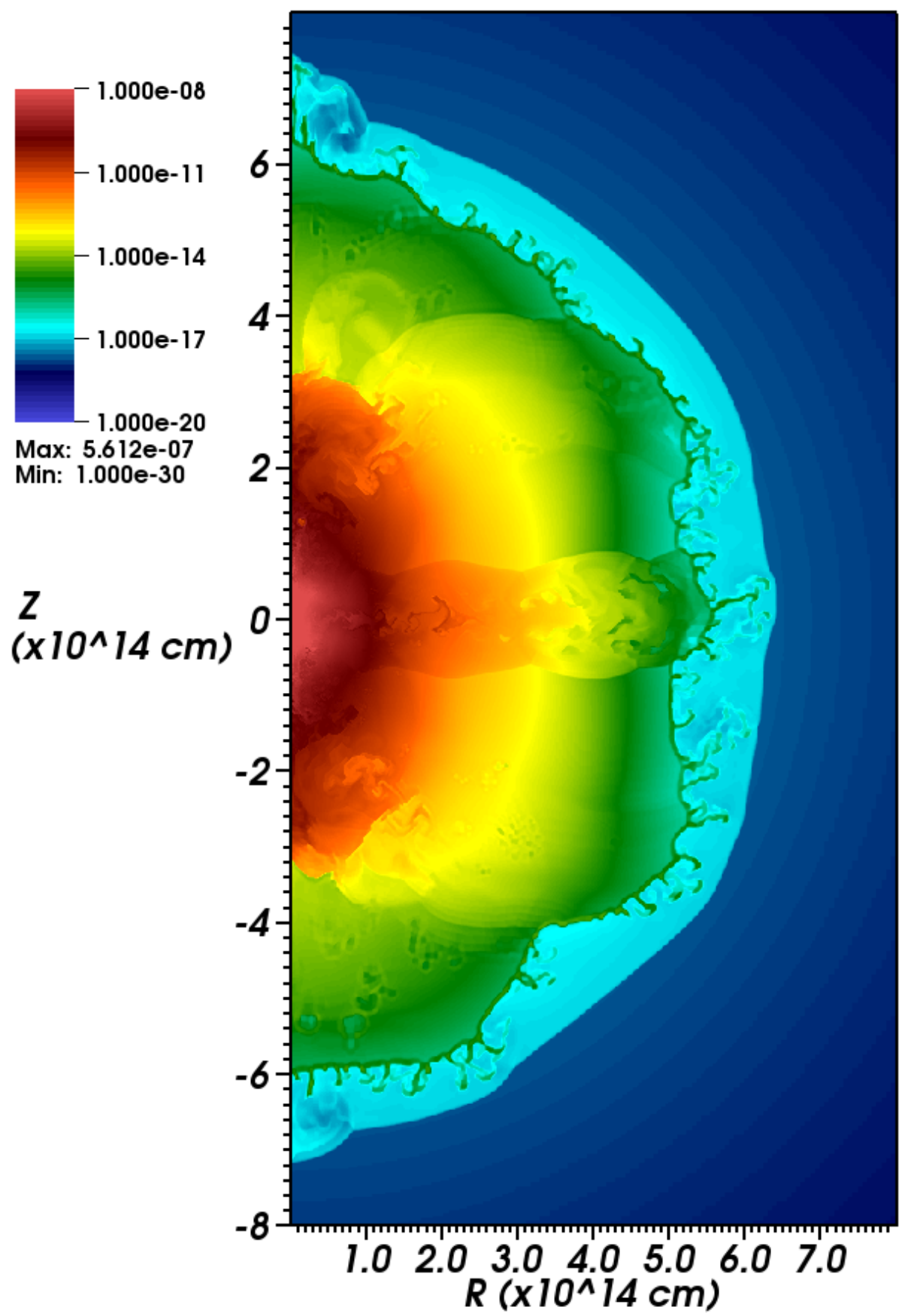}
\end{tabular}
\caption{Density plots for model m7r6cold at 8, 228, and 10$^5$ seconds.  The left panel shows m7r6cold at the time jet injection has ceased.  The bipolar shocks have crossed in the equator creating a dense pancake of ejecta.  The lower-density jet material has already been largely shredded by instabilities.  The middle panel shows m7r6cold at the time of initial shock eruption.  Instabilities in the jet material have created high-velocity fingers that have impinged upon and distorted the shock structure.  Upon shock eruption, the shocks sweep out into the wind at speeds around $0.3c$ and cross again in the equatorial plane. The right panel shows m7r6cold at the end of the simulation.  As in the simulations in the smaller progenitor, a dense, unstable shell has formed at the contact between the wind and ejecta.  The overall shock structure is very slightly prolate.}
\label{fig:m7r6cold}
\end{figure*}

\begin{figure*}
\centering
\begin{tabular}{ccc}
\includegraphics[width=2in]{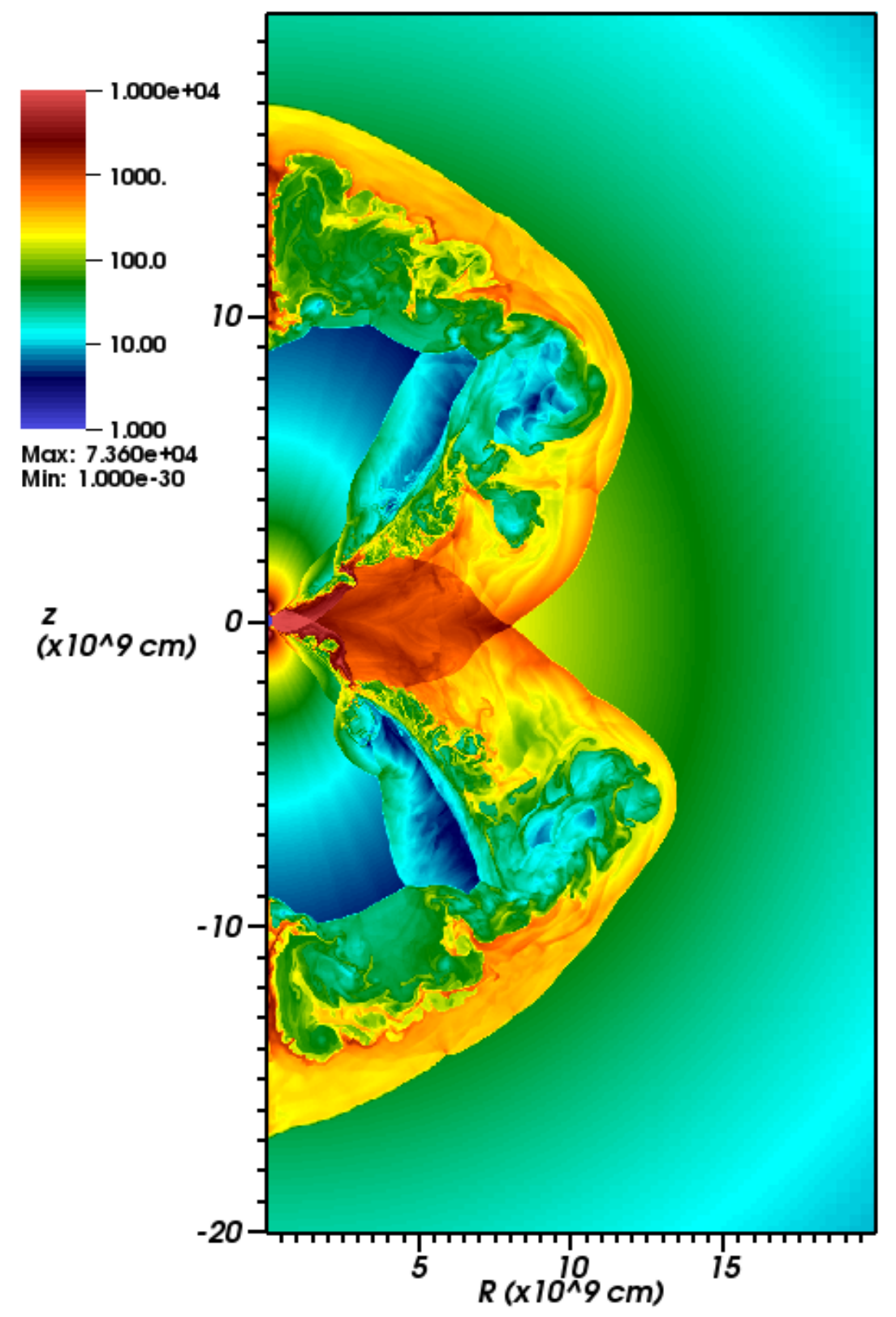} &
\includegraphics[width=2in]{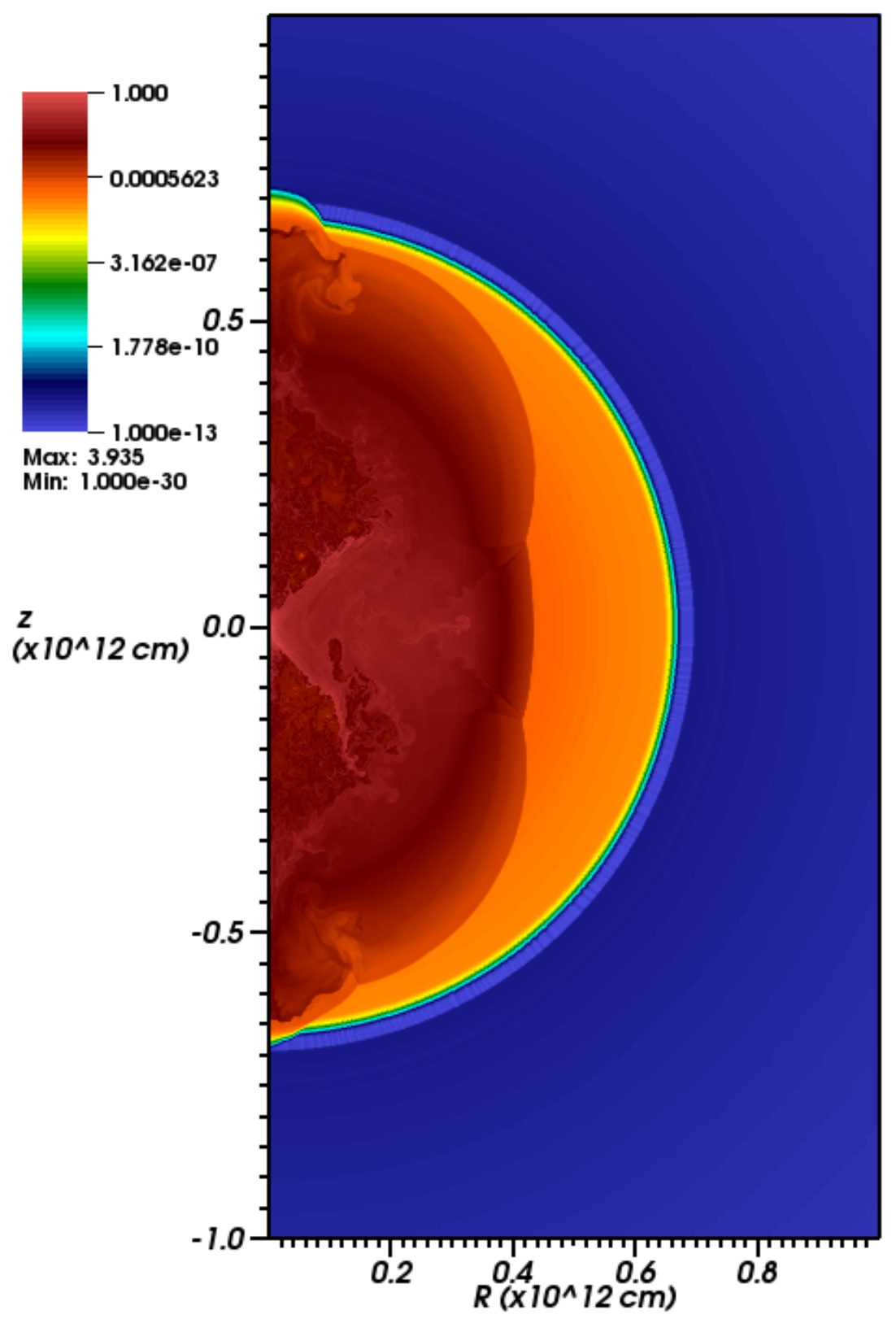} &
\includegraphics[width=2in]{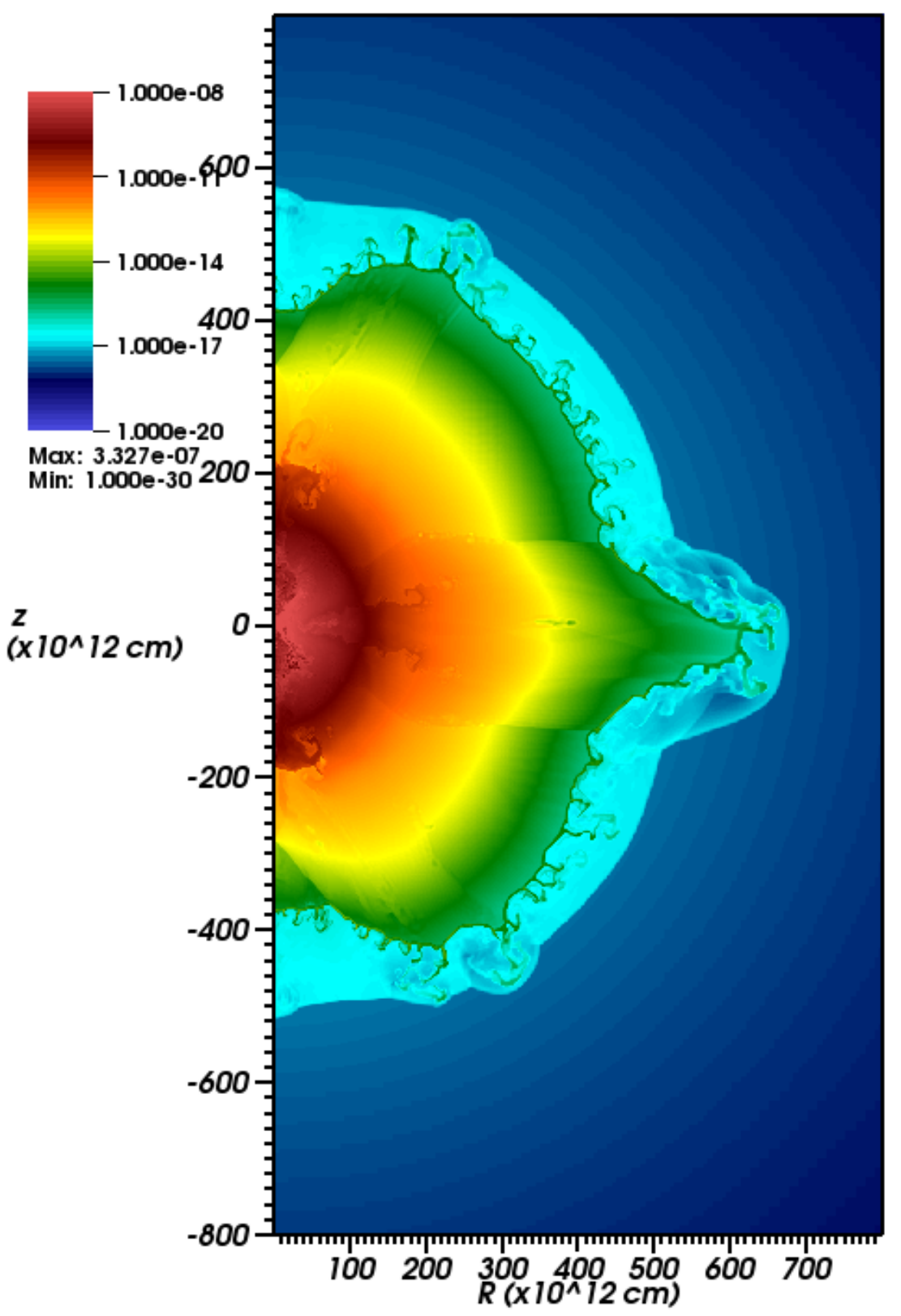}
\end{tabular}
\caption{Density plots for model m7r6hot at 8 seconds, 390 s, and $10^5$ s, from left to right.  The left panel shows model m7r6hot at the moment jet injection is shut off.  Highly unstable and turbulent flow is already evident.  The bipolar shocks have crossed in the equator creating a dense toroidal outflow there.  The middle panel shows this model at the moment of polar shock breakout.  This model is significantly less prolate than its counterpart, m7r6cold (Fig. \ref{fig:m7r6cold}), leading to an equatorial shock breakout that is relatively close in time to the polar shock breakout.  The right panel shows the density field at the end of the simulation.  The final structure is slightly oblate.  A thin, unstable shell is evident at the contact discontinuity between the supernova ejecta and swept-up wind material.}
\label{fig:m7r6hot}
\end{figure*}

Figures \ref{fig:m2r1cold} - \ref{fig:m7r6hot} show show density plots of the four jet-driven explosion simulations at three epochs:  when jet injection stops, initial shock breakout, and the end of the simulation.  In each jet explosion simulation, the jets drive bipolar shocks that expand out from the jet injection sites along the cylindrical axis.  The shocks cross in the equatorial plane and establish a dense, hot pancake of unbound material.  The shocks in all cases erupt from the surface of the progenitor stars first at the poles.  The shocks accelerate into the low-density wind region and sweep around the surface of the progenitor and cross again on the equatorial plane.  This happens just before the original equatorial shock structure erupts from the progenitor surface.  The prolate shock structure evolves toward sphericity in the wind region as the reverse shock, established by the outgoing shock colliding with the wind, sweeps up an unstable shell of ejecta.  %Phase velocity?

The explosions in the smaller progenitor, models m2r1hot and m2r1cold, reach the surface of the progenitor approximately 50 seconds after the start of the simulations.  For explosion m2r1cold, the shocks take about 30 seconds to cross the surface of the progenitor and collide along the equatorial plane.  The shock surface-crossing time is only 20 seconds in model m2r1hot because the shock structure is more spherical than in m2r1cold.  The shocks reach peak speeds of about $1.4\times10^{10}\ {\rm cm\ s^{-1}}$ immediately following eruption from the progenitor surface and then begin to slow in the wind.  At the end of the simulations, around one day after shock eruption, the pole to equator axis ratio for m2r1cold is 1.2 and for m2r1hot is 1.0. 

The explosions in the larger progenitor take significantly different amounts of time to reach the surface.  Model m7r6cold takes about 225 seconds to erupt from the progenitor poles, while model m7r6hot takes about 390 seconds to do the same.  The time it takes the shocks to sweep across the progenitor surface is also different, taking  225 seconds in model m7r6cold and only 125 seconds in model m7r6hot.  As is the case for the smaller progenitor simulations, this is because the hot-jet model is less prolate than the cold.   In the larger progenitor, the shocks reach peak speeds after breakout of about $1\times10^{10}$ cm s$^{-1}$. 

\begin{figure*}
\centering
\begin{tabular}{ccc}
\includegraphics[width=2in]{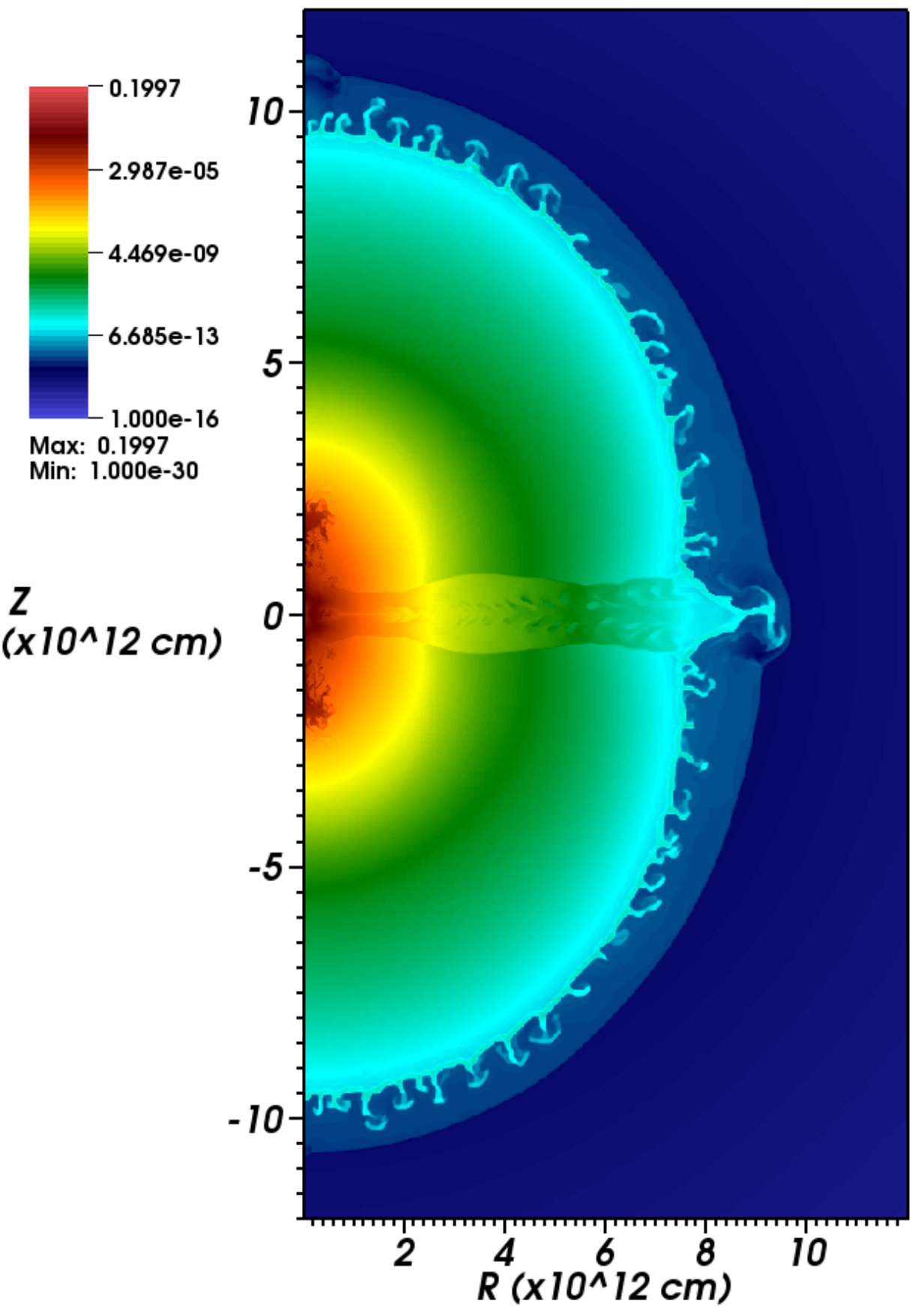} &
\includegraphics[width=2in]{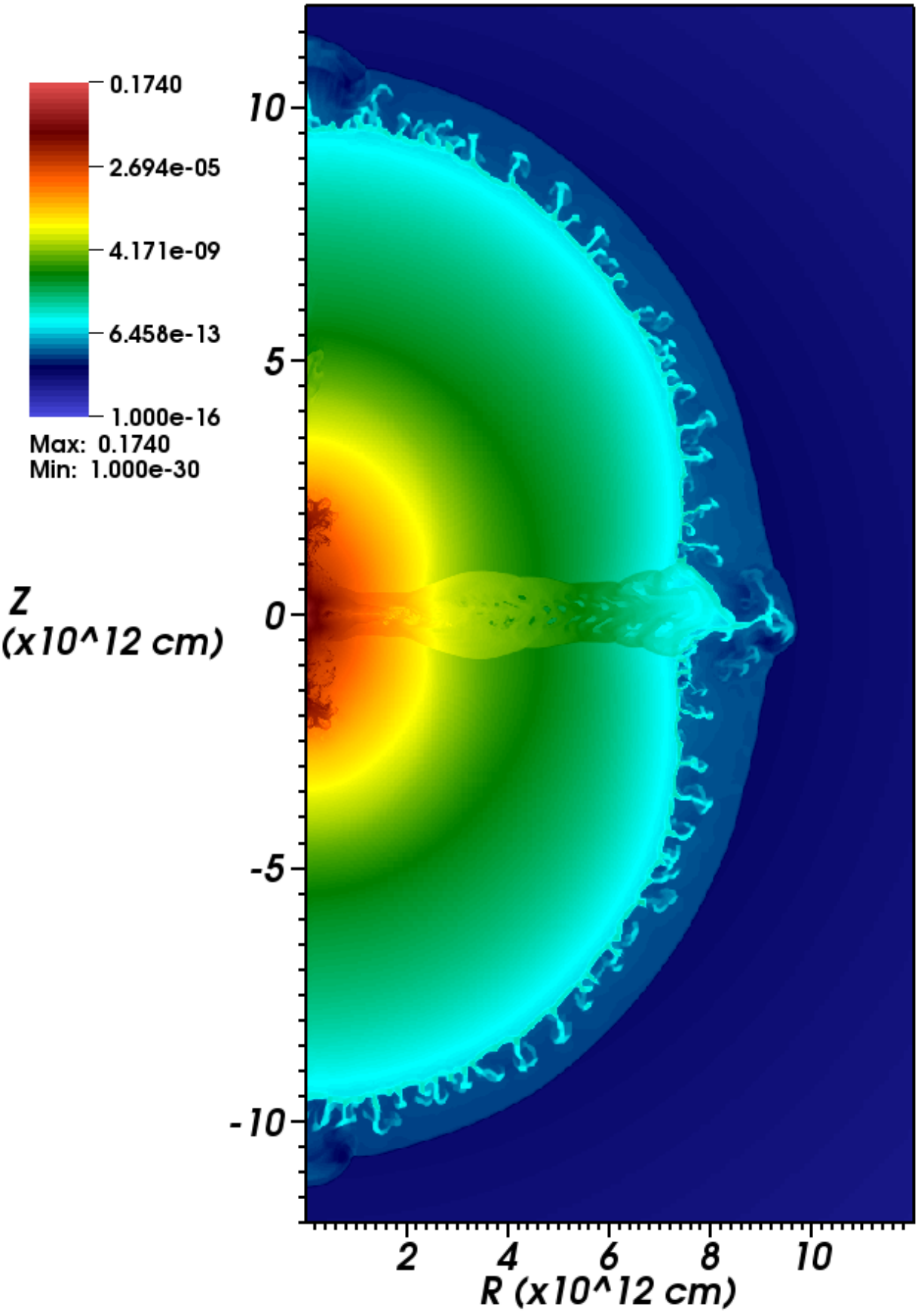} &
\includegraphics[width=2in]{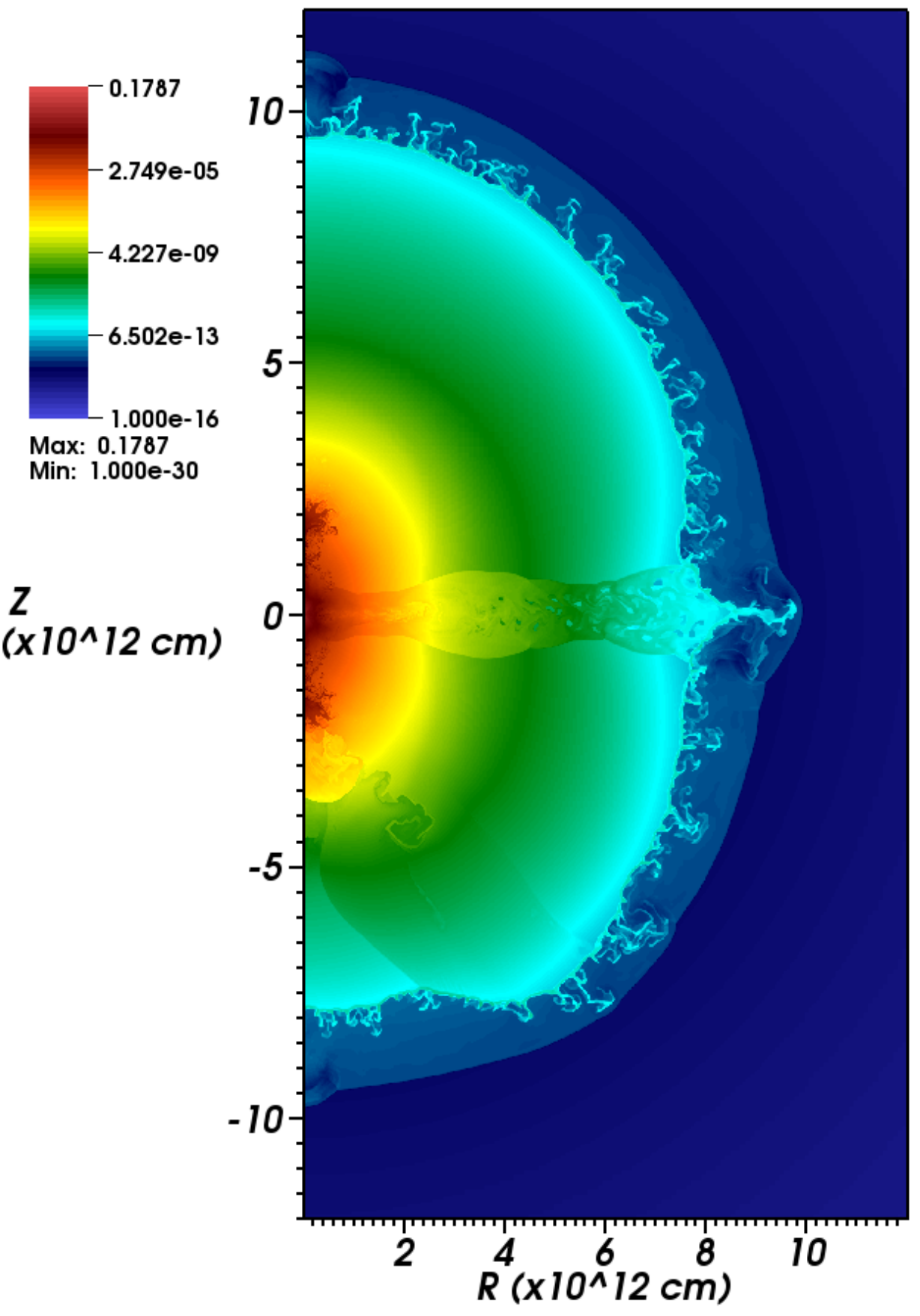}
\end{tabular}
\caption{Density plots for model m2r1cold at three different resolutions at 1000 seconds.  The extent of the shock structures are very similar for all three resolutions.  The amount of North-South asymmetry is greater in the high-resolution simulation, due in large part to the influence of the symmetry axis (see text). }
\label{fig:resolution}
\end{figure*}

We have carried out a resolution study using jet explosion model m2r1cold.  We have run two additional simulations at resolutions of $\eta N_x^{-1} = \pi/788$ and $\eta N_x^{-1} = \pi/1331$.  The results, compared with those of the fiducial simulation with $\eta N_x^{-1} = \pi/1024$, are shown in Figure \ref{fig:resolution}.  The higher resolution simulation shows large-scale of north-south asymmetry.  This is due to small-scale north-south asymmetries near the cylindrical axis early-on that then propagate to large scales as the simulation proceeds to later times.  Artificially accelerated growth of instabilities near the axis is a well-known problem in Eulerian calculations carried-out in curvilinear coordinate systems and higher-resolution simulations are more susceptible to these artificial instabilities \citep[see, e.g.][]{Fryxell:91}.  Additionally, it has been documented that the directionally-split piecewise parabolic method for Eulerian hydrodynamics does not conserve symmetries in small-scale structures \citep[see, e.g.,][]{Liska:03, Almgren:06}.  The higher resolution simulation produces slightly higher shock velocities momentarily during shock breakout, reaching a maximum speed of $1.5\times10^{10}$ cm s$^{-1}$.  Within the accuracy of velocity measurements, the low-resolution simulation attains breakout shock speeds equivalent to those of the medium-resolution simulations, about $1.4\times10^{10}$ cm s$^{-1}$. The shock velocities in the different resolution simulations quickly become equivalent, as is evident by the similar extent of the shock structures shown in Figure \ref{fig:resolution}.  Because of the influence of the numerical artifacts that appear in the high-resolution simulation, the results found in the low- and intermediate-resolution simulations are more reliable.  In fact, in terms of calculating the X-ray emission from the simulations, there is negligible difference between the low- and intermediate simulations, as shown in Figure \ref{fig:m2r1cold_res}, indicating that the gross dynamics have effectively converged at the intermediate resolution.  The small-scale differences between the low- and intermediate-resolution simulations have little impact on the resulting X-ray emission.

\section{Shock Breakout Emission Calculation}
\label{sec:breakout}

In order to understand the observational effects that aspherical shock breakout would produce, we simulate the X-ray emission from our simulations.  We calculate the X-ray spectra at each output time from our simulations and produce X-ray light curves and integrated spectra.  We also compute the estimated X-ray counts as would be detected by the {\it Swift} XRT, accounting for the XRT detector response function and photon absorption along the line of sight.  In this section we discuss the details of our approach.  In Section \ref{sec:radModels} we present the simulated spectra and light curves for the explosion simulations and discuss the general observable characteristics of aspherical shock breakout. 

%\subsection{Emission Calculations}
\label{sec:radCode}

\begin{figure*}
\centering
\plottwo{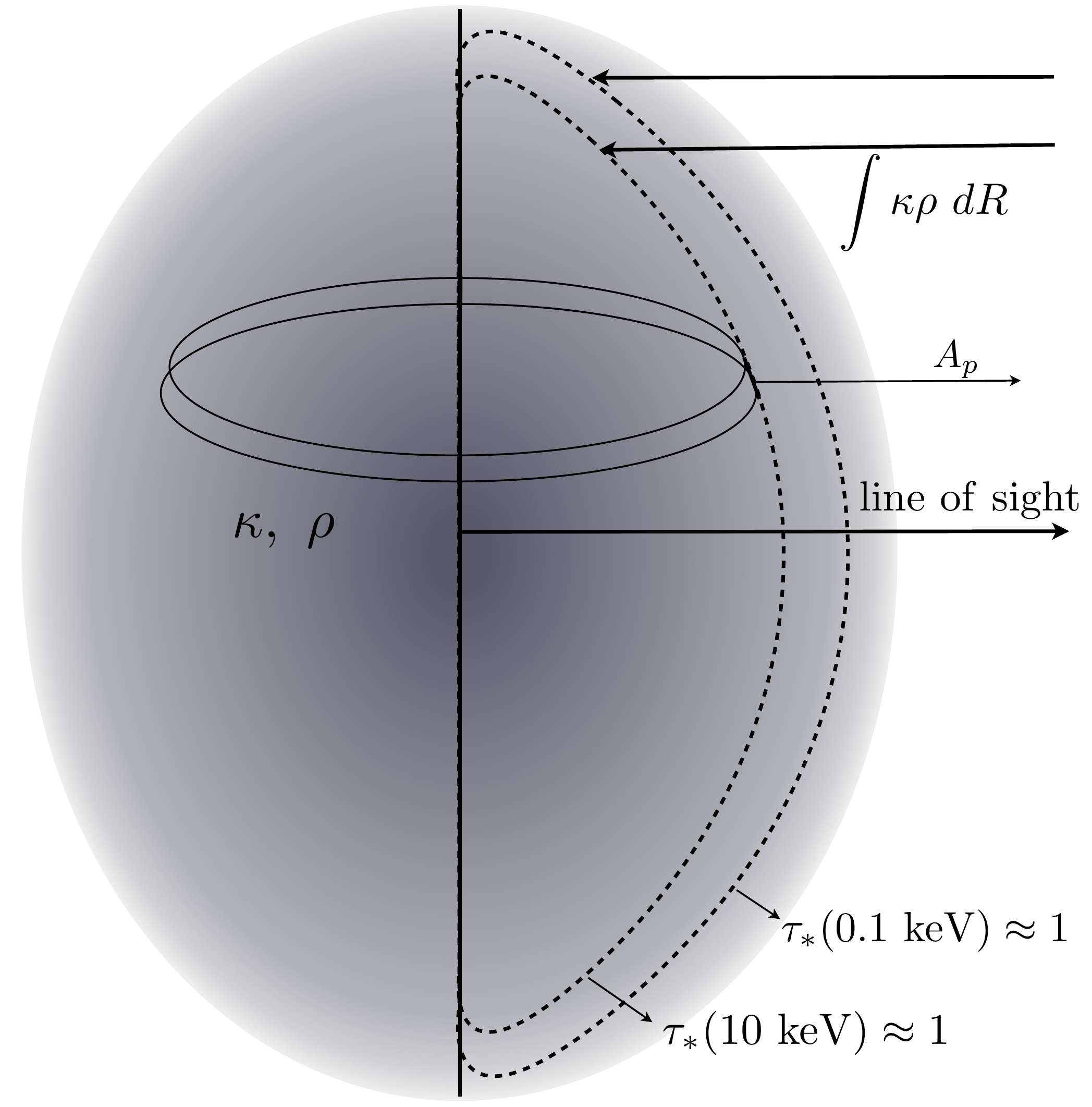}{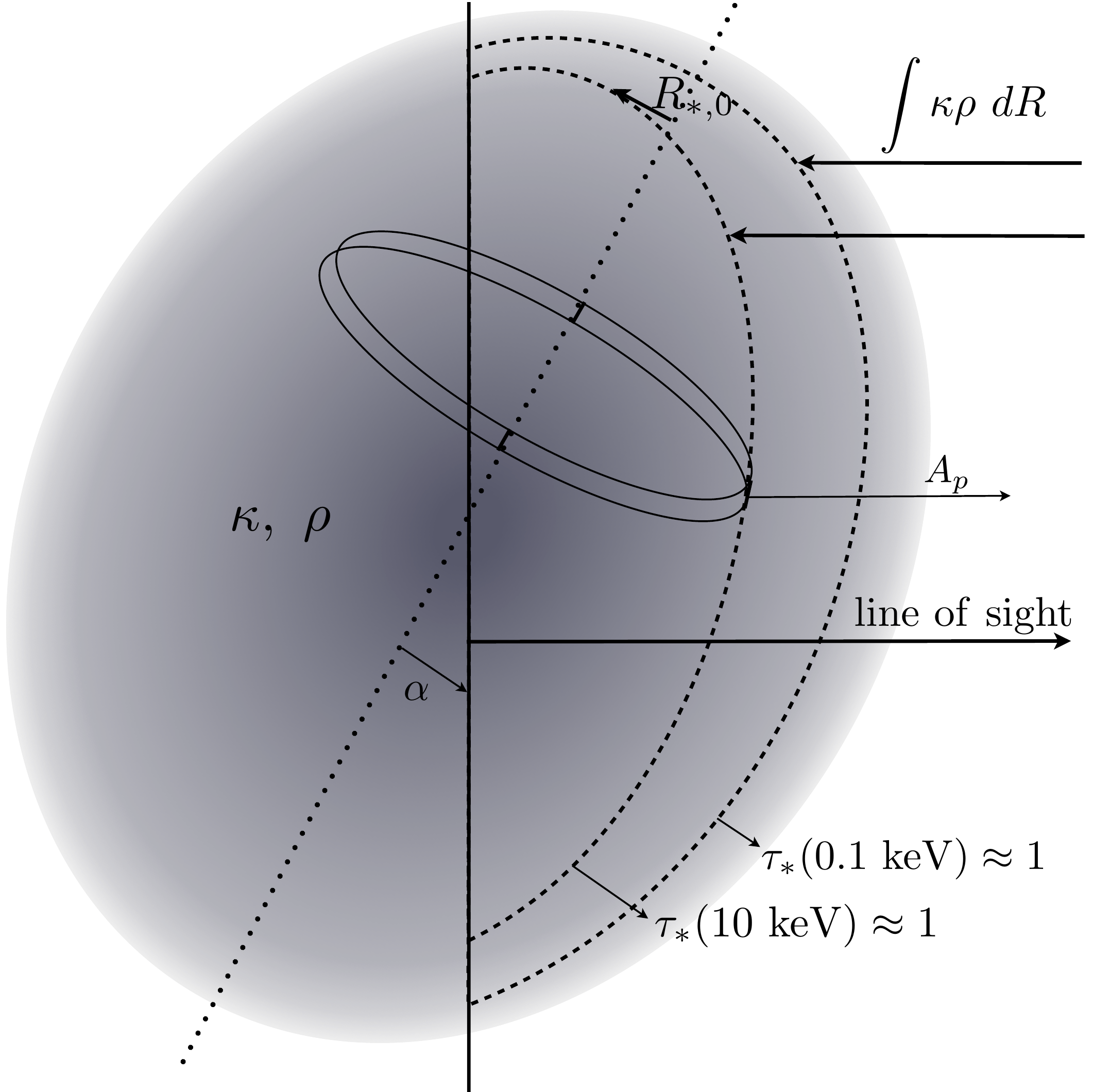}
\caption{Schematic diagram of the integration to find the optical depths.  The gray-scale gradient represents the density and opacity fields.  The integration of the optical depths is carried out along lines-of-sight from the observer, indicated by the top arrows in the diagram.  The pair of dashed lines demonstrate the photon energy dependence of the effective optical depth, $\tau_*$.  The right panel shows the integration geometry for a viewing angle not perpendicular to the simulation symmetry axis.  The shapes of the emitting surface will be different in the rotated case than in the non-rotated case.  The right panel also demonstrates how $R_{*,0}$ can be negative.}
\label{fig:opticalDepth}
\end{figure*}

We calculate the spectra of our simulation results in the {\it Swift} XRT bandpass (0.1 to 10 keV) as a post-processing step.  The AMR data are first merged onto a uniformly-spaced 2D cylindrical grid using volume-weighted averaging.  In order to calculate the emission as seen from multiple viewing angles, the data are rotated prior to calculating the optical depths (see Figure \ref{fig:opticalDepth}). The electron scattering and absorption optical depths, $\tau_{\rm es}$ and $\tau_{\rm abs}$, are then calculated via integration along rays directed from the assumed location of the observer.   We assume the observer's line of sight to be along the cylindrical $R$-coordinate in the post-rotated data.  This line of sight is then not normal to the simulation data symmetry axis for non-zero viewing angles, as shown in Figure \ref{fig:opticalDepth}.  We define the thermalization depth, where the radiation and matter temperatures equilibrate, to be where the effective optical depth $\tau_{*} = \sqrt{3 \tau_{\rm abs} \tau_{\rm tot}} = 2/3$, and $\tau_{\rm tot} = \tau_{\rm es} + \tau_{\rm abs}$ \citep[see, e.g.,][]{Rybicki:86, Ensman:92}.  We assume a black-body emission spectrum is formed with  temperature equal to the matter temperature at the thermalization depth.  The emission directed toward the observer is then the black body flux from the thermalization depth times the surface area of the thermalization depth projected toward the observer.

The thermalization depth, and hence the emissivity, is strongly dependent on the scattering and absorptive opacities.  We use multi-group opacities obtained from the TOPS database maintained by Los Alamos National Lab \citep{Magee:95}.  These opacities are temperature- and density-dependent and use 32 photon energy groups spaced logarithmically in the XRT bandpass.  We assume a chemical mixture comprised mostly of helium, as would be relevant to the Type Ib SN 2008D, with a solar mix of metals at half of the solar metal abundance (i.e., $X = 0$, $Y = 0.992$, and $Z = 0.008$).  The inclusion of metals is critically important to obtaining accurate values for the absorptive opacities, as we discuss in Section \ref{sec:metal_abs}.  The TOPS database gives values for the number of free electrons per atom, $N_e$, the Rosseland mean opacity, and the Planck mean opacity at each temperature, density, and photon energy.  We assume the total absorptive opacity, $\kappa_{\rm abs}(E_\gamma)$, to be the Planck opacity.  The electron scattering opacity is $\kappa_{\rm es} = 0.1 N_e$ cm$^2$ g$^{-1}$, for a predominately helium gas.  The optical depths are then
\beq
\tau_{\rm [es,abs]}(E_\gamma,z) = \int^{R}_\infty \kappa_{\rm [es,abs]}(E_\gamma,R',z) \rho(R',z)\ dR', 
\eeq
where $R$ and $z$ are the cylindrical radius and height, and $E_\gamma$ is photon energy.  A schematic diagram of the optical depth integration is shown in Figure \ref{fig:opticalDepth}.  The radius of the thermalization depth as a function of photon energy and $z$ is then
\beq
R_{\rm *}(E_\gamma, z) = R({\rm where}\ \tau_*(E_\gamma,z) = 2/3), 
\eeq
and the temperature at the thermalization depth is $T_*(E_\gamma, z) = T(R_*)$.  

The intensity of the X-ray emission is assumed to have a black-body spectral energy distribution with a color temperature equal to $T_*(E_\gamma, z)$, i.e.,
\beq
I(E_\gamma, z) = \frac{2E_\gamma^3}{h^2 c^2} \frac{1}{e^{E_\gamma/k T_*}-1}.
\eeq
The emergent flux is then the intensity multiplied by the area projected toward the observer.  The projected area in cylindrical geometry for a line of sight perpendicular to the symmetry axis is just $A_p (E_\gamma,z) = 2 R_{*,0}(E_\gamma,z) dz$, where $R_{*,0}$ is the transformation of $R_*$ into the frame in which the simulation symmetry axis corresponds with the cylindrical axis.  For other viewing angles the projected area becomes
\beq
A_p(E_\gamma, z) = \pi \sin^2{\alpha}\ \abs{R_{*,0}} dR + 2 f \cos^2{\alpha}\ R_{*,0} dz,
\eeq
where $R_{*,0}$ is always measured from the symmetry axis (see Figure \ref{fig:opticalDepth}), $f$ is 0 for $R_{*,0} < 0$ and 1 otherwise, and $\alpha$ is the angle between the symmetry axis and the $z$-axis, or equivalently the angle between the line of sight and the normal to the symmetry axis.  In this way, a surface area element is treated as a cylindrical ring and we account for shadowing effects.  We have assumed that the effective optical depth, $\tau_*$, is axisymmetric.  This is only approximately correct as it neglects limb-darkening effects at latitudes away from the symmetry axis.  The right panel of Figure \ref{fig:opticalDepth} shows a graphical representation of the integration for a non-zero viewing angle $\alpha$.  

The total specific luminosity projected toward the observer is
\beq
L_p(E_\gamma) = \sum_z I(E_\gamma, z) A_p(E_\gamma,z).
\label{eq:Lp}
\eeq
The total luminosity directed toward the observer is $\int L_p(E_\gamma)\ dE_\gamma$.  In the calculation of $L_p(E_\gamma)$, we correct for light-travel time effects.  For the arbitrary geometries we consider, this is accomplished by assuming that the observed time cadence is the same as the simulation output cadence, $t_n^{\rm obs} = t_n^{\rm sim} = t_n$.  Each emitting surface area element is assumed to have a constant luminosity in the interval $dt_n$.  The arrival time of the energy emitted by a surface area element in a given time interval, i.e., $L_p(E_\gamma)dt_n$, is calculated based on the geometry of that area element, and the amount of energy the observer would {\it see} in time $dt_n$ is summed-up.  The observer's measured luminosity at time $t_n$ is then this energy divided by the time interval $dt_n$.  This approach to correcting for light travel time is applicable to arbitrary geometries of emitting surfaces.

The emitting regions of our simulations are typically sampled by about 1000 lines of sight along which the specific luminosities are calculated.  We restrict our analysis to the {\it Swift} XRT bandpass, 0.1 keV to 10 keV.  Since the absorptive opacity is photon energy dependent, the thermalization depth is different for each photon energy group.   The total spectrum is thus not a single temperature black body but the superposition of many black bodies at different temperatures and with different emission areas.  In order to make a more direct comparison with the XRT observations, we convolve our model X-ray spectra with the XRT detector response function and account for X-ray absorption due to neutral matter along the line of sight.  We assume a distance to SN 2008D of 31 Mpc.

\section{Simulated Spectra and Light Curves}
\label{sec:radModels}

\begin{figure}
\centering
\plotone{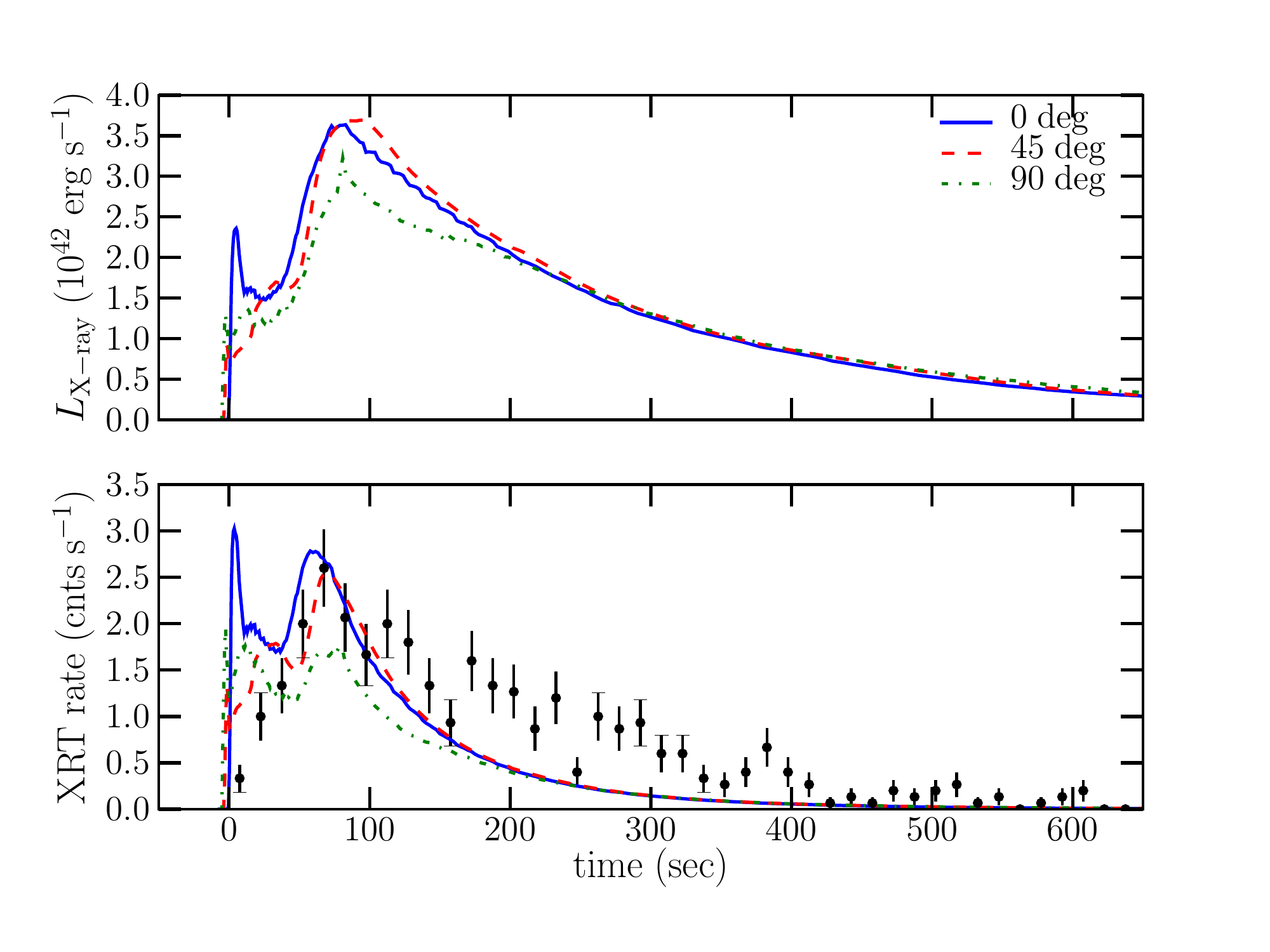}
\caption{XRT band (0.1 - 10 keV) light curves for model m2r1cold at viewing angles of 0, 45, and 90 degrees.  The top panel shows the X-ray luminosity light curves and the bottom panel shows the XRT count rate light curves corrected for detector response and absorption along the line of sight.  The observed XRT light curve of XRO 080109 is also shown for comparison.  The light curves are double-peaked in shape, with the first peak corresponding to emission from the spot where the shock is erupting from the progenitor surface and the second peak corresponding to enhanced emission from the equator as the shocks cross.  The second peak, which has a much larger luminosity than the first, is characterized by softer emission relative to the first and so results in a similar X-ray count rate to the first peak, once corrections for absorption are made.  The light curve for a viewing angle of 90 degrees (along the axis of symmetry) is generally less luminous than the other viewing angles, even early on.  This is because once the shocks erupt from the poles, they expand quickly and obscure the very bright ring on the surface of the progenitor where the shock continues to erupt. }
\label{fig:lc_m2r1cold}
\end{figure}

\begin{figure}
\centering
\plotone{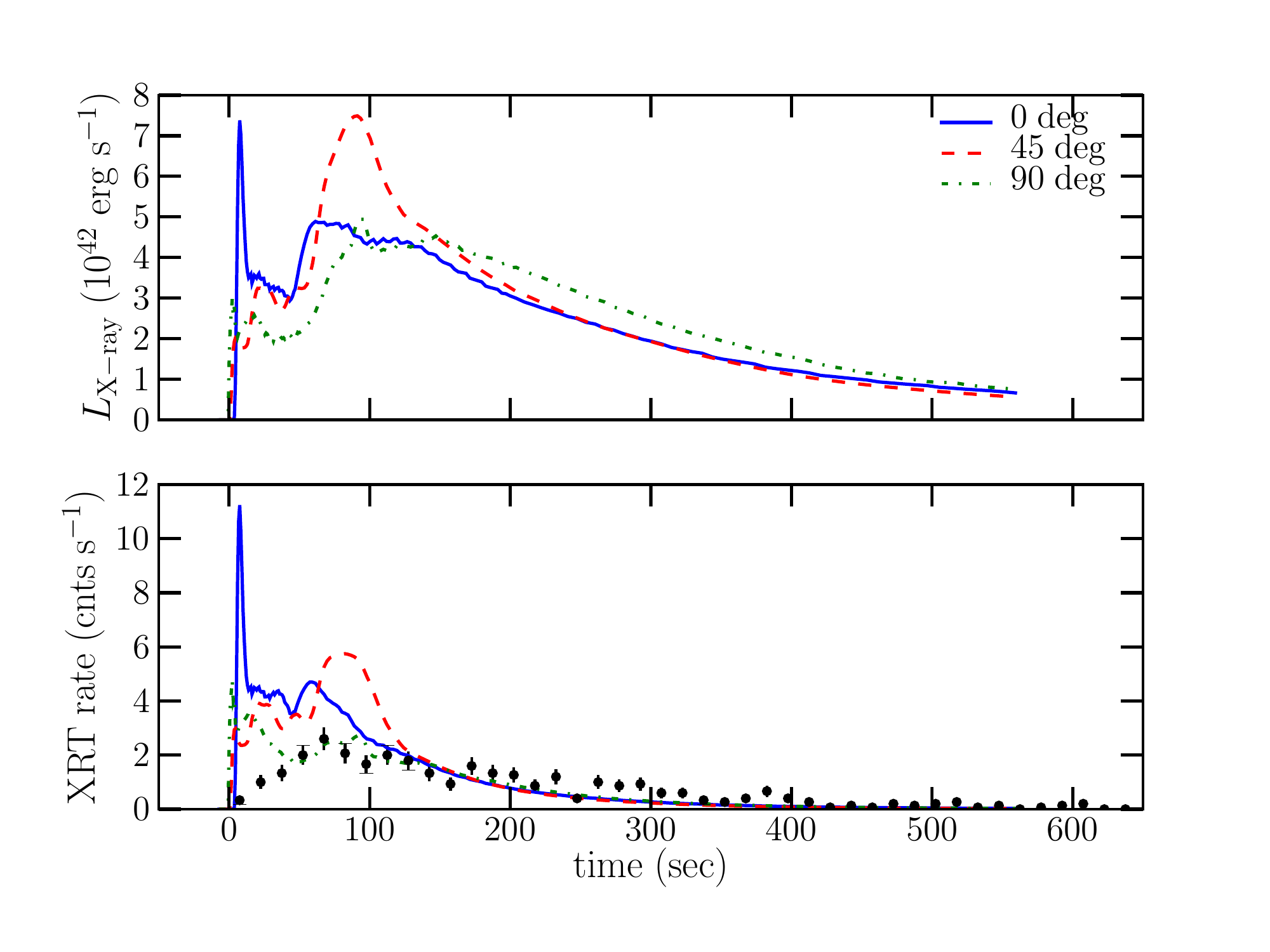}
\caption{XRT band (0.1 - 10 keV) light curves for m2r1hot at viewing angles of 0, 45, and 90 degrees.  The top panel shows the X-ray luminosity as a function of time and the bottom panel shows the predicted XRT count rate accounting for detector response and absorption due to neutral matter.  The observed XRT light curve of XRO 080109 is also shown for comparison.   }
\label{fig:lc_m2r1hot}
\end{figure}

\begin{figure}
\centering
\plotone{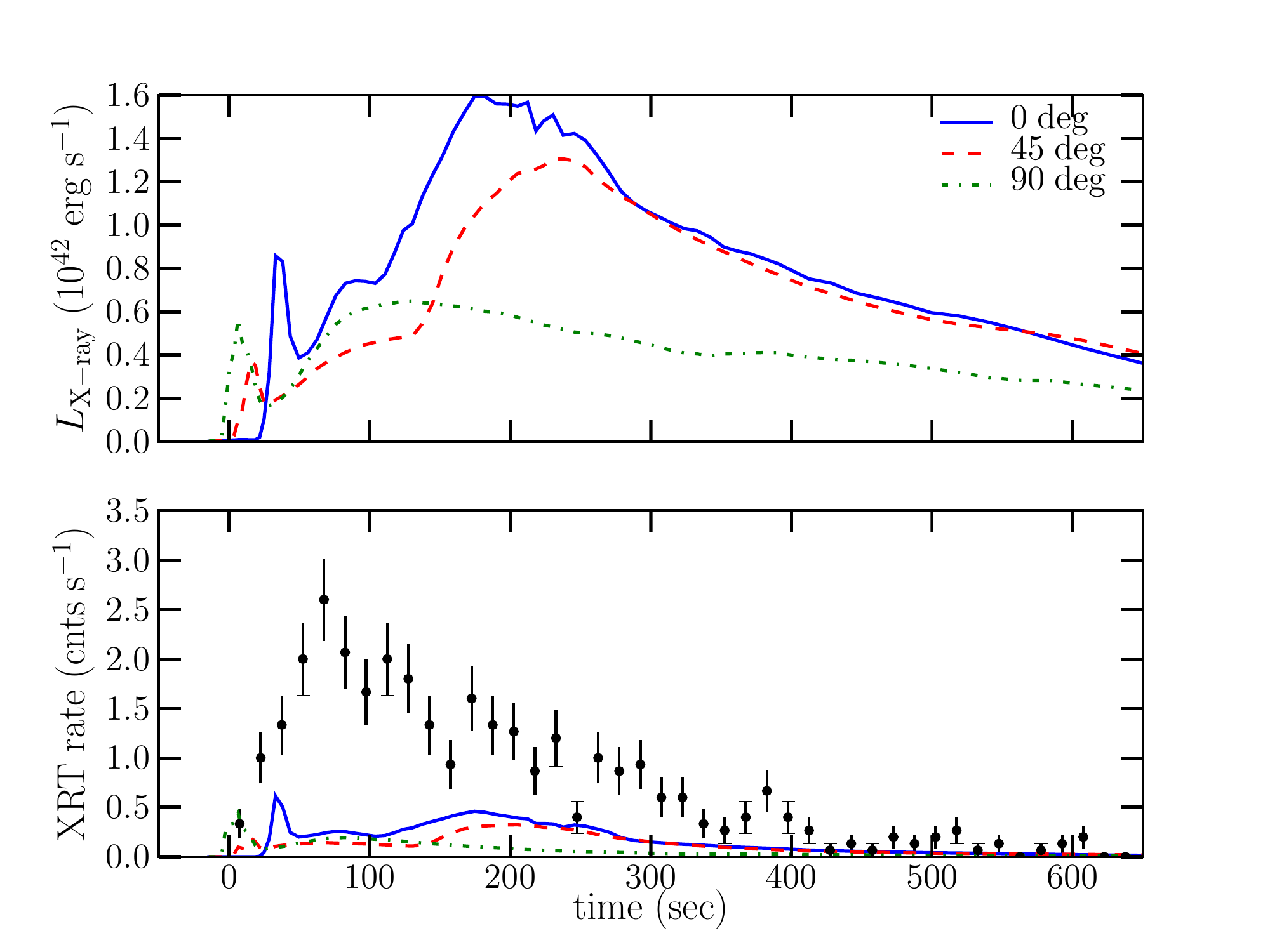}
\caption{XRT band (0.1 - 10 keV) light curves for m7r6cold at viewing angles of 0, 45, and 90 degrees.  The top panel shows the X-ray luminosity as a function of time and the bottom panel shows the predicted XRT count rate accounting for detector response and absorption due to neutral matter.  The observed XRT light curve of XRO 080109 is also shown for comparison. }
\label{fig:lc_m7r6cold}
\end{figure}

\begin{figure}
\centering
\plotone{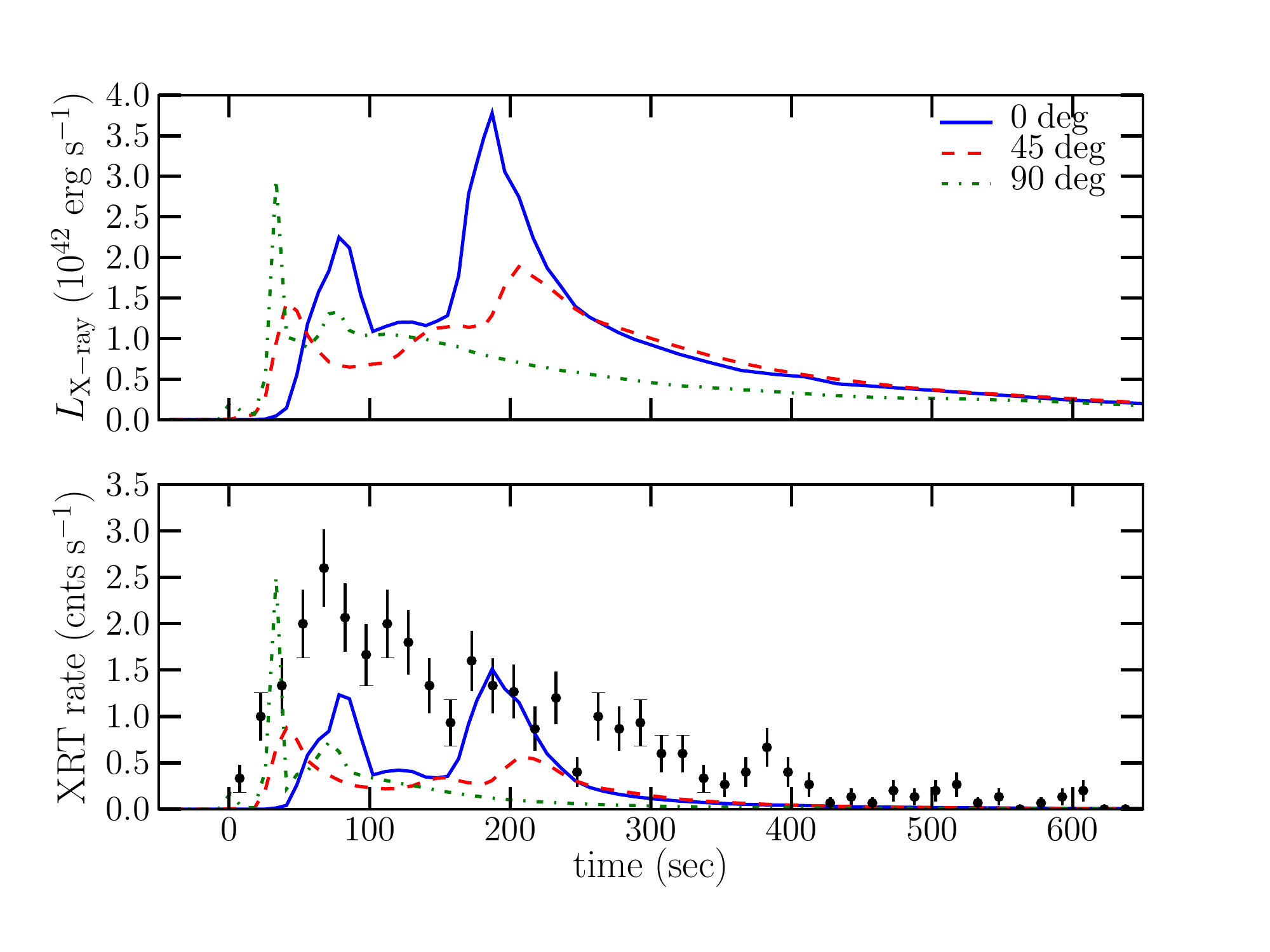}
\caption{XRT band (0.1 - 10 keV) light curves for m7r6hot at viewing angles of 0, 45, and 90 degrees.  The top panel shows the X-ray luminosity as a function of time and the bottom panel shows the predicted XRT count rate accounting for detector response and absorption due to neutral matter.  The observed XRT light curve of XRO 080109 is also shown for comparison. }
\label{fig:lc_m7r6hot}
\end{figure}

\begin{figure}
\centering
\plotone{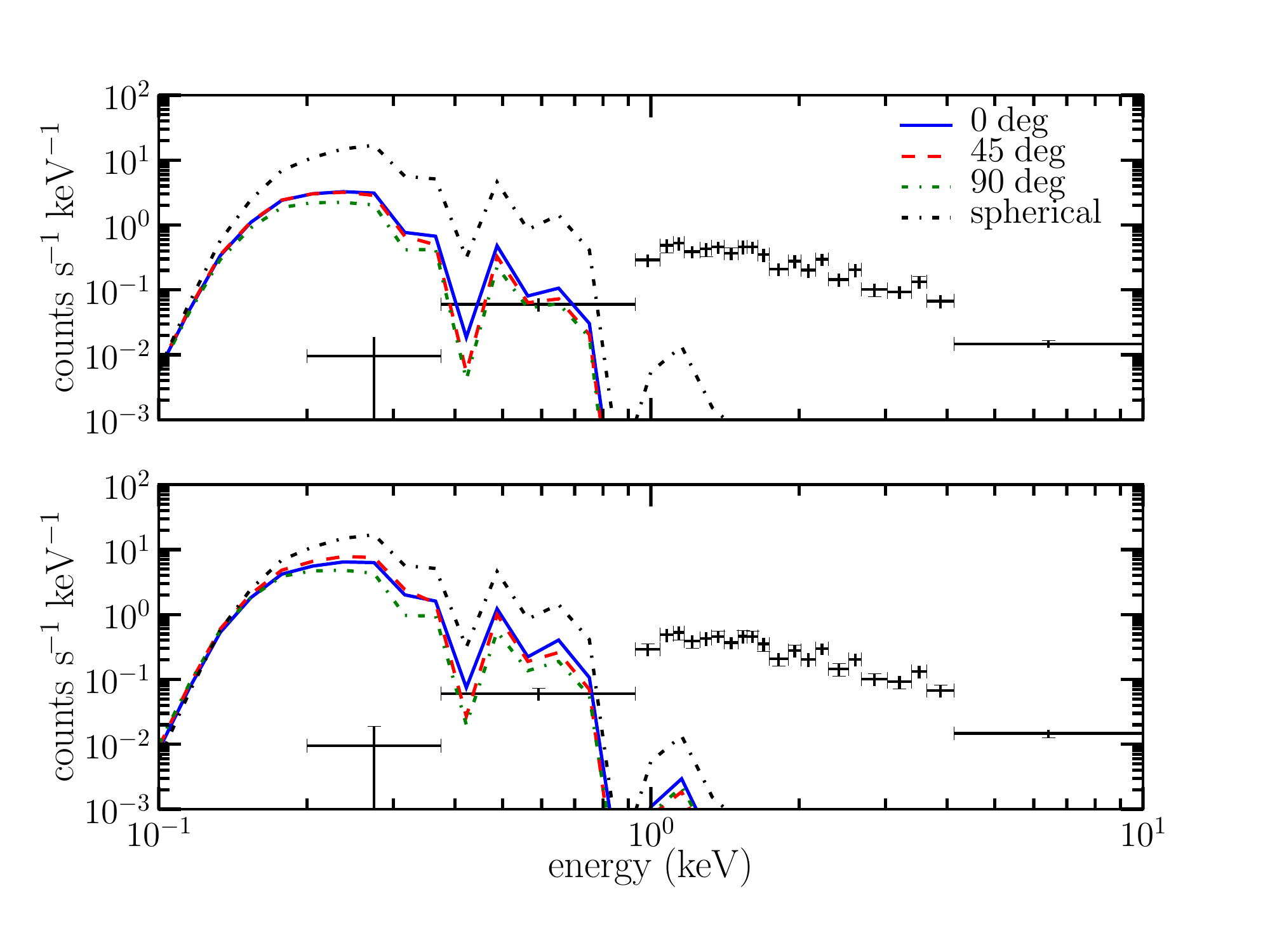}
\caption{Time-integrated X-ray spectra for the smaller progenitor explosion models, m2r1cold (top) and m2r1hot (bottom) at different viewing angles.  The spectra are corrected for detector response and absorption along the line of sight, assuming a neutral matter column depth of $N_{\rm H} = 1.7\times10^{20}\ {\rm cm^{-2}}$.  For comparison, we also plot the spectrum for the spherical explosion, m2r1sph, and the observed XRT spectra of XRO 080109.}
\label{fig:spec_m2r1}
\end{figure}

\begin{figure}
\centering
\plotone{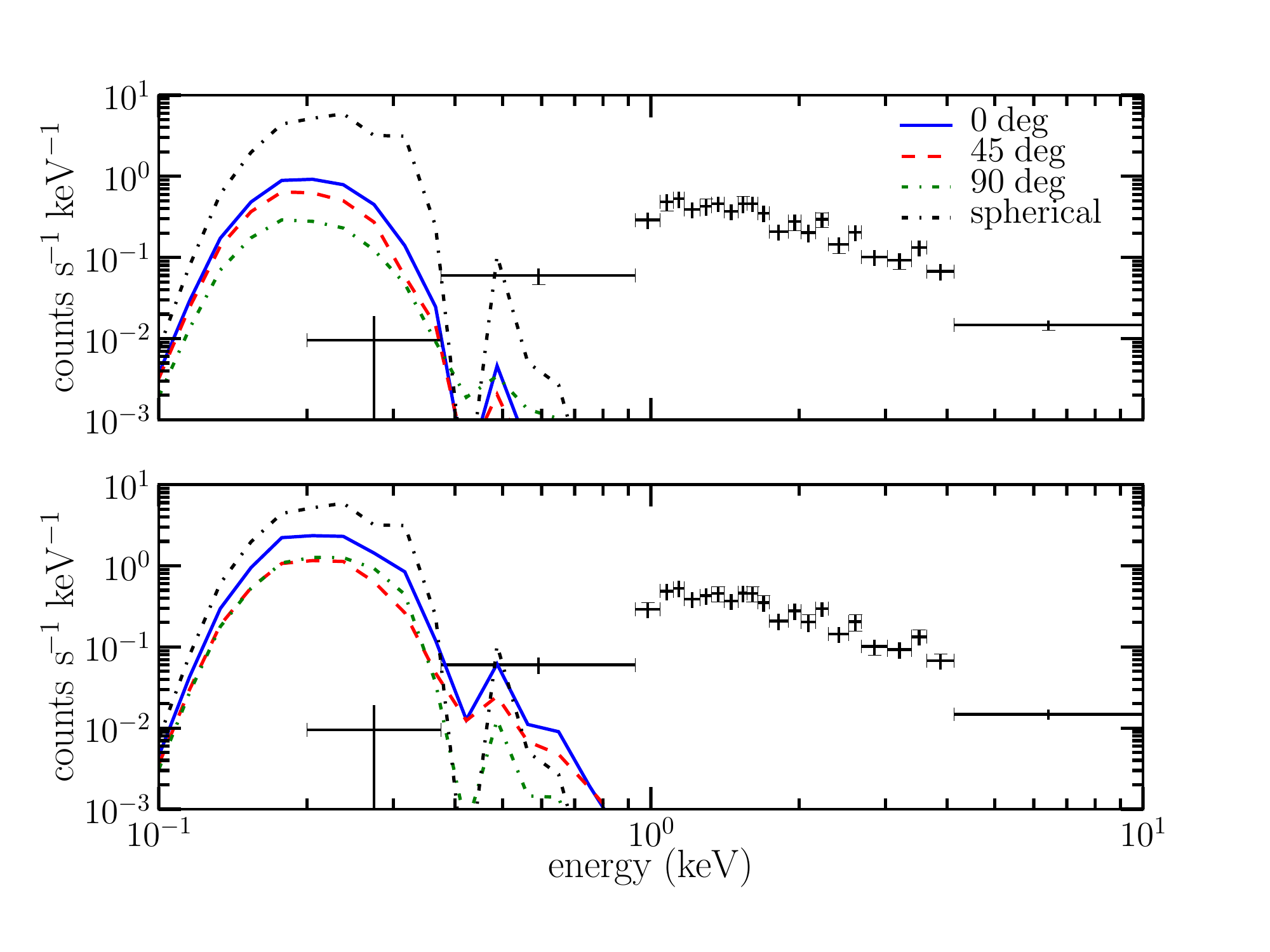}
\caption{Time-integrated X-ray spectra for the larger progenitor explosion models, m7r6cold (top) and m7r6hot (bottom) at different viewing angles.  The spectra are corrected for detector response and absorption along the line of sight, assuming a neutral matter column depth of $N_{\rm H} = 1.7\times10^{20}\ {\rm cm^{-2}}$.  For comparison, we also plot the spectrum for the spherical explosion, m7r6sph,  and the observed XRT spectra of XRO 080109.}
\label{fig:spec_m7r6}
\end{figure}

We have calculated shock breakout X-ray spectra and light curves at various observer viewing angles for the four jet-driven explosion models.  The light curves are presented in Figures \ref{fig:lc_m2r1cold} - \ref{fig:lc_m7r6hot} and the time-averaged spectra in Figures \ref{fig:spec_m2r1} - \ref{fig:spec_m7r6}.  All emission models are calculated assuming a neutral matter column depth along the line of sight, $N_{\rm H}$, of $1.7\times10^{20}$ cm$^{-2}$, corresponding to the Galactic value in the direction of SN 2008D \citep{Dickey:90}.  X-ray photon absorption due to neutral matter could occur both in the host galaxy (but well beyond the scales we simulate) and locally in the Milky Way.  All models are calculated using opacities appropriate for a helium gas with a solar mix of metals with metal abundances that are half of the solar values  (i.e., $Z=0.5Z_\odot$ = 0.008).  The influence of varying both $N_{\rm H}$ and $Z$ are discussed in Section \ref{sec:metal_abs}.  The general light curve characteristics are given in Table \ref{table:lc}.  

\begin{figure}
\centering
\plotone{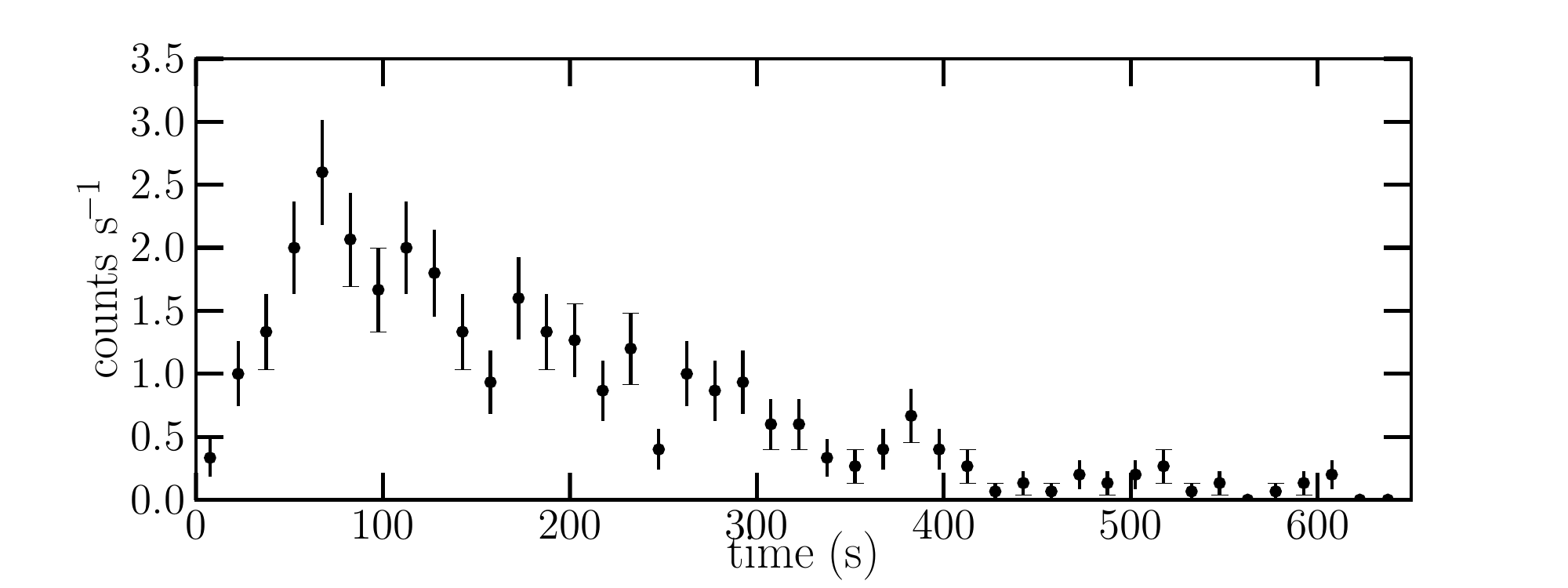}
\caption{The XRT light curve of XRO 080109.  The light curve is constructed using a photon binning length of 15 seconds.  The FWHM of the light curve is about 100 seconds and the rise time to maximum is about 70 seconds.  The X-ray burst lasts about 600 seconds.}
\label{fig:sn08Dlc}
\end{figure}

\begin{figure}
\centering
\plotone{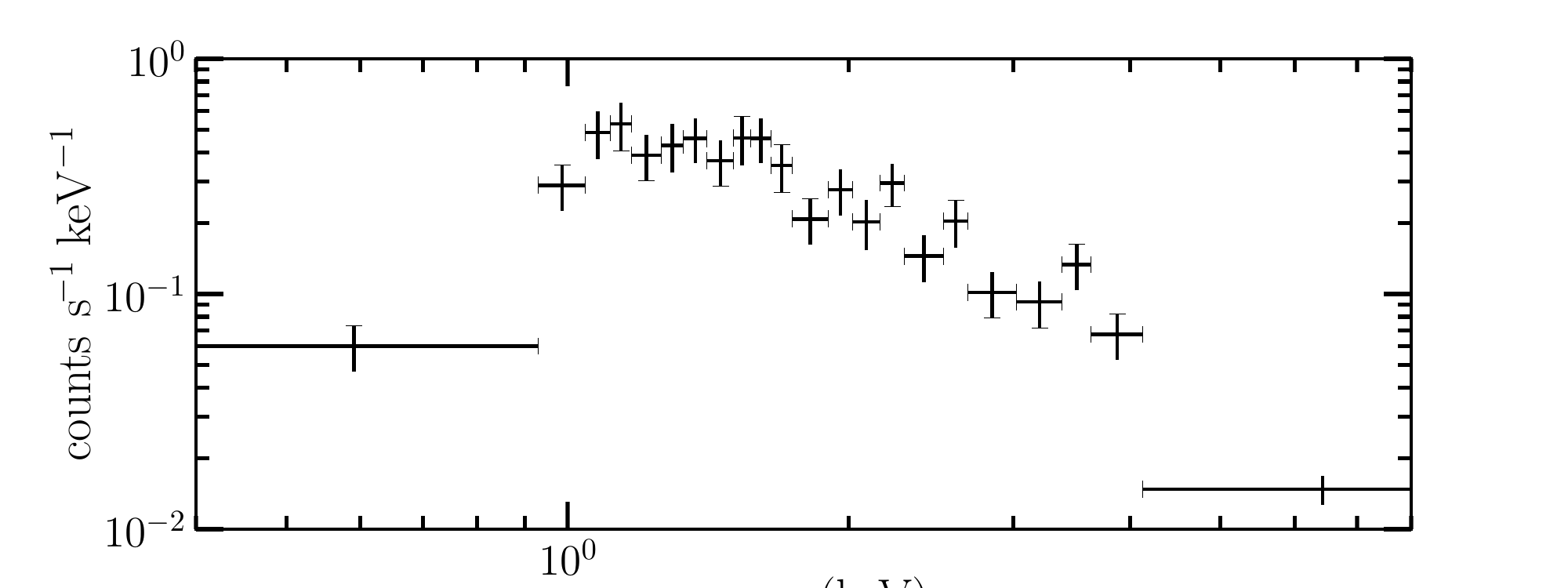}
\caption{XRT spectrum of XRO 080109 integrated over the total burst time of 600 seconds.}
\label{fig:sn08Dspec}
\end{figure}

We show the XRT data of XRO 080109 in Figures \ref{fig:sn08Dlc} and \ref{fig:sn08Dspec}.  To prepare these data, we downloaded {\it Swift} observation 00031081002 from the High Energy Astrophysics Science Archive Research Center \footnote{\url{http://heasarc.gsfc.nasa.gov/}} and reprocessed the XRT data with the ``xrtpipeline'' task using the latest calibration files.  We extracted events from SN 2008D using xselect and produced a spectrum and light curve.  Because SN 2008D was piled up during this observation, we used an annular extraction region of 7\farcs07 inner radius and 40\farcs81 outer radius, corresponding to the 40\% encircled energy radius of the PSF and the 90\% encircled energy radius, respectively.  The response files were produced using the ``xrtmkarf'' task, which took into account the fact that we extracted events from only 50\% of the PSF.  These response files were also used in the post-processing of our simulation results to compare the models directly to the 2008D data.  Therefore, the simulation light curves and spectra can be thought of as having been observed with only $\sim$1/2 of the XRT effective area.

\begin{deluxetable}{lccccc}
\tablewidth{0pt}
\tablecaption{Light Curve Characteristics}
\tablehead{
\colhead{Model} & \colhead{Angle \tablenotemark{a}} & \colhead{$L_{\rm X, max}$ \tablenotemark{b}} & \colhead{FWHM \tablenotemark{c}} & \colhead{$\Delta t$ \tablenotemark{d}}& \colhead{Energy \tablenotemark{e}}
}
\startdata
m2r1sph & \nodata & $12.7$ & 7.5 & 137.3 & $0.52$ \\
m2r1cold & 0 & 3.6 & 103.6 & 581.4  & 0.90 \\
m2r1cold & $\pi/4$ & 3.7 & 123.2 & 584.8 & 0.91 \\
m2r1cold & $\pi/2$ & 3.2 & 117.8 & 658.4 & 0.85 \\
m2r1hot & 0 & $7.4$ & 5.9 & 518.4 & $1.32$ \\
m2r1hot & $\pi/4$ & $7.5$ & 111.6 & 476.8 & $1.37$ \\
m2r1hot & $\pi/2$ & $4.9$ & 96.8 & 559.2 & $1.41$  \\
m7r6sph & \nodata & $9.4$ & 27.9 & 245.1 & $0.64$ \\
m7r6cold & 0 & 1.6 & 223.5 & 825.4 & 0.60 \\
m7r6cold & $\pi/4$ & 1.3 & 280.7 & 924.0 & 0.53 \\
m7r6cold & $\pi/2$ & 0.65 & 16.9 & 1104.3 & 0.34 \\
m7r6hot & 0 & 3.8 & 154.9 & 235.6 & 0.56 \\
m7r6hot & $\pi/4$ & 1.9 & 199.8 & 623.2 & 0.49 \\
m7r6hot & $\pi/2$ & 3.0 & 7.5 & 383.8 & 0.29 \\
\enddata
\tablenotetext{a}{Viewing angles of 0 represents a line of sight along the equator and viewing angles of $\pi/2$ represents a line of sight along the axis of symmetry.}
\tablenotetext{b}{Maximum X-ray luminosity in units of $10^{42}$ ergs s$^{-1}$.}
\tablenotetext{c}{Full width at half maximum X-ray count rate of the light curve in seconds.}
\tablenotetext{d}{Total time over which the X-ray count rate was greater than 10\% of maximum.}
\tablenotetext{e}{Total radiated energy integrated over $\Delta t$ in units of $10^{45}$ ergs.}
\label{table:lc}
\end{deluxetable}

Figures \ref{fig:lc_m2r1cold} - \ref{fig:lc_m7r6hot} show both the total X-ray luminosity and the predicted X-ray count rate, corrected for detector response, X-ray absorption, and distance.  As can be seen, the shapes of the count rate curve and the luminosity curve are not the same.  This is because the underlying spectrum is varying in time.  At times when the spectra are softer, more photons are produced at lower energies where the effects of X-ray absorption and detector response are stronger.  Thus it is not appropriate to use a constant count rate-to-luminosity conversion factor at all times.  

\begin{figure}
\centering
\plotone{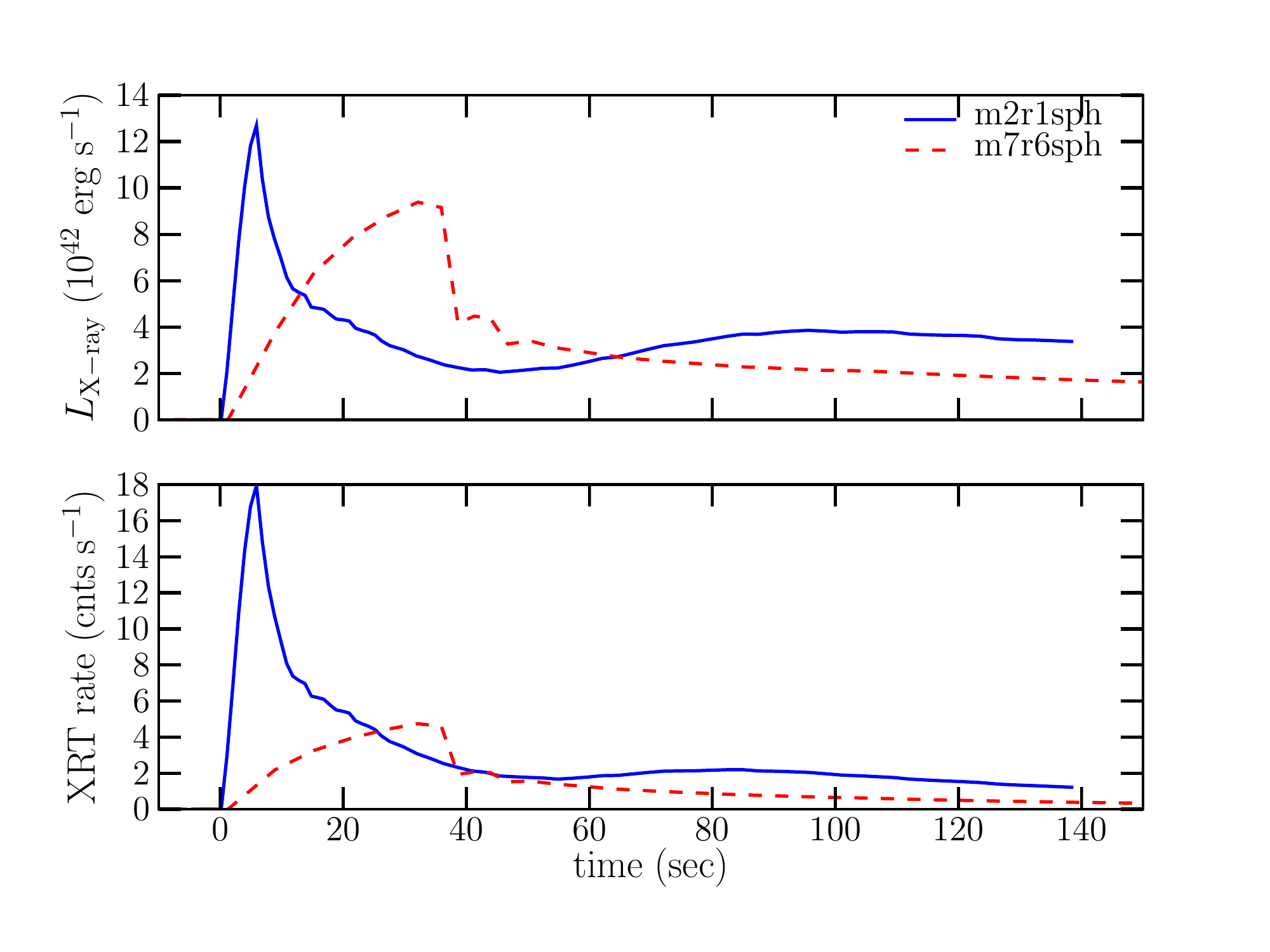}
\caption{XRT band (0.1 - 10 keV) light curves for the spherical explosion models, m2r1sph (solid blue line) and m7r6sph (dashed red line).  The time coordinates have been adjusted so that the peak of both light curves lies at $t=0$.  The top panel shows the X-ray luminosity and the bottom panel shows the XRT count rates corrected for detector response and absorption along the line of sight.  The width of the light curves is set by the light crossing time of the progenitors.  The light curve of m2r1sph is comprised of significantly harder photons than that of m7r6sph which is the reason the count rate light curves show more disparity than the luminosity curves. }
\label{fig:lc_sph}
\end{figure}

\begin{figure}
\centering
\plotone{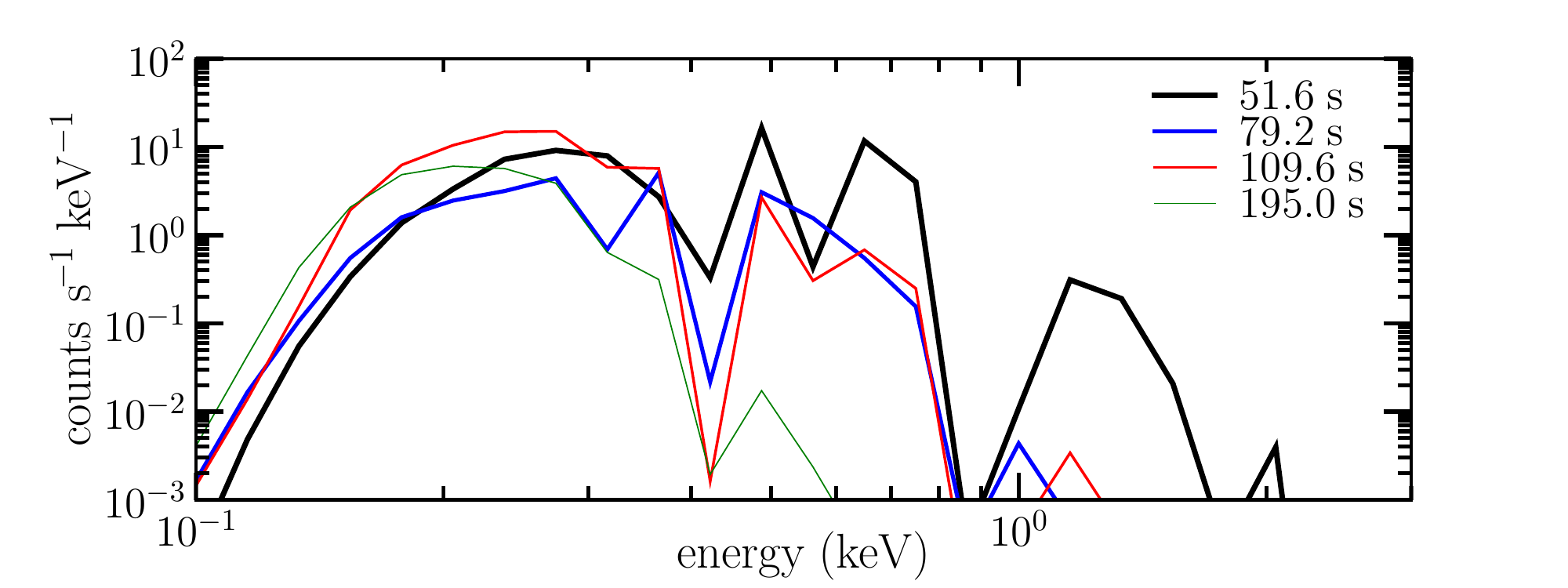}
\caption{ Instantaneous XRT spectra for model m2r1cold at four different times during the first 150 seconds of breakout emission. The times of the spectra are indicated in the legend. The spectra show significant evolution to softer emission as the burst proceeds.}
\label{fig:spec_m2r1time}
\end{figure}

In each explosion model, the spectrum varies significantly throughout the burst.  This is true even in the spherical explosions (see Figure \ref{fig:lc_sph}), though to a lesser extent, because as the radius of the thermalization depth increases and the temperature there cools adiabatically, the reduction in emissivity is somewhat balanced by the increased emitting surface area.  Figure \ref{fig:spec_m2r1time} demonstrates the temporal variability of the spectrum for model m2r1cold.  This figure shows the instantaneous spectrum of m2r1cold at four times during the first 150 seconds of breakout emission.  The thermalization layer in each explosion lies in the region between the forward and reverse shocks.  At later parts of the X-ray bursts, the thermalization layer coincides roughly with the contact discontinuity between the swept-up wind and the reverse-shocked ejecta, where the density increases significantly.  Therefore, the nature of the breakout emission can depend significantly on the character of the wind.  The absence of a progenitor wind (or the limit of a very low-density wind) may result in the thermalization layers being driven deeper into the ejecta, possibly below the contact discontinuity.  Since the temperature is highly discontinuous at the contact discontinuity, this could dramatically change the emission.

\begin{figure*}
\centering
\begin{tabular}{ccc}
\includegraphics[width=2in]{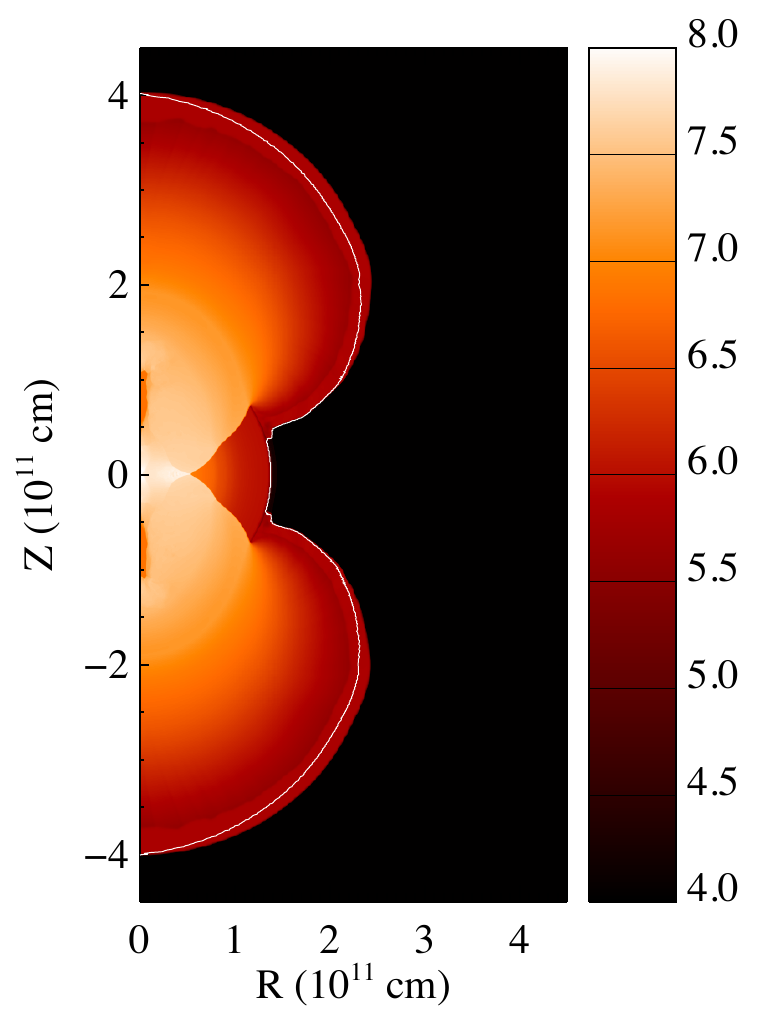}
\includegraphics[width=2in]{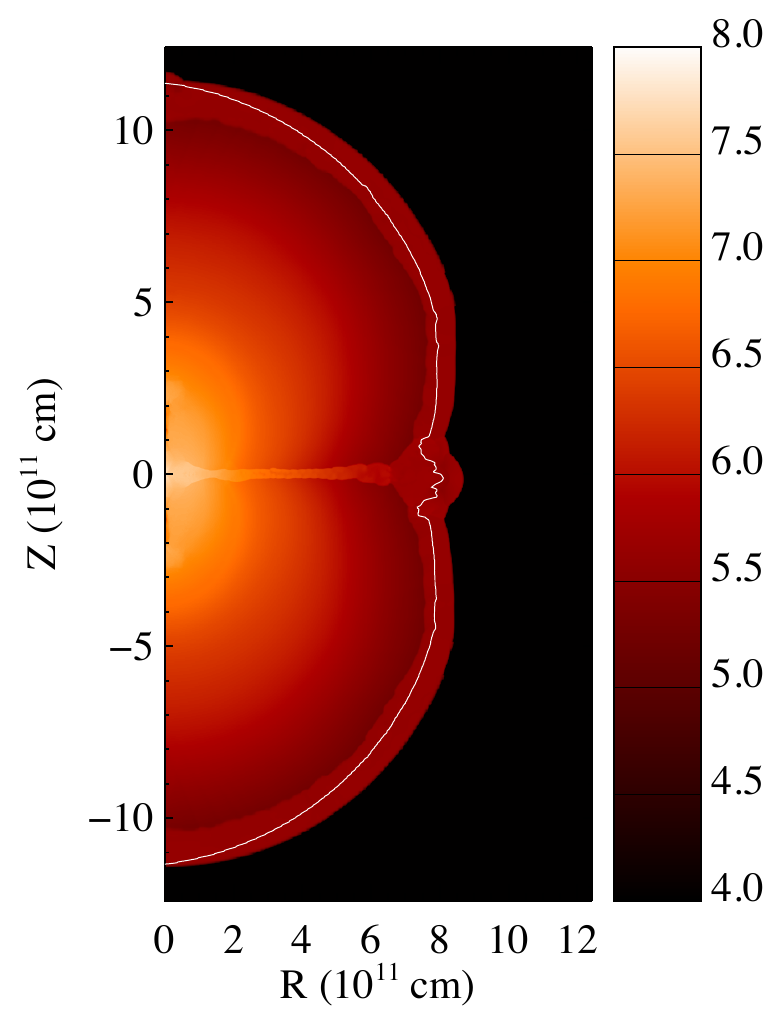}
\includegraphics[width=2in]{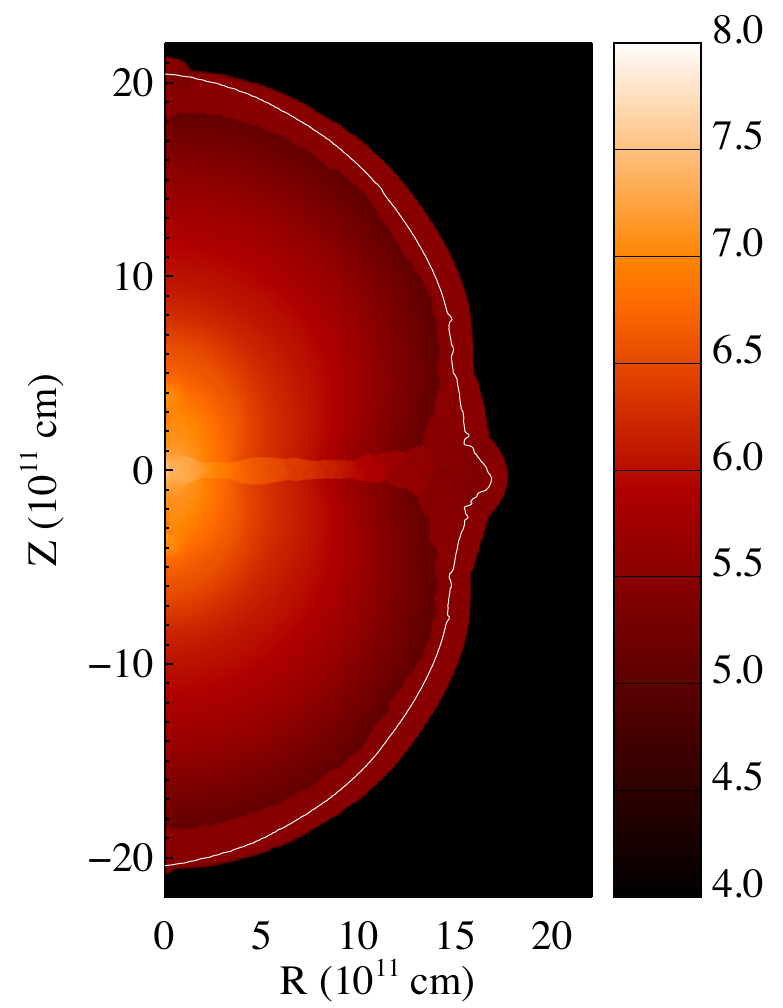}
\end{tabular}
\caption{The logarithmic temperature in Kelvin for model m2r1cold at three post-breakout times:  68.2 s, 124.9 s, 202.3 s, from left to right.  The white contour shows the location of the thermalization depth at 0.75 keV.  The line-of-sight is along the equator.  Prior to shock eruption, the thermalization depth is located in the transition layer between the wind and the progenitor model.  Post-shock-breakout, the thermalization depth is in the shocked wind, ahead of the reverse shock.}
\label{fig:thermDepth}
\end{figure*}

The light curves for these jet-driven explosions may be generally described as having two peaks.  The first is associated with the initial shock breakout at the poles.  The dip in the light curves following this first peak is due to the decreasing temperature of the thermalization layer as the shock moves out further into the wind.  The second peak is attributed to emission from the equator as the bipolar shocks cross there and create a pancake of very hot, twice-shocked gas.  As the explosions evolve more toward sphericity at later times, the light curves decay exponentially.  The time scales of the light curves are related to the shock crossing times of the progenitors.  Figure \ref{fig:thermDepth} shows the post-breakout evolution of the temperature and thermalization depth at 0.75 keV for model m2r1cold.  Before shock breakout the location of the thermalization layer is the transition region between the wind and the progenitor star, where the density increases rapidly.  Emission from these regions is, however, negligible in the XRT band due to the relatively low temperatures there.  After shock breakout the thermalization layer lies in between the forward and reverse shocks, the region comprise of shocked, accelerated wind material.

Light curves of explosion models m2r1cold and m2r1hot (the smaller progenitor) are shown in Figures \ref{fig:lc_m2r1cold} and \ref{fig:lc_m2r1hot} for three viewing angles:  along the equator (0 degrees), 45 degrees, and along the axis of symmetry (90 degrees).  The widths of the light curves are roughly 100 seconds at all angles for both models and the total X-ray burst time is around 500 to 600 seconds.  These values are very close to what was observed for SN 2008D, shown in Figure \ref{fig:sn08Dlc}.  Examining the shape of the model light curves shows that for 0 degree viewing angles the light curves rise very quickly.  This is due to two effects.  The first is simply because the emission from both poles is visible.  The second is due to the nature of an aspherical breakout and the effects of light travel corrections.  As the bipolar shocks continue to erupt from the progenitor surface, the brightest emission is coming from where the shocks are just reaching the surface (essentially two rings moving across the stellar surface from the poles to the equator).  Thus, the brightest emitting regions are moving rapidly {\it toward} the observer causing a pile up of emission once light travel time corrections are made.  This effect is reduced, or eliminated, at higher viewing angles because the brightly emitting rings are no longer moving so much toward the observer.  

Table \ref{table:lc} lists the resulting light curve characteristics for our models m2r1cold and m2r1hot, as well as m2r1sph for comparison.  As expected, the spherical explosion has a higher peak luminosity, but a much shorter FWHM and overall burst time, $\Delta t$.  The width of the light curve for m2r1sph is set by the light crossing time of the progenitor. The simulated XRT count rates for models m2r1hot and, especially, m2r1cold are very similar to those we find for XRO 080109/SN 2008D.  The light curve rise time and FWHM for model m2r1cold at a viewing angle of 45 degrees (red dashed curve in Figure \ref{fig:lc_m2r1cold}) is a good match to SN 2008D.

The X-ray spectra for models m2r1cold and m2r1hot are shown in Figure \ref{fig:spec_m2r1}.  These spectra are corrected for the detector response function and for X-ray absorpotion (assuming $N_{\rm H} = 1.7\times10^{20}$).  They are also averaged over the burst time, $\Delta t$, as was done for the observations of XRO 080109/SN 2008D \citep[see Figure \ref{fig:sn08Dspec};][]{Soderberg:08, Modjaz:09}.  The shapes of the spectra at the different viewing angles are very similar.  Indeed there is not much difference between the two different models.  The time-averaging washes away the major differences that are apparent in the light curves.  The shape of the spherical explosion spectra are also very similar.  They are generally brighter, but this is due to a shorter averaging time.  The spectra in each case are softer than the spectrum of XRO 080109 (see Figure \ref{fig:sn08Dspec}).

The light curves for the simulations in the larger progenitor (Figures \ref{fig:lc_m7r6cold} and \ref{fig:lc_m7r6hot}) are characterized by much longer time scales and overall less bright emission.  The FWHM for models m7r6cold and m7r6hot range from around 100 to 200 seconds at various angles while the total burst times are from 300 to 1000 seconds (see Table \ref{table:lc}).  The peak luminosities and total radiated energies are less than in the analogous explosions in the smaller progenitor.  While the ratio of explosion energies to ejecta masses are roughly equivalent across all simulations, the reduced luminosities in the larger progenitor can be explained by a slightly lower shock velocity during the burst and a lower wind density at the radius of shock breakout (i.e., the radius of the progenitor star).  As we discuss in more detail in Section \ref{sec:wind}, the density of the wind plays an important role in the strength of the X-ray emission because the thermalization depths during the bursts lie in the shocked wind.  The light curve for m7r6cold is similar in shape to m2r1cold and m2r1hot but with longer timescales.  Model m7r6hot, however, exhibits a dramatically double-peaked light curve.

\begin{figure}
\centering
\plotone{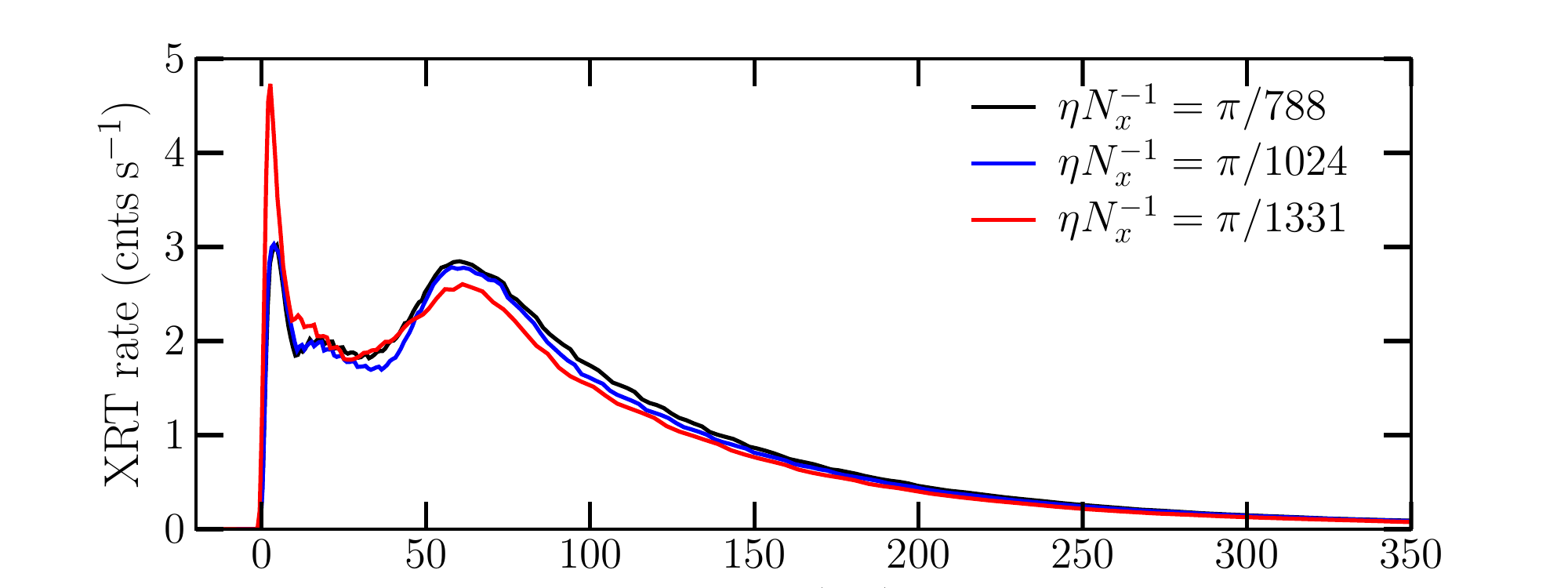}
\caption{XRT count rate light curves for model m2r1cold at three different resolutions. }
\label{fig:m2r1cold_res}
\end{figure}

As discussed in Section \ref{sec:simulations}, we ran simulations of model m2r1cold at three different resolutions.  Figure \ref{fig:m2r1cold_res} shows the X-ray light curves for these three simulations plotted together.  The light curves of the low- and fiducial resolution simulations are extremely similar.  The high-resolution case varies only slightly from the lower-resolution light curves.  The early peak is brighter, due to the slightly greater shock speeds just after breakout (see \ref{sec:wind}).  The second peak is slightly dimmer, due to the lesser degree of extension of the southern shock structure in the high-resolution simulation, leading to a smaller emitting area as seen from a viewing angle of 0 degrees.   

\subsection{Dependence on Metallicity and X-ray Absorption}
\label{sec:metal_abs}

\begin{figure}
\centering
\plotone{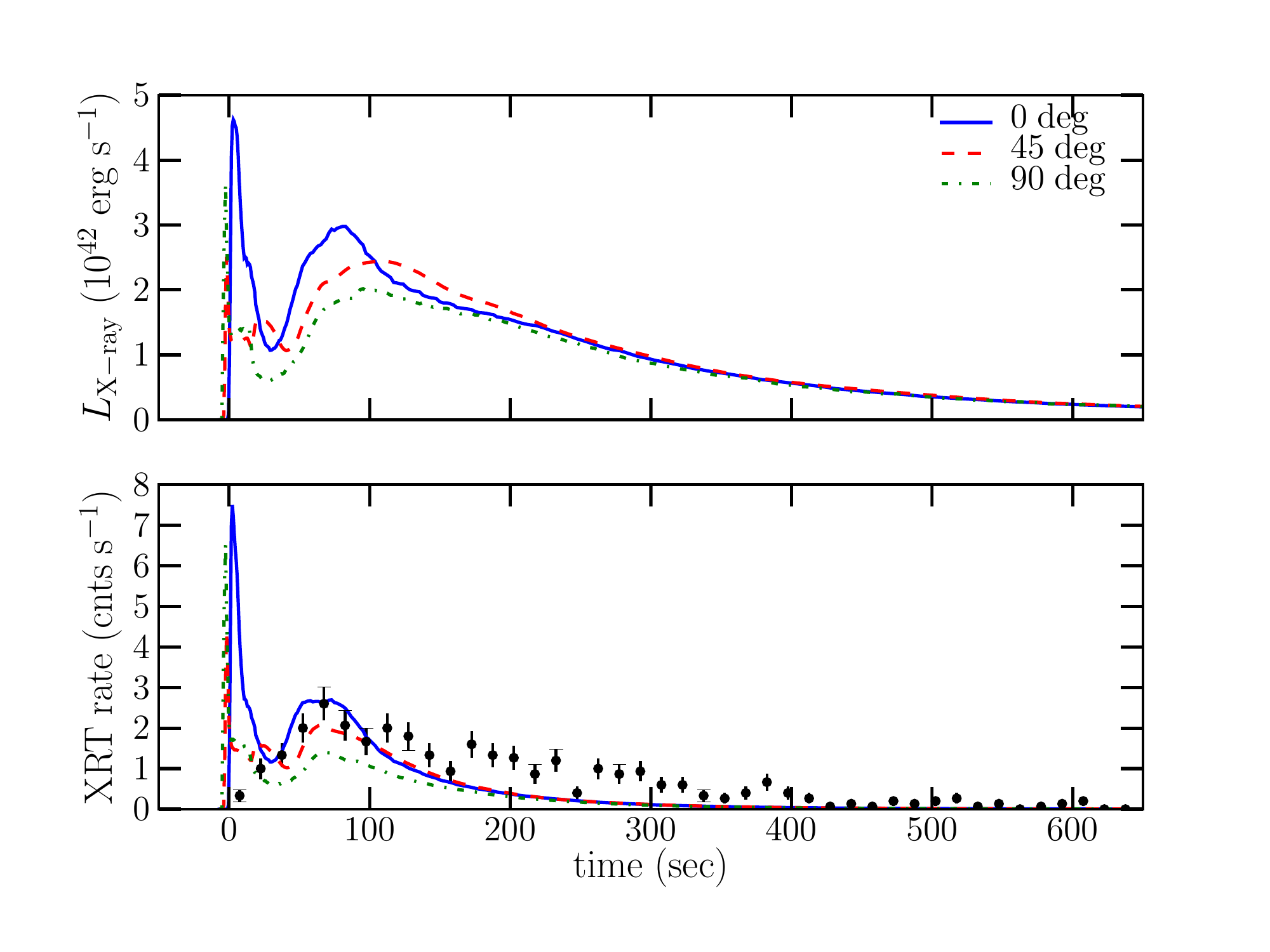}
\caption{XRT band (0.1 - 10 keV) light curves for m2r1cold at viewing angles of 0, 45, and 90 degrees calculated assuming a metallicity of $Z=0.1Z_\odot$.  The top panel shows the X-ray luminosity as a function of time and the bottom panel shows the predicted XRT count rate accounting for detector response and absorption due to neutral matter. The observed XRT light curve of XRO 080109 is also shown for comparison. }
\label{fig:lc_m2r1cold_heTenthSolar}
\end{figure}

\begin{figure}
\centering
\plotone{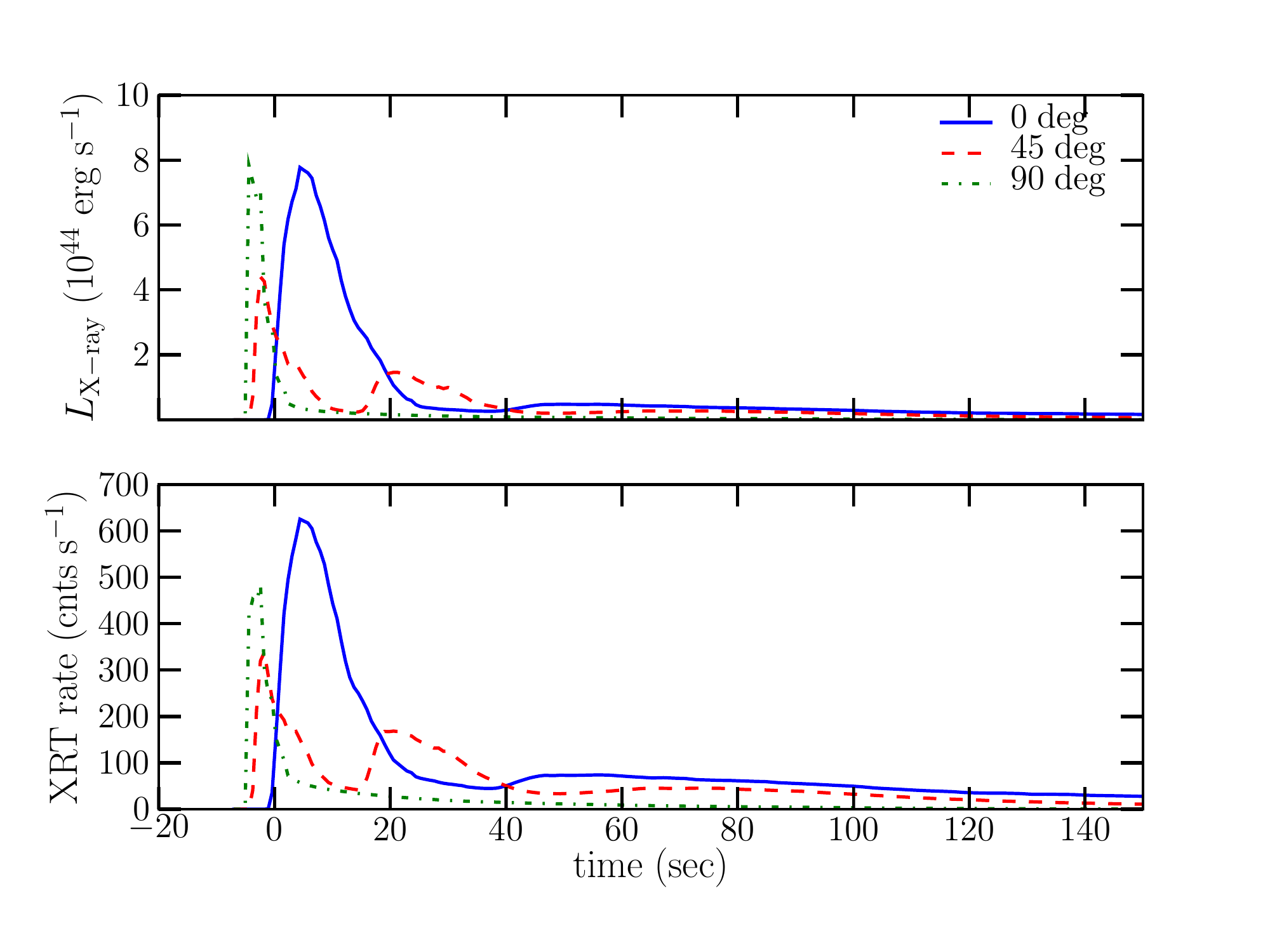}
\caption{XRT band (0.1 - 10 keV) light curves for m2r1cold at viewing angles of 0, 45, and 90 degrees calculated assuming a metallicity of $Z=0Z_\odot$, metal-free gas.  The top panel shows the X-ray luminosity as a function of time and the bottom panel shows the predicted XRT count rate accounting for detector response and absorption due to neutral matter. }
\label{fig:lc_m2r1cold_heNew}
\end{figure}

\begin{figure}
\centering
\plotone{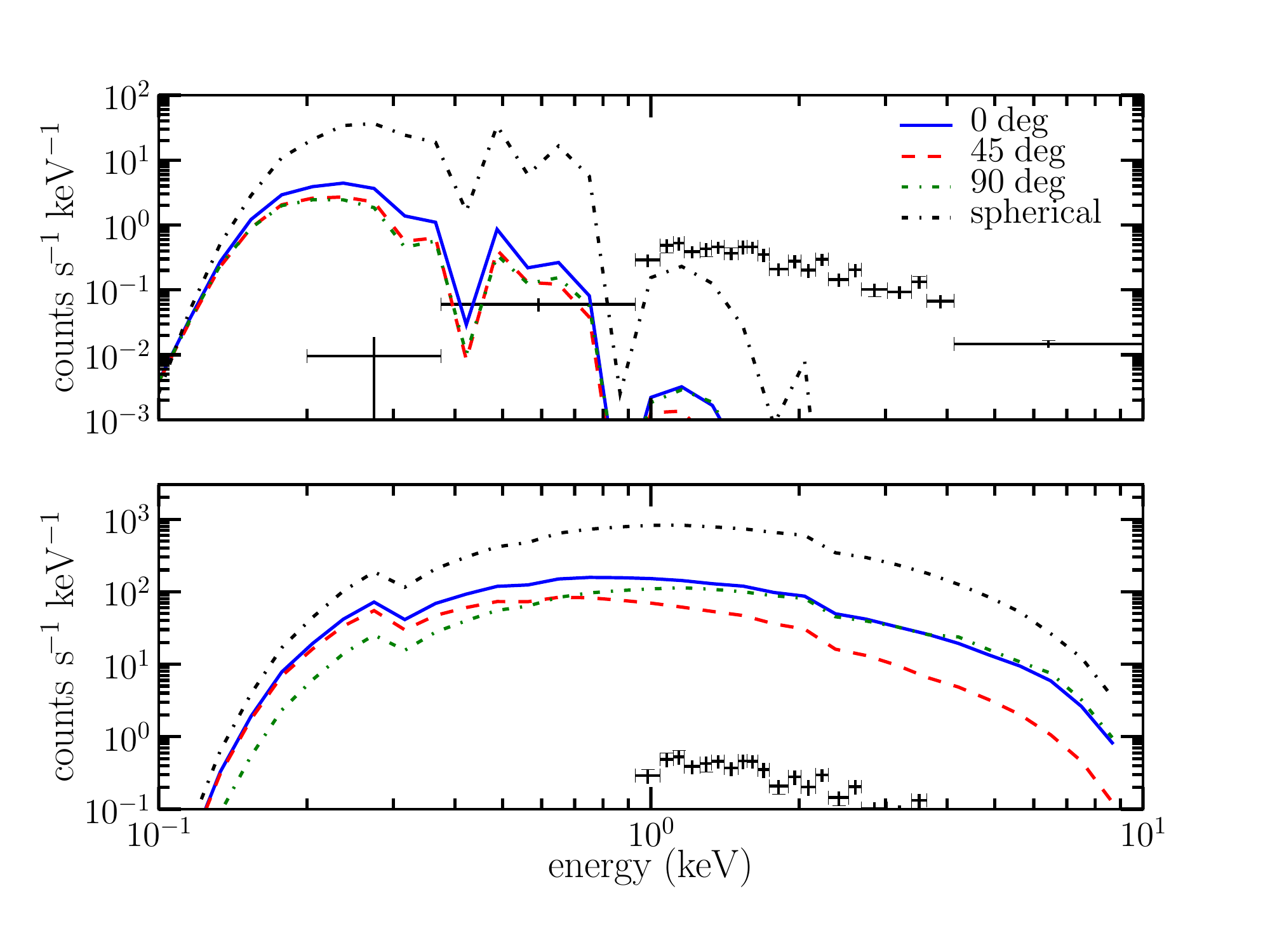}
\caption{Time-integrated X-ray spectrum for model m2r1cold using $Z=0.1Z_\odot$ (top) and $Z=0$ (bottom) along with the spectrum for the spherical explosion at the respective metallicities.  The observed XRT spectrum of XRO 080109 is also plotted.}
\label{fig:spec_m2r1cold_metals}
\end{figure}

The absorptive opacities depend strongly on the metallicity of the absorbing gas.  To illustrate this we have calculated simulated spectra and light curves of our explosion models using metallicity values of $0.1 Z_\odot$ and $0.0 Z_\odot$ (i.e., metal-free).  Figures \ref{fig:lc_m2r1cold_heTenthSolar} and \ref{fig:lc_m2r1cold_heNew} show the light curves for model m2r1cold for metallicities of $Z=0.1Z_\odot$ and $Z=0$.  The spectra for these cases are shown in Figure \ref{fig:spec_m2r1cold_metals}.  The lower metallicity drives the thermalization depth deeper into the explosion where the temperatures are higher, resulting in significantly increased emission (note the difference in scale in Figure \ref{fig:lc_m2r1cold_heNew}).  For the case of metal-free gas, this increase is dramatic.  Due to the significantly reduced absorptive opacities, the thermalization depths are pushed down below the reverse shock and into the deep, very hot regions of the ejecta.  The calculated emission for this metallicity is orders of magnitude greater than the other models and the observations of SN 2008D.  The difference in the emission characteristics between our fiducial models with $Z=0.5Z_\odot$ and the models with $Z=0.1Z_\odot$ is less drastic.  This is because for $Z=0.1 Z_\odot$, the thermalization depths are still above the reverse shock and the temperature in between the forward and reverse shocks does not vary greatly.  The metal-free case spectra also noticeably lack the deep `absorption' features at $\sim0.4$ keV and $\sim0.9$ keV.  These `absorption' features are caused by a significant increase in the absorptive opacities at these energies (due to the presence of metals), pushing the thermalization depths to larger, cooler radii.  We note that our fiducial metallicity of $Z=0.5Z_\odot$ is consistent, within the accuracies, with three independent measurements of the metallicity of the region around SN 2008D: \citet{Soderberg:08, Thoene:09, Modjaz:10}.

\begin{figure}
\centering
\plotone{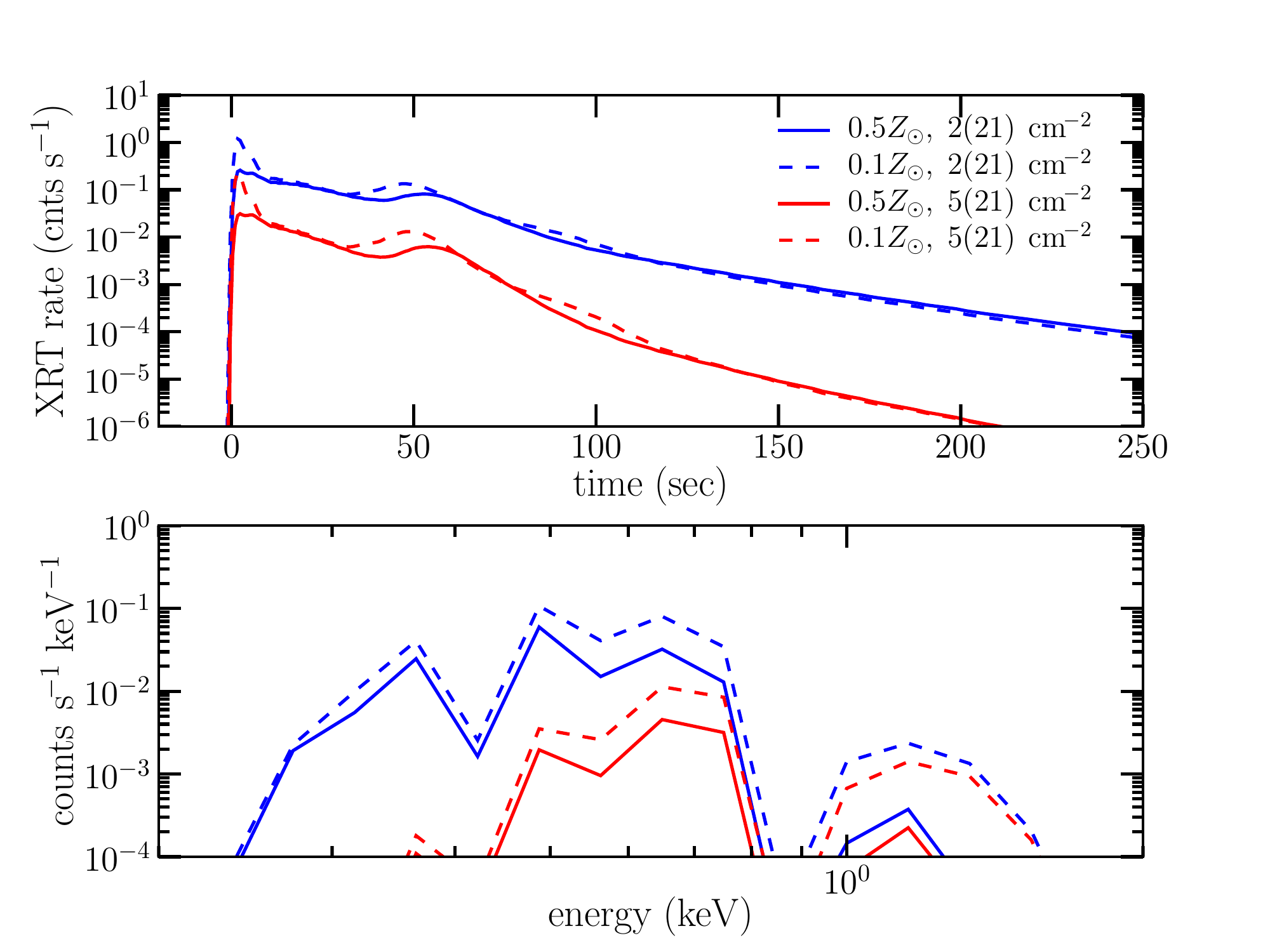}
\caption{XRT count rate light curve (top) and time-integrated XRT spectrum (bottom) for m2r1cold calculated using metallicities of $0.1Z_\odot$ and $0.5Z_\odot$ and neutral matter column depth of $2\times10^{21}$ cm$^{-2}$ and $5\times10^{21}$ cm$^{-2}$.}
\label{fig:m2r1cold_nh}
\end{figure}

\begin{figure}
\centering
\plotone {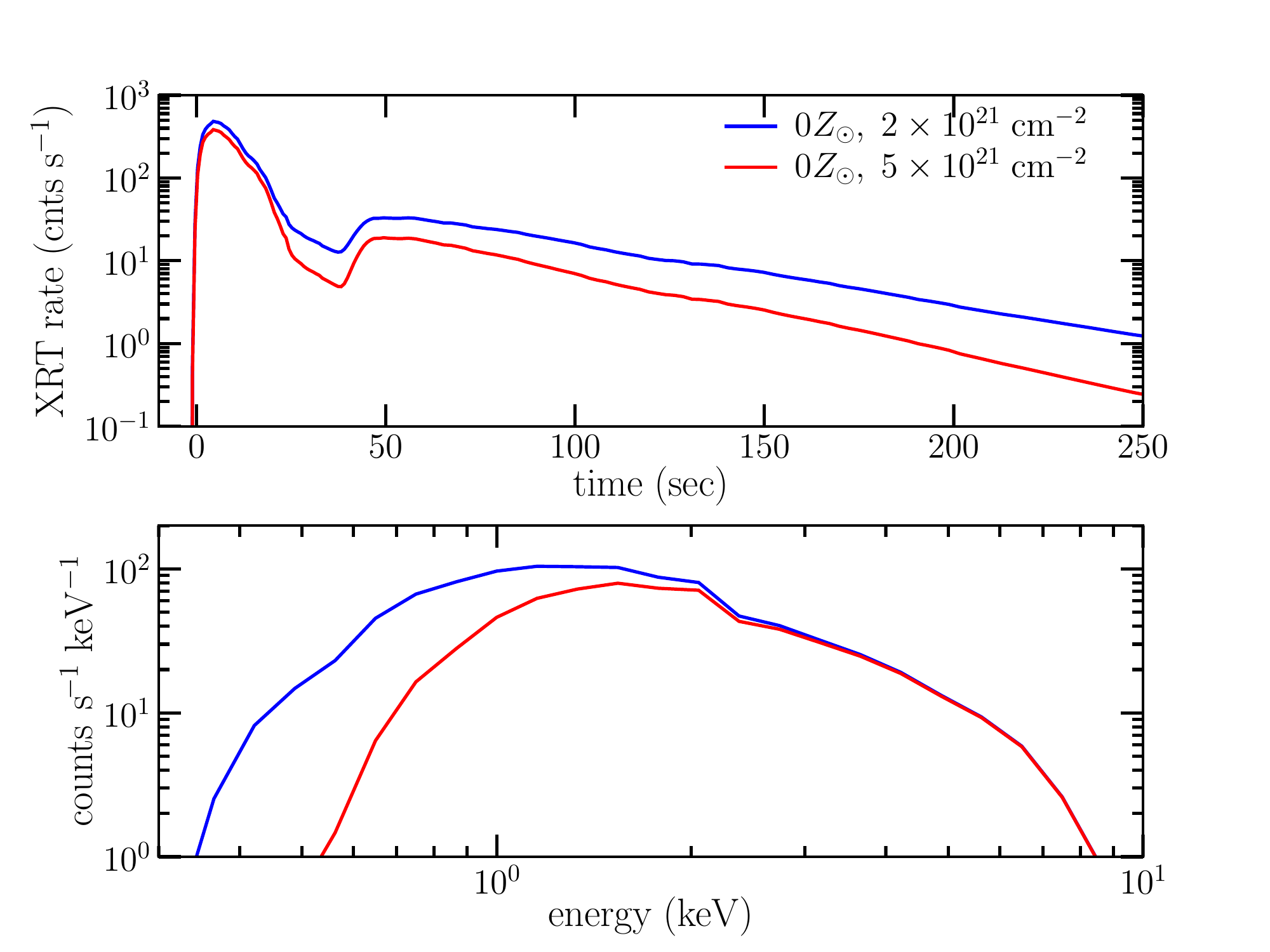}
\caption{XRT count rate light curve (top) and time-integrated XRT spectrum (bottom) for m2r1cold calculated for metal-free gas and neutral matter column depth of $2\times10^{21}$ cm$^{-2}$ and $5\times10^{21}$ cm$^{-2}$.}
\label{fig:m2r1cold_nhnew}
\end{figure}

We plot XRT count rates and spectra for model m2r1cold using different values of $N_{\rm H}$ in Figures \ref{fig:m2r1cold_nh} and \ref{fig:m2r1cold_nhnew}.  If the location of SN 2008D in its host galaxy were particularly dense, X-ray absorption by neutral matter atoms in the host galaxy may be significant, warranting a $N_{\rm H}$ beyond the Galactic value of 1.7$\times10^{20}$ cm$^{-2}$.  Increased hydrogen column depth increases the absorption of lower-energy X-ray photons.  Since our simulated spectra for the non-zero metallicity cases are dominated by emission from below about 1 keV, increasing $N_{\rm H}$ dramatically reduces the resultant count rates.  Due to a significantly greater amount of hard emission in the $Z=0$ model, the light curve and spectrum are effected very little by an increase in $N_{\rm H}$.  It is possible to increase the neutral matter column depth so that the peak count rates for the metal-free model are similar to those observed for XRO 080109 ($< 10$ cnts s$^{-1}$).  This requires, however, an unrealistically high value of $N_{\rm H}$, greater than $10^{23}$ cm$^{-2}$ and results in a very narrow light curve as the softer emission after the first 20 seconds is almost entirely absorbed.

\subsection{Color Temperature Enhancement}
\label{sec:wind}

In our simulations, the thermalization depths where the X-ray spectrum is formed, lie in the region of shocked wind, behind the forward shock and ahead of the reverse shock.  Since we assume local emission with a black-body spectrum, the resultant luminosity is a strong function of the temperature used in calculating the spectrum.  In the region in between the forward and reverse shock, this temperature is dependent on the forward shock velocity and the wind density.  To see this, assume that enthalpy flux is conserved at the shock front such that, in the frame of the shock, 
\beq
\rho_1 v_1 ( \onehalf v_1^2 + \epsilon_1+P_1/\rho_1) = \rho_2 v_2 ( \onehalf v_2^2 + \epsilon_2+P_2/\rho_2),
\label{eq:eshock}
\eeq
where $\rho$ is gas density, $v$ is gas velocity, $\epsilon$ is the specific internal energy of the gas, $P$ is the gas pressure, and subscript 1 denotes pre-shock values and subscript 2, post-shock values.  We can assume that the post-shock internal energy is dominated by contributions from radiation, $\epsilon_2 = 3P_{2}/\rho_2 = a_{\rm rad} T_2^4 / \rho_2$, and the density jump at the shock is $\rho_2 = 7 \rho_1$, for a strong shock.  So long as the shock is strong, the upstream internal energy is negligible, $\epsilon_1 \approx P_1 \approx 0$.  Conservation of mass flux at the shock also gives $v_2 = v_1  \rho_1 / \rho_2 = v_1/7$, in the shock frame.  In the frame of the progenitor star, $v_1$ is the shock speed, $v_{\rm sh}$, as long as the shock is moving fast relative to the pre-shock gas.  Substituting these relations into equation (\ref{eq:eshock}) and solving for the post-shock temperature yields
\beq
T_2 \approx 7.6\times10^{5}\  \rho_{-11}^{1/4}\ v_{\rm sh, 10}^{1/2}\ {\rm K},
\eeq
where $\rho_{-11}$ is the wind density in units of $10^{-11}$ g cm$^{-3}$ and $v_{\rm sh,10}$ is the shock velocity in units of $10^{10}$ cm s$^{-1}$, appropriate for shock breakout into a Wolf-Rayet wind.  Thus, the post-shock temperature may be increased by enhancing the wind density or the energy of the explosion (which, in turn, increases $v_{\rm sh}$).

In order to demonstrate the influence that an increased post-shock temperature has on our simulated spectra and light curves, we present the spectrum and light curve for model m2r1cold calculated by using a temperature that had been enhanced by a factor of 1.8 above the temperature found in our hydrodynamic simulations.  Such an increase in temperature would result from a factor of ten increase in $\rho_1 v_{\rm sh}^2$.  \citet{Tanaka:09} find that one dimensional explosion models with kinetic energy to ejecta mass ratios of 1.4 - 1.7  $(10^{51}\ {\rm erg} / M_\sun)$ fit the late time spectrum and light curve of SN 2008D best.  This ratio is about a factor of 2 greater than what we have used in our most simulations, including m2r1cold.  This alone could account for a ten-fold increase in the shock ram pressure $\rho_1 v_{\rm sh}^2$ because the mass averaged-velocity will scale roughly as $\bar v_{\rm ej}^2 \propto E_{\rm K} / M_{\rm ej}$ and the shock speed will far exceed $\bar v_{\rm ej}$.  

\begin{figure}
\centering
\plotone{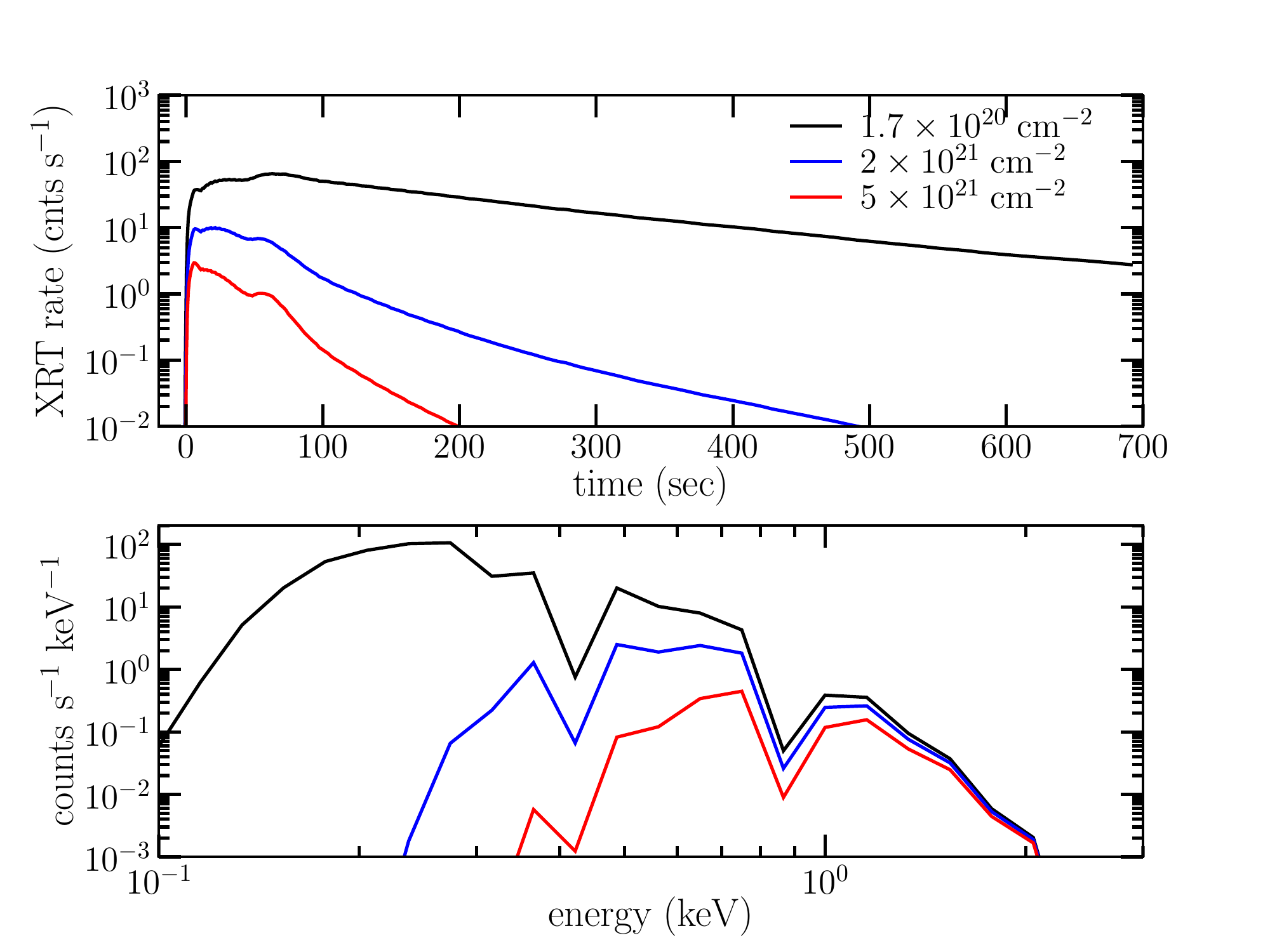}
\caption{XRT count rate light curves (top) and time-integrated XRT spectra (bottom) for m2r1cold calculated using a temperature enhancement factor of 1.8 at various values of $N_{\rm H}$.}
\label{fig:m2r1cold_ht}
\end{figure}

\begin{figure}
\centering
\plotone{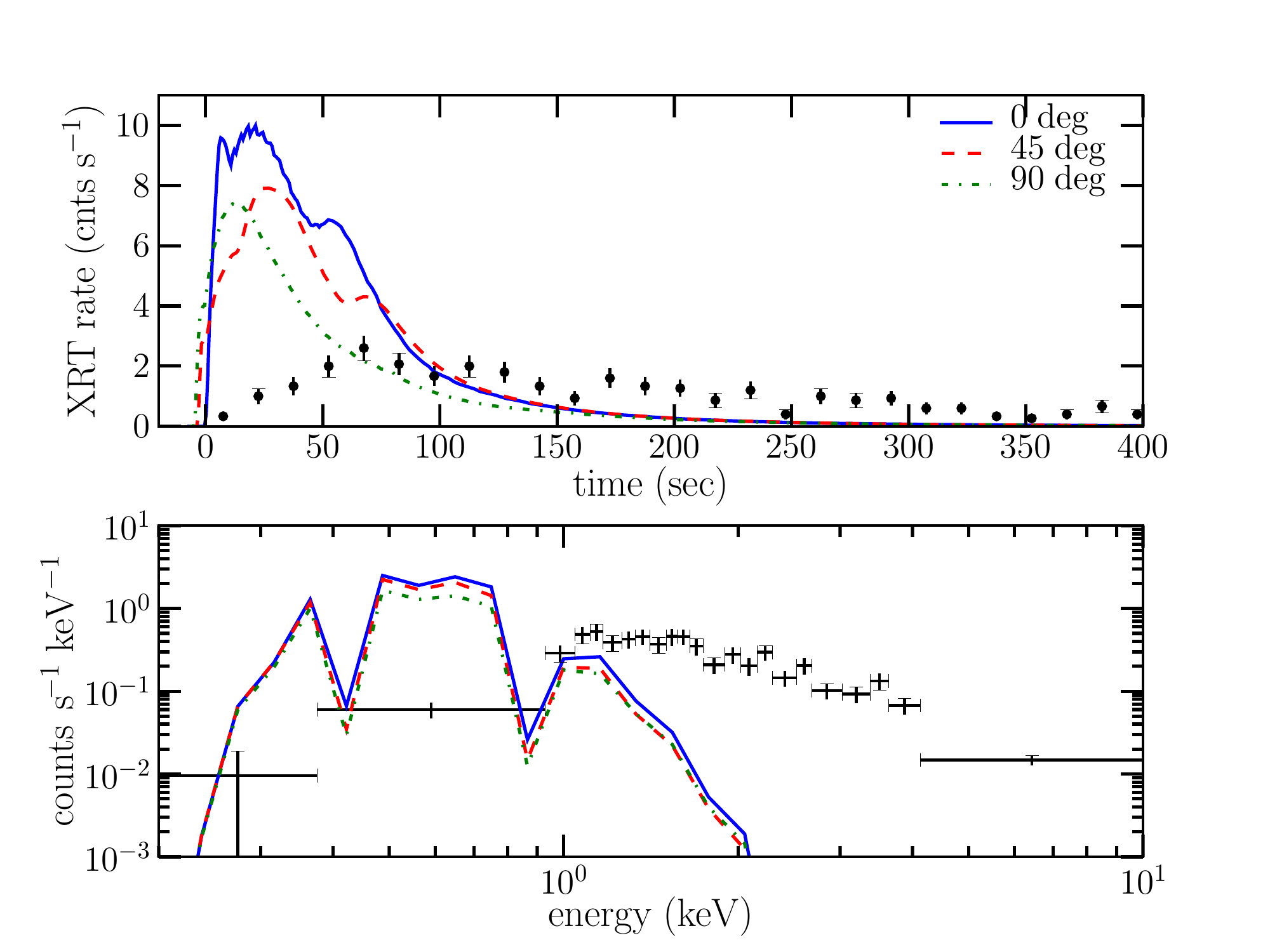}
\caption{XRT count rate light curve (top) and time-integrated XRT spectra (bottom) for the enhanced temperature version of m2r1cold at various viewing angles assuming a neutral matter column depth of $2\times10^{21}\ {\rm cm^{-2}}$.  Also shown are the observed XRT light curve (top) and spectrum (bottom).}
\label{fig:m2r1cold_ht2}
\end{figure}

The results of this enhanced temperature calculation for model m2r1cold are shown in Figure \ref{fig:m2r1cold_ht}.  The simulated light curves and spectra are all for a metallicity of $0.5 Z_\odot$ and a viewing angle of 0 degrees.  Figure \ref{fig:m2r1cold_ht} shows the behavior of the simulated emission with increased $N_{\rm H}$.  The enhanced temperature calculation yields a peak X-ray luminosity of $4.9\times10^{43}$ erg s$^{-1}$ and radiates a total of $2.0\times10^{46}$ ergs.  For the case of $N_{\rm H} = 1.7\times10^{20}\ {\rm cm^{-2}}$, the X-ray count rate peaks at 65 s$^{-1}$, much higher than was observed for XRO 080109.  Increasing the column depth brings down the peak count rate, narrows the width of the light curve, and hardens the spectrum.  For $N_{\rm H} = 2\times10^{21}\ {\rm cm^{-2}}$, the peak count rate is 10 s$^{-1}$, slightly greater than the observed value.  The light curve at this column depth is somewhat narrower and rises more quickly than that of XRO 080109.  The spectrum is also still too soft and devoid of significant count rates above 2 keV.  In Figure \ref{fig:m2r1cold_ht2} we show the angular dependence of the simulated light curve and spectra for the enhanced temperature version of m2r1cold calculated with a metallicity of $0.5Z_\odot$ and $N_{\rm H} = 1.7\times10^{20}\ {\rm cm^{-2}}$.  The spectrum is harder than the fiducial models, computed without temperature enhancement, but is still softer than that of XRO 080109.

The color temperature of the emission may also be enhanced by non-LTE effects. The radiation escaping from the shock during breakout from a Wolf-Rayet star is not in thermal equilibrium with the gas at $\tau_* = 2/3$ \citep{Katz:09, Nakar:10}.  In fact, during the early parts of the shock breakout the radiation temperature may be orders of magnitude greater than the gas temperature at  $\tau_* = 2/3$ \citep{Nakar:10}.  Over the course of the first tens of seconds of the X-ray burst, the radiation temperature will quickly drop down closer to the gas temperature.  This time-dependent enhancement of the radiation temperature, if accounted for in our calculations, could produce the hard X-rays that are missing in our spectra while not dramatically increasing the luminosity at later times in the light curve.

\subsection{SED Shape}

\begin{figure}
\centering
\plotone{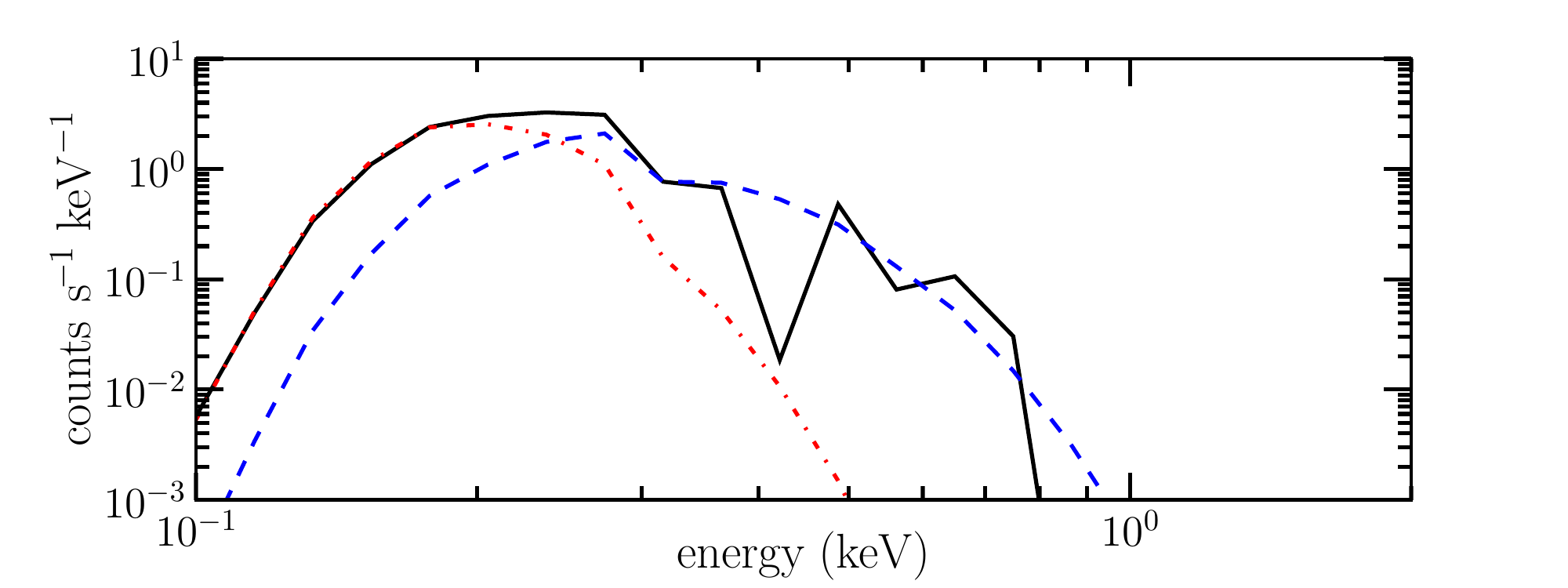}
\caption{X-ray spectrum of jet explosion model m2r1cold (black line) plotted with two spherical, single temperature and radius black-body SEDs, corrected for XRT detector response and X-ray absorption. The red dash-dot line is a black-body SED with spherical radius of $1.2\times10^{12}$ cm and temperature $3\times10^5$ K.  The blue dashed line is a black-body SED with radius of $8\times10^{10}$ cm and temperature $7\times10^5$ K.  No single spherical black-body SED fits the entire spectrum of m2r1cold well.}
\label{fig:bbSpec_m2r1cold}
\end{figure}

\begin{figure}
\centering
\plotone{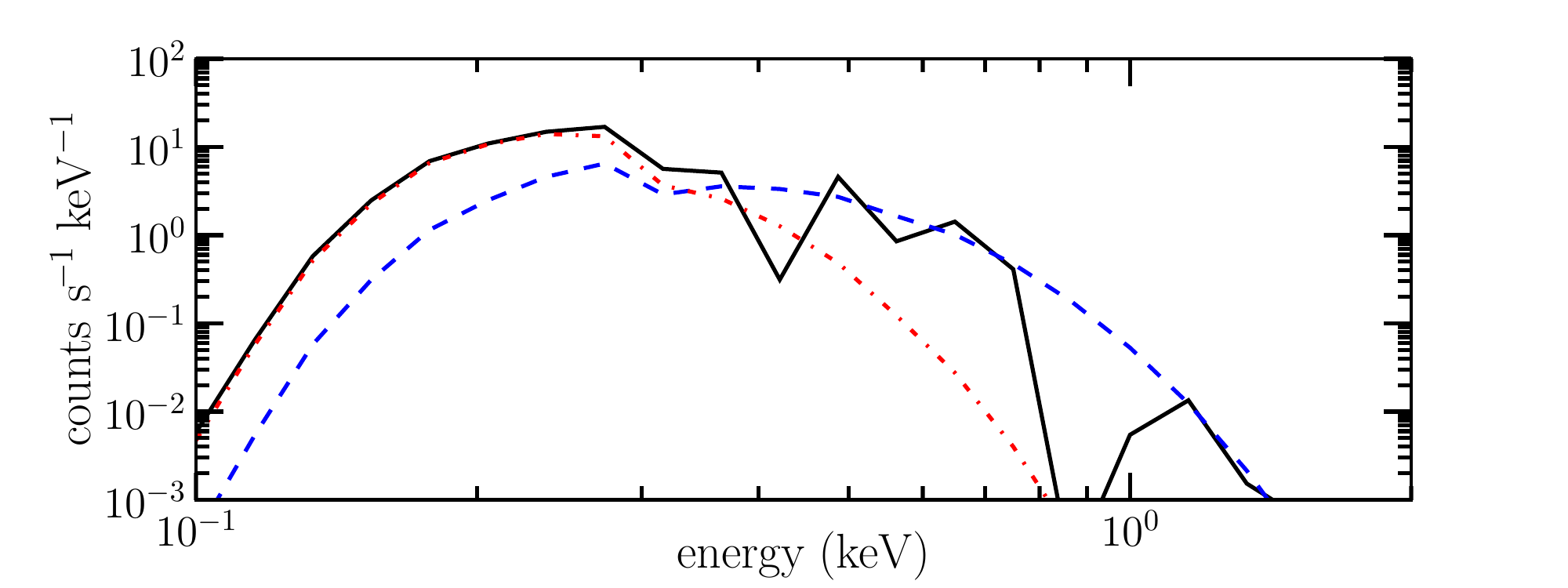}
\caption{X-ray spectrum of spherical explosion model m2r1sph (black line) plotted with two spherical, single temperature and radius black-body SEDs.  The red dash-dot line is a black-body SED with a spherical radius of $5\times10^{11}$ cm and temperature $5\times10^5$ K.  The blue dashed line is a black-body SED with a radius of $7\times10^{10}$ cm and temperature $1\times10^6$ K.  Despite being a spherical explosion, no single spherical black-body SED fits the spectrum well.  This is because the radius of the emitting region in the explosion model is changing rapidly and no single color temperature can describe the emission accurately.}
\label{fig:bbSpec_sph}
\end{figure}

The shapes of the spectral energy distributions of our simulated shock breakout models are not well-fit by single- temperature and radius spherical black-bodies.  This is the case for both the jet-driven explosion simulations and the spherical explosion models.  Figures \ref{fig:bbSpec_m2r1cold} and \ref{fig:bbSpec_sph} show the X-ray spectra of models m2r1cold and m2r1sph along with spherical single-temperature and radius black-body SEDs corrected for XRT detector response and X-ray absorption.  As is shown, no single temperature and radius black-body can fit both the soft and hard parts of the spectrum in either the jet-driven or spherical explosions.  This demonstrates that, regardless of shock break out geometry, such simple black-body models are not applicable.  Shock breakout is too dynamic a process to be modeled by a single average radius and a single average temperature black-body, even for spherical breakouts.  Simple dynamical models with radii expanding at constant velocities and temperatures cooling adiabatically could provide vast improvements over simple static models and could provide a better relation to the physical process of shock breakout.

\section{Discussion and Conclusions}
\label{sec:discussion}

We have modeled the emission arising from the breakout of an aspherical supernova shock assuming that the emission is thermal in nature.  Our hydrodynamic simulations do not account for radiative effects such as disparate ion and radiation temperatures, radiative pre-acceleration of gas ahead of the shock, escape of photons from the shock at optical depths greater than 1, etc.  Despite the limitations of our simulations, we have demonstrated that in the more general case of a non-spherical explosion, the breakout emission is dramatically different from that expected from a spherically-symmetric shock breakout.  One of the most important results of this work is that the timescales of the light curves for aspherical shock breakout are not set by the light travel time across the progenitor star, but are instead related to the much longer shock crossing time of the progenitor.  Thus, shock breakout light curves contain information about the geometry of the shock structure as well as the radius of breakout.  We also show that for aspherical shock breakouts, the observer's viewing angle can play an important role in determining the shape of the observed light curve.  These general results apply to any, arbitrarily aspherical supernova, not just a jet-driven, bipolar supernova.  

The fiducial emission models are generally under-luminous and have soft spectra compared to the actual observations.  \citet{Tanaka:09} advocate explosions with higher energies than what we have simulated and \citet{Soderberg:08} posit that the wind around the progenitor of SN 2008D was optically-thick, thus very dense.  Either an increased explosion energy or increased wind density could increase the X-ray luminosity of shock breakout; however, even if the wind and explosion energies were increased to give luminosities commensurate with those inferred for XRO 080109, the spectrum would still remain lacking in sufficient luminosity above 2 keV to match the observed spectrum (see Figure \ref{fig:m2r1cold_ht}).  This seems to indicate the need for a scattering of the thermal X-ray photons to higher energies, as prescribed by \citet{Soderberg:08} and \citet{WangX:08}, or other, non-thermal processes \citep{Katz:09,Nakar:10}.  This could serve to harden our simulated spectra and also lengthen the burst time since, in our models, the later parts of the light curve are made up of a larger fraction of soft, highly-absorbed photons.

The simulated spectra may also be hardened by relaxing the assumption of thermal equilibrium between the matter and radiation.  During shock breakout from a massive, compact Wolf-Rayet progenitor the matter and radiation are not in LTE \citep{Katz:09, Nakar:10}.  Our method does not account for non-LTE effects.  As discussed by \citet{Nakar:10}, if the matter and radiation are not in LTE, the observed color temperature could be much greater than the matter temperature at $\tau_* \approx 2/3$.  Because the internal energy would not be similarly enhanced above the LTE case, the bolometric luminosities for non-LTE breakouts will not depend strongly on the coupling between matter and radiation, as the color temperature does.  The color-temperature enhancing effects in non-LTE breakouts discussed by \citet{Nakar:10} are strongly time-dependent with the greatest difference in the temperatures of radiation and matter being at the instant of shock breakout.  If non-LTE effects, such as described by \citet{Nakar:10}, were included in our simulations, the spectra would be significantly hardened, especially at early times in the breakout, and the X-ray luminosities modestly enhanced.  This could account for the missing hard X-ray emission in our simulated spectra and increase the XRT count rates without the need to invoke a very dense wind or a greatly enhanced explosion energy.  We note, however, that \citet{Nakar:10} do not include the effects of a wind surrounding the progenitor.  In our calculations, the presence and character of the wind is an integral factor in the breakout emission formation since the thermalization depths in our simulations lie at, or ahead of, the contact discontinuity between the ejecta and the wind.

\begin{figure}
\centering
\plotone{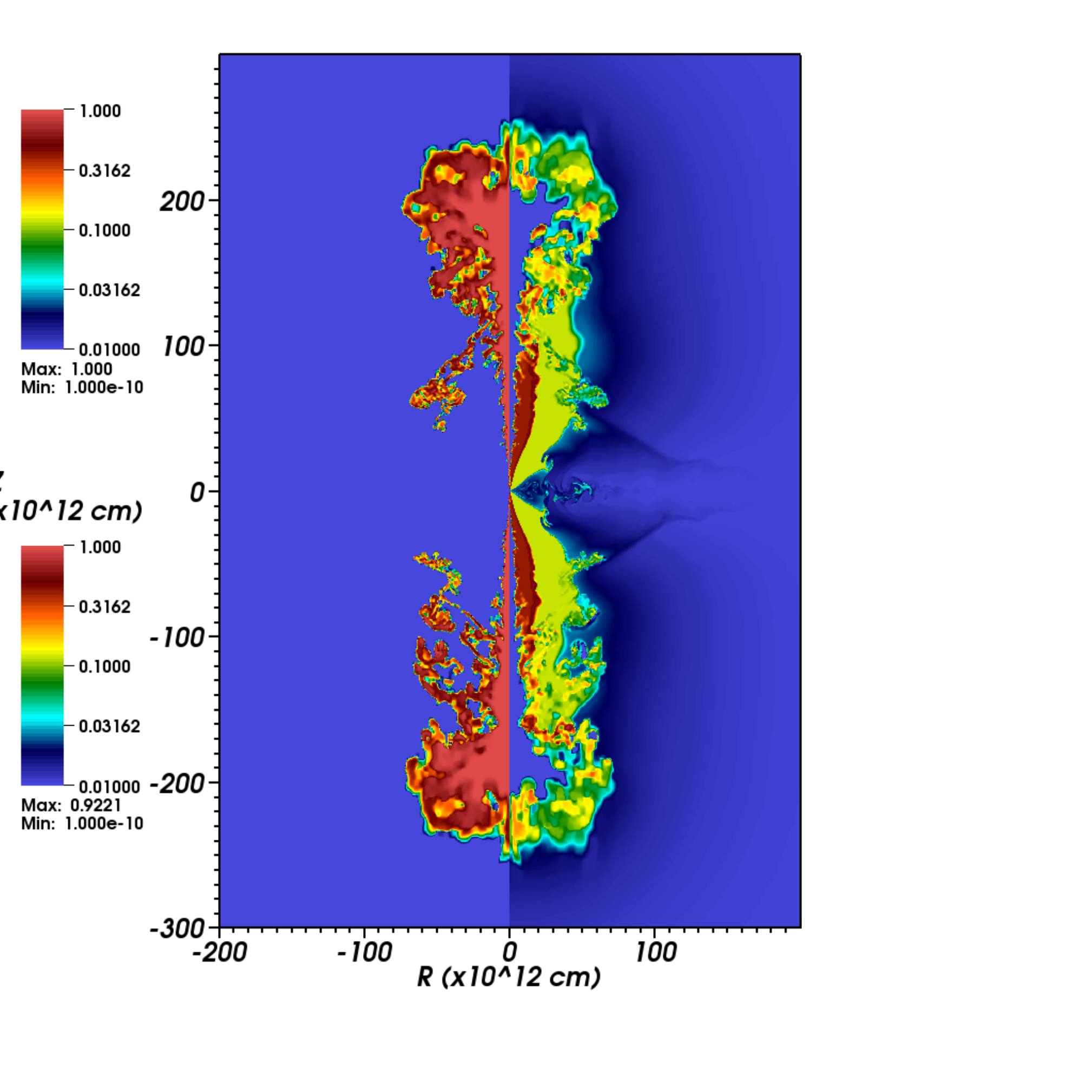}
\caption{Nickel (left) and oxygen (right) mass fractions for model m2r1cold at the end of the simulation ($10^5$ seconds).  Both distributions show extension along the symmetry axis with a great deal of small scale, turbulent structure.  The oxygen distribution also shows cavities caused by the jets.}
\label{fig:m2r1cold_NiO}
\end{figure}

The presence of a double-peaked oxygen line in the spectra of SN 2008D at about 109 days may be evidence for asphericity in the explosion \citep{Modjaz:09}.  \citet{Modjaz:09} suggest that this may be evidence for a ring-like distribution of oxygen, similar to the conclusions reached by \citet{Modjaz:08} for several other Type Ib/c supernovae with double-peaked oxygen lines.\footnote{This interpretation of double-peaked oxygen lines is somewhat controversial.  \citet{Milisavljevic:10} argue that this feature is caused by an oxygen doublet at these wavelengths.}  A double-peaked oxygen line may also result from an asymmetric distribution of radioactive cobalt powering the excitation of oxygen \citep{Modjaz:08, Gerardy:00}.  If the cobalt distribution were aspherical, however, other radioactively-excited lines would be expected to show a similar double-peaked behavior, which is not the case for SN 2008D or the Type Ib/c SNe discussed in \citet{Modjaz:08}.  \citet{Tanaka:09} find that a fraction of nickel must be mixed outward to adequately model the spectra of SN 2008D, further indicating asphericity of the explosion.  Figure \ref{fig:m2r1cold_NiO} shows the nickel and oxygen mass fractions from our model m2r1cold at $10^5$ seconds.  In our simulations, both the nickel and oxygen distributions are aspherical, as is shown in Figure \ref{fig:m2r1cold_NiO}.  This may be able to account for the double-peaked oxygen line in the spectra of SN 2008D, however detailed spectral synthesis calculations are needed to be sure.  Note that our oxygen distribution is prolate, not ring-like as \citet{Modjaz:08} recommend for explaining double-peaked oxygen lines in SNe.

The distribution of intermediate-mass elements such as oxygen we find in our simulations differs from what some other groups find in similar studies of aspherical SNe.  \citet{Maeda:02} present hydrodynamic simulations of aspherical CCSNe targeted to explaining the observations of SN 1998bw.  They show that the intermediate elements are ejected from the explosion in an equatorial torus.  We find that the intermediate elements are ejecta in a bipolar geometry, as shown in Figure \ref{fig:m2r1cold_NiO}.  The difference in the two results comes from the manner in which the explosions are initiated.  In \citeauthor{Maeda:02}, the explosions are started by depositing kinetic energy asymmetrically in the center of the progenitor.  This pushes the intermediate elements outward while simultaneously the more energetic material near the poles compresses them into a toroidal geometry.  In our simulation the explosions are driven entirely by the bipolar jets.  The intermediate-mass material near the progenitor's equator is accreted into the central engine.  Intermediate elements are entrained in the jets and carried out into a configuration that resembles that of the jets themselves.  The equatorial torus in our simulations is comprised primarily of helium. The final distribution of intermediate mass elements thus depends on the mode of asymmetric energy input. Determining the distribution observationally may thus help to constrain models.

\citet{Maund:09} present early spectropolarimetric observations of SN 2008D.  They find that the continuum polarization is relatively small, indicating that the supernova photosphere may be only slightly aspherical.  They also find that there is significant polarization in certain spectral lines, indicating that the line-forming regions of various elements are markedly aspherical.  This is in qualitative agreement with the results of our simulations.  The photospheres at late times are nearly round, while the detailed composition structure is dramatically asymmetric, as shown in Figure \ref{fig:m2r1cold_NiO}.  A late-time shock structure that is nearly round, or at least not dramatically aspherical, is also consistent with the radio measurements of \citet{Bietenholz:09}.
 
The number of observed supernova shock breakouts has been increasing, and this trend is likely to continue, opening a new window for exploring core-collapse supernova through observation.  A multidimensional interpretation of these observations will be critically important to gaining an accurate understanding of supernova shock breakout.  We find that aspherical shock breakout can account for the light curve time scales observed for XRO 080109/SN 2008D without requiring an extremely dense wind or abnormally large Wolf-Rayet progenitor star.  Our models with the smaller progenitor, m2r1cold and m2r1hot, have light curves with FWHMs of about 100 seconds, roughly matching XRO 080109.  We find that our models are generally under-luminous and have spectra too soft to match XRO 080109; however this is likely due to our assumption of LTE.  The shock breakout from a WR progenitor is not in LTE and the radiation temperature in the non-LTE case could be significantly enhanced above the matter temperature \citep{Nakar:10}.  Accounting for non-LTE effects, then, could brighten our simulated light curves and harden the simulated spectra.  Inclusion of non-LTE effects will be presented in future work.  

\acknowledgements
The authors thank Peter H\"oflich and Ehud Nakar for helpful discussion concerning the physics of shock breakout.  S.M.C. is very grateful for support provided through NASA Earth and Space Science Fellowship program.  J.C.W. is supported in part by NSF grant AST-0707769 and in part by DOE funding to the Center for Laser Experimental Research (CLEAR) through the University of Michigan.  The software used in this work was in part developed by the DOE-supported ASC/Alliance Center for Astrophysical Thermonuclear Flashes at the University of Chicago.  The authors acknowledge the Texas Advanced Computing Center (TACC) at The University of Texas at Austin for providing high-performance computing, visualization, and data storage resources that have contributed to the research results reported within this paper.

\bibliography{References}

\end{document}